\documentclass{aa}

\pdfoutput=1
\usepackage{natbib}
\bibpunct{(}{)}{;}{a}{}{,} 

\usepackage{latexsym,exscale,amssymb,amsmath}
\usepackage{eurosym}
\usepackage{color}
\usepackage{graphicx}
\usepackage[percent]{overpic}
\usepackage{hyperref}

\mathchardef\mhyphen="2D

\begin{document}

\title{\bf [C\,{\sc ii}] emission from L1630 in the Orion B molecular cloud}
\author{C. H. M. Pabst\inst{\ref{inst1}} \and J. R. Goicoechea\inst{\ref{inst2}} \and D. Teyssier\inst{\ref{inst3}} \and O. Bern\'{e}\inst{\ref{inst4}} \and B. B. Ochsendorf\inst{\ref{inst5}} \and M. G. Wolfire\inst{\ref{inst6}} \and R. D. Higgins\inst{\ref{inst7}} \and D. Riquelme\inst{\ref{inst10}} \and C. Risacher\inst{\ref{inst10}} \and J. Pety\inst{\ref{inst8},\ref{inst9}} \and F. Le Petit\inst{\ref{inst9}} \and E. Roueff\inst{\ref{inst9}} \and E. Bron\inst{\ref{inst2},\ref{inst9}} \and A. G. G. M. Tielens\inst{\ref{inst1}} }

\institute{Leiden Observatory, Leiden University, P.O. Box 9513, 2300 RA Leiden, Netherlands\label{inst1}; \href{mailto:pabst@strw.leidenuniv.nl}{\texttt{pabst@strw.leidenuniv.nl}}
\and ICMM-CSIC, Calle Sor Juana Ines de la Cruz 3, 28049 Cantoblanco, Madrid, Spain \label{inst2} 
\and Herschel Science Center, ESA/ESAC, P.O. Box 78, Villanueva de la Ca\~{n}ada, 28691 Madrid, Spain \label{inst3}
\and CNRS, IRAP, 9 Av. colonel Roche, BP 44346, 31028 Toulouse Cedex 4, France \label{inst4}
\and Department of Physics and Astronomy, The Johns Hopkins University, 3400 North Charles Street, Baltimore, MD 21218, USA\label{inst5}
\and Department of Astronomy, University of Maryland, College Park, MD 20742, USA\label{inst6}
\and I. Physikalisches Institut der Universit\"{a}t zu K\"{o}ln, Z\"{u}lpicher Strasse 77, 50937 K\"{o}ln, Germany\label{inst7}
\and Max-Planck-Institut f\"{u}r Radioastronomie, Auf dem H\"{u}gel 69, 53121 Bonn, Germany\label{inst10}
\and IRAM, 300 rue de la Piscine, 38406 Saint Martin d'H\`{e}res, France \label{inst8}
\and LERMA, Observatoire de Paris, PSL Research University, CNRS, Sorbonne Universit\'{e}s, UPMC Univ. Paris 06, F-75014, Paris, France \label{inst9}
}
\date{Received 28 March 2017 / Accepted 12 June 2017}

\abstract{{\bf\sf Context}: L1630 in the Orion B molecular cloud, which includes the iconic Horsehead Nebula, illuminated by the star system $\sigma$ Ori, is an example of a photodissociation region (PDR). In PDRs, stellar radiation impinges on the surface of dense material, often a molecular cloud, thereby inducing a complex network of chemical reactions and physical processes.\\
{\bf\sf Aims}: Observations toward L1630 allow us to study the interplay between stellar radiation and a molecular cloud under relatively benign conditions, that is, intermediate densities and an intermediate UV radiation field. Contrary to the well-studied Orion Molecular Cloud 1 (OMC1), which hosts much harsher conditions, L1630 has little star formation. Our goal is to relate the [C\,{\sc ii}] fine-structure line emission to the physical conditions predominant in L1630 and compare it to studies of OMC1.\\
{\bf\sf Methods}: The [C\,{\sc ii}] $158\,\mu\mathrm{m}$ line emission of L1630 around the Horsehead Nebula, an area of $12\arcmin \times 17\arcmin$, was observed using the upgraded German Receiver for Astronomy at Terahertz Frequencies (upGREAT) onboard the Stratospheric Observatory for Infrared Astronomy (SOFIA).\\
{\bf\sf Results}: Of the [C\,{\sc ii}] emission from the mapped area 95\%, $13\,L_{\sun}$, originates from the molecular cloud; the adjacent H\,{\sc ii} region contributes only 5\%, that is, $1\,L_{\sun}$. From comparison with other data (CO$\,(1\mhyphen 0)$-line emission, far-infrared (FIR) continuum studies, emission from polycyclic aromatic hydrocarbons (PAHs)), we infer a gas density of the molecular cloud of $n_{\mathrm{H}}\sim 3\cdot 10^3\,\mathrm{cm^{-3}}$, with surface layers, including the Horsehead Nebula, having a density of up to $n_{\mathrm{H}}\sim 4\cdot 10^4\,\mathrm{cm^{-3}}$. The temperature of the surface gas is $T\sim 100\,\mathrm{K}$. The average [C\,{\sc ii}] cooling efficiency within the molecular cloud is $1.3\cdot 10^{-2}$. The fraction of the mass of the molecular cloud within the studied area that is traced by [C\,{\sc ii}] is only $8\%$. Our PDR models are able to reproduce the FIR-[C\,{\sc ii}] correlations and also the CO$\,(1\mhyphen 0)$-[C\,{\sc ii}] correlations. Finally, we compare our results on the heating efficiency of the gas with theoretical studies of photoelectric heating by PAHs, clusters of PAHs, and very small grains, and find the heating efficiency to be lower than theoretically predicted, a continuation of the trend set by other observations.\\
{\bf\sf Conclusions}: In L1630 only a small fraction of the gas mass is traced by [C\,{\sc ii}]. Most of the [C\,{\sc ii}] emission in the mapped area stems from PDR surfaces. The layered edge-on structure of the molecular cloud and limitations in spatial resolution put constraints on our ability to relate different tracers to each other and to the physical conditions. From our study, we conclude that the relation between [C\,{\sc ii}] emission and physical conditions is likely to be more complicated than often assumed. The theoretical heating efficiency is higher than the one we calculate from the observed [C\,{\sc ii}] emission in the L1630 molecular cloud.
\vspace{0.5cm}
}

\titlerunning{[C\,{\sc ii}] emission from L1630}
\maketitle

\section{Introduction}

One of the main challenges of astronomy and cosmology is to model, and reach an understanding, of the evolution of galaxies and large-scale structure. The star-formation rate (SFR) is a crucial parameter in these models. In order to measure the SFR in distant galaxies, several possible tracers have been and are being studied: ultraviolet (UV) radiation, infrared (IR) radiation, emission from polycyclic aromatic hydrocarbons (PAHs), atomic and molecular lines \citep[e.g.,][]{Kennicutt1998, Kennicutt2012}. With the advent of the Atacama Large (sub)Millimeter Array (ALMA), it has become popular to use the [C\,{\sc ii}] $158\,\mu\mathrm{m}$ line as an indicator of the SFR over cosmic history \citep[e.g.,][]{Herrera-Camus, Vallini, Pentericci}. However, the origin of [C\,{\sc ii}] emission on a galactic scale is still unclear.

Intuitively, the SFR is expected to depend on the local conditions in the interstellar medium (ISM), the gas and dust that form the environment of stars. The ISM comes in different phases, diffuse gas being the most prevalent. These phases are the cold neutral medium (CNM) with moderate gas densities, $n\sim 30\,\mathrm{cm}^{-3}$, and moderate gas temperatures, $T\sim 100\,\mathrm{K}$, the warm neutral and warm ionized medium (WNM and WIM) with low densities and high temperatures, $n\sim 0.3\,\mathrm{cm}^{-3}$ and $T\sim 8000\,\mathrm{K}$, and the hot ionized medium (HIM) with very low densities and very high temperatures, $n\sim 3\cdot 10^{-3}\,\mathrm{cm}^{-3}$ and $T\sim 10^6\,\mathrm{K}$. Most of the gas of the ISM is in the neutral phase. Other ubiquitous components of the ISM are H\,{\sc ii} regions around massive stars with densities ranging from $n\sim 1\,\mathrm{cm}^{-3}$ to $n\sim 10^5\,\mathrm{cm}^{-3}$ and $T\sim 10^4\,\mathrm{K}$, and molecular clouds with high density and low temperatures, $n\sim 10^3\mhyphen10^8\,\mathrm{cm}^{-3}$ and $T\sim 10\mhyphen30\,\mathrm{K}$ \citep{HollenbachTielens}. These phases are not isolated from each other, but there is an exchange of matter between them, particularly driven by ionization, winds, and explosions of massive stars. Molecular clouds especially are the birthplaces of new (massive) stars and thereby of vital interest. \cite{Meyer} provide a review of star formation in L1630. At the interface between an H\,{\sc ii} region, ionized by a massive star, and a parental molecular cloud, a photodissociation region (PDR) is formed, where intense stellar UV radiation impinges on the surface of the dense cloud. At the surface of these clouds, the gas is atomic; deeper inside the cloud, the molecular fraction increases. The study of PDRs reveals much about the interplay between stars (including hosts of newly formed stars) and the ISM, thereby yielding valuable insight into the process of star formation (see \cite{HollenbachTielens} for a review of PDRs).

The ISM is mainly heated by stellar radiation, specifically by far-ultraviolet (FUV) radiation with energies between 6 and 13.6 eV. The characteristics of the gas cooling allow us to infer the amount and, possibly, the source of the heating. One of the main coolants of the cold neutral medium is the [C\,{\sc ii}] $^2P_{3/2}\mhyphen^2P_{1/2}$ fine-structure line at $\lambda\simeq 158\,\mu\mbox{m}$, that is, $\Delta E/k_B \simeq 91.2\,\mathrm{K}$. The [C\,{\sc ii}] line is also one of the brightest lines in PDRs, carrying up to 5\% of the total far-infrared (FIR) luminosity, the other 95\% mainly stemming from UV irradiated dust. Carbon has an ionization potential of 11.3 eV, hence $\mathrm{C}^+$ traces the transition from H$^+$ to H and H$_2$. Another important coolant is the [O\,{\sc i}] line at $\lambda\simeq63\,\mu\mbox{m}$ ($\Delta E/k_B \simeq 228\,\mathrm{K}$). The ratio of those two main coolants depends on the temperature and density of the gas. For $T=100\,\mathrm{K}$, the [O\,{\sc i}] cooling efficiency overtakes the [C\,{\sc ii}] cooling efficiency at $n\simeq 3\cdot 10^4\,\mathrm{cm}^{-3}$; at $n=3\cdot 10^3\,\mathrm{cm}^{-3}$, the [O\,{\sc i}] contribution to the total cooling is about 5\% \citep[cf.][]{Tielens}.

The [C\,{\sc ii}] line has been studied in a variety of environments. Important contributions may come from diffuse clouds (CNM), dense PDRs, surfaces of molecular clouds, and (low-density) ionized gas including the WIM (e.g., \cite{Wolfire1995}, \cite{Ossenkopf}, \cite{Gerin}). \cite{Langer} identify warm CO-dark molecular gas in Galactic diffuse clouds by means of [C\,{\sc ii}] emission. \cite{Jaffe} conducted an earlier study observing the extended [C\,{\sc ii}] emission from the Orion B molecular cloud (L1630). This study is preceded by a [C\,{\sc ii}] survey of the Orion Molecular Cloud 1 (OMC1) in Orion A by \cite{Stacey1989}. \cite{Goico} present a velocity-resolved [C\,{\sc ii}] map toward OMC1, observed by the Heterodyne Instrument for the Far-Infrared (HIFI) onboard the {\it Herschel} satellite in 2012. Velocity-resolved [C\,{\sc ii}] and [$^{13}$C\,{\sc ii}] emission from the star-forming region NGC 2024 in L1630 was observed in 2011 using the GREAT (German Receiver for Astronomy at Terahertz Frequencies) instrument onboard the airborne Stratospheric Observatory for Infrared Astronomy (SOFIA) and analyzed by \cite{Graf}; the neighboring reflection nebula NGC 2023 was observed in 2013/14 using the same instrument. \cite{Sandell2015} discussed the physical conditions, morphology, and kinematics of that region. A theoretical study on collisional excitation of the [C\,{\sc ii}] fine-structure transition was performed by \cite{Goldsmith2012}. The GOT C$^+$ survey (Galactic Observations of Terahertz C$^+$) survey \citep{Pineda}, also a Herschel/HIFI study, investigated specifically the relationship between [C\,{\sc ii}] luminosity and SFR. This study found a good correlation on Galactic scales. This was also established by \cite{Stacey2010} and \cite{Herrera-Camus} at low and high redshift.

On December 11, 2015, a part of the Orion B molecular cloud, including the Horsehead Nebula, was observed in [C\,{\sc ii}] with the upGREAT instrument, the first multi-pixel extension of GREAT, onboard SOFIA, as presented and described in \cite{Risacher}. The survey was conducted "to demonstrate the unique and important scientific capabilities of SOFIA, and to provide a publicly available high-value SOFIA data set".\footnote{\texttt{https://www.sofia.usra.edu/science/proposing-and-\\observing/proposal-calls/sofia-directors-discretionary-\\time/horsehead-nebula.}} It allows us to study [C\,{\sc ii}] emission and its correlations with other astrophysical tracers under moderate conditions (intermediate density and moderate UV-radiation field), as opposed to the high density and intense UV-radiation field in OMC1.

In the present study, we analyze the [C\,{\sc ii}] emission from a $12\arcmin\times 17\arcmin$ area of the L1630 molecular cloud in Orion B that is illuminated by the nearby star system, $\sigma$ Ori. Our distance to the star system is approximately $334\,\mathrm{pc}$, which we also assume to be the distance to the molecular cloud. The projected distance between the star system and the molecular cloud is $3.2\,\mathrm{pc}$ (\cite{Ochsendorf} and references therein). Part of the mapped area, in which star formation is low, is the well-known Horsehead Nebula. The star-forming regions NGC 2023 and NGC 2024 are adjacent to the mapped area, but not included. We compare the velocity-resolved [C\,{\sc ii}] SOFIA/upGREAT observations with new CO$\,(1\mhyphen 0)$ observations of the molecular gas obtained with the $30\,\mathrm{m}$ telescope \citep{Pety2017} at the Institut de Radioastronomie Millim\'{e}trique (IRAM), with {\it Spitzer}/Infrared Array Camera (IRAC) studies of the PAH emission from the PDR surfaces, H$\alpha$ observations of the ionized gas, and with existing far-infrared continuum studies using {\it Herschel}/Photoconductor Array Camera and Spectrometer (PACS) and Spectral and Photometric Imaging Receiver (SPIRE) data to determine dust properties and trace the radiation field. This wealth of data allows us to separate emission from the ionized gas, neutral PDR, and molecular cloud, in order to derive global heating efficiencies and their dependence on the local conditions, and to make detailed comparisons to PDR models.

This paper is organized as follows. In Section 2, we present the observations. In Section 3, we divide the surveyed area into regions with specific characteristics. Furthermore, we study the kinematics of the gas as revealed by the SOFIA/upGREAT observations of [C\,{\sc ii}] emission and the correlation of the various data sets with each other. Section 4 contains a discussion of the results obtained in Section 3 and we derive column densities and other gas properties. We conclude with a summary of our results and an outlook for the future in Section 5.

\section{Observations}

\subsection{[C\,{\sc ii}] Observations} 
The [C\,{\sc ii}] emission in Orion B (L1630) was observed on December 11, 2015 using the upGREAT instrument onboard SOFIA. The region was observed using the upGREAT optimized on-the-fly mapping mode. The region was split into four tiles, each covering an area of $363\arcsec\times 508.2\arcsec$. In this mode, the array is rotated $19.1^{\circ}$ on the sky and an on-the-fly (OTF) scan is undertaken. By performing a second scan separated by $5.5\arcsec$ perpendicular to the scan direction, it is possible to fully sample a region $72.6\arcsec$ wide along the scan direction (cf. \cite{Risacher} for details). By combining OTF scans in the RA and Dec direction, it is possible to cover the map region with multiple pixels. Each tile was made up of ten $x$-direction OTF scans and 14 $y$-direction OTF scans. A spectrum was recorded every $6\arcsec$. Scans in the $x$-direction had an integration time of 0.4 s, while those in the $y$-direction had an integration time of 0.3 s. Since the $y$-scan length is longer than the $x$-scan length, the integration time was reduced. This is due to Allan variance stability time limits of 30 seconds. A reference position at about $12\arcmin$ to the west of the map was observed, which was verified to be free of $^{12}$CO$\,(2\mhyphen 1)$ and $^{13}$CO$\,(2\mhyphen 1)$ emission with the James Clerk Maxwell Telescope (JCMT) (G. Sandell, priv. comm.). The reference position was checked to be free of [C\,{\sc ii}] emission to a $1\,\mathrm{K}$ level. A supplementary {\sc off} contamination check was undertaken whereby the reference position was calibrated using the internal {\sc hot} reference measurement; these spectra also showed no evidence of {\sc off} emission to a $1\,\mathrm{K}$ level. The [C\,{\sc ii}] map itself, showing no "absorption" features anywhere, confirms that there cannot be notable [C\,{\sc ii}] emission at the reference position. An {\sc off} measurement is ideally taken after 30 s of {\sc on} source integration to avoid drift problems in the calibrated data. For this observing run, each tile was observed twice in the $x$- and $y$-directions, resulting in a total integration time per map pixel of 1.4 s. For a spectral resolution of $0.19\,\mathrm{km\,s}^{-1}$, this results in a noise rms in the final data cube of $2\,\mathrm{K}$ in the velocity channels free of emission.

\begin{figure}[ht]
\includegraphics[width=0.5\textwidth, height=0.5\textwidth]{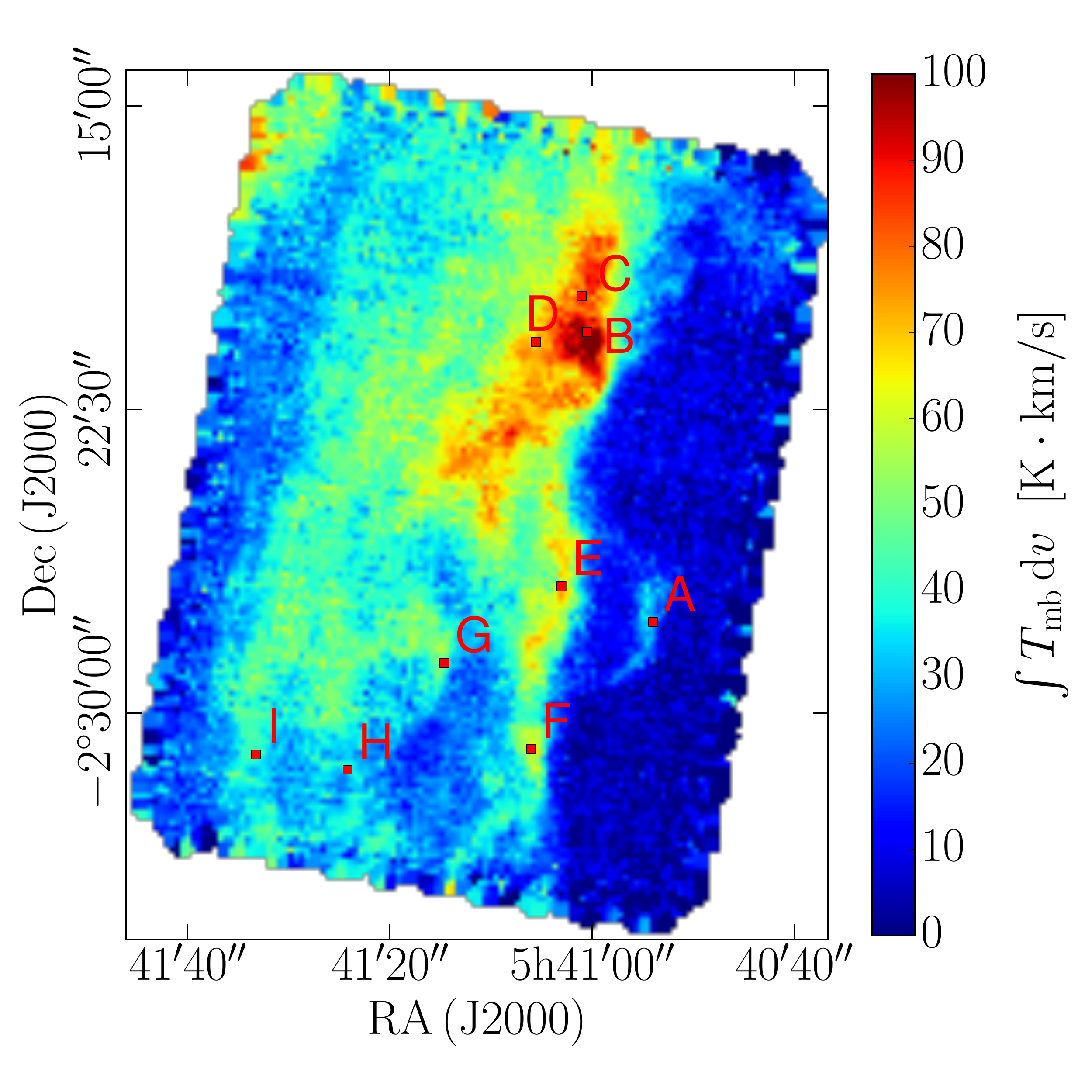}
\caption{[C\,{\sc ii}] line-integrated intensity; points indicate positions where individual spectra are extracted for illustrative purposes (see Sec. \ref{sec.spectra}).}
\label{Fig.Cplus}
\end{figure}

The data cube provided by the SOFIA Science Center was processed using the Grenoble Image and Line Data Analysis Software\footnote{See \texttt{http://www.iram.fr/IRAMFR/GILDAS} for more information about the GILDAS softwares~\citep{pety05a}.}/Continuum and Line Analysis Single-dish Software (GILDAS/CLASS). We subtract a baseline of order one from the spectra. The spectral data were integrated over the velocity range (with respect to the Local Standard of Rest, LSR) $v_{\mathrm{LSR}} = 6\mhyphen20\,\mathrm{km\,s^{-1}}$ to obtain the line-integrated intensity, which is shown in Fig. \ref{Fig.Cplus}. Channel maps are shown in Fig. \ref{Fig.channels}. The spatial resolution of our final maps is $15.9\arcsec$. For comparison with other tracers, we use a Gaussian kernel for convolution. At the rim of the map, the [C\,{\sc ii}] signal suffers from noise and we ignore an outer rim of $45\arcsec$ in our analysis.

\begin{figure*}[htb]
\begin{minipage}{0.49\textwidth}
\includegraphics[width=\textwidth, height=\textwidth]{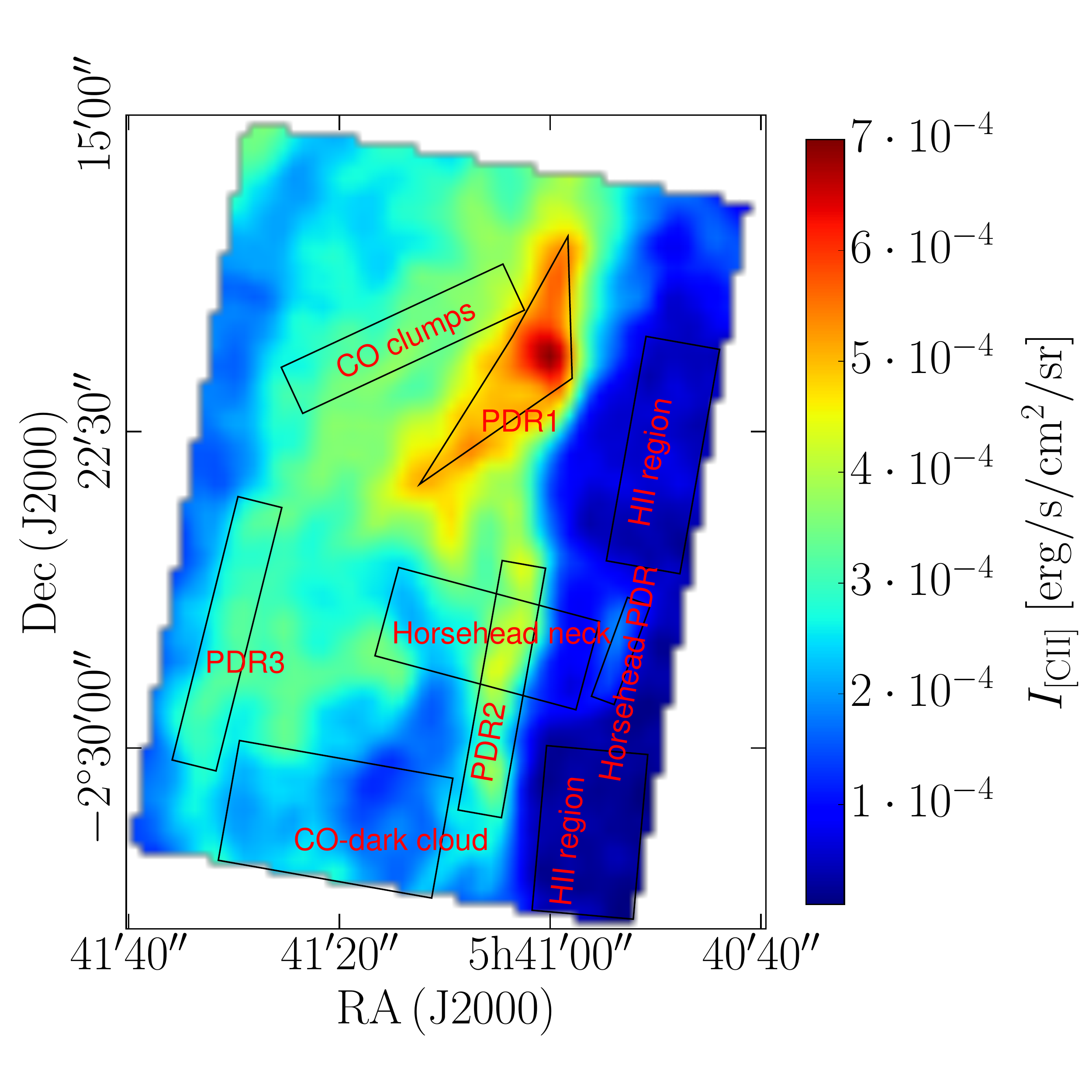}
\caption{[C\,{\sc ii}] line-integrated intensity convolved to $36\arcsec$ resolution with selected regions (see Sec. \ref{sec.GlobMorph}) indicated.}
\label{Fig.Cplusrepro}
\end{minipage}
\begin{minipage}{0.49\textwidth}
\includegraphics[width=\textwidth, height=\textwidth]{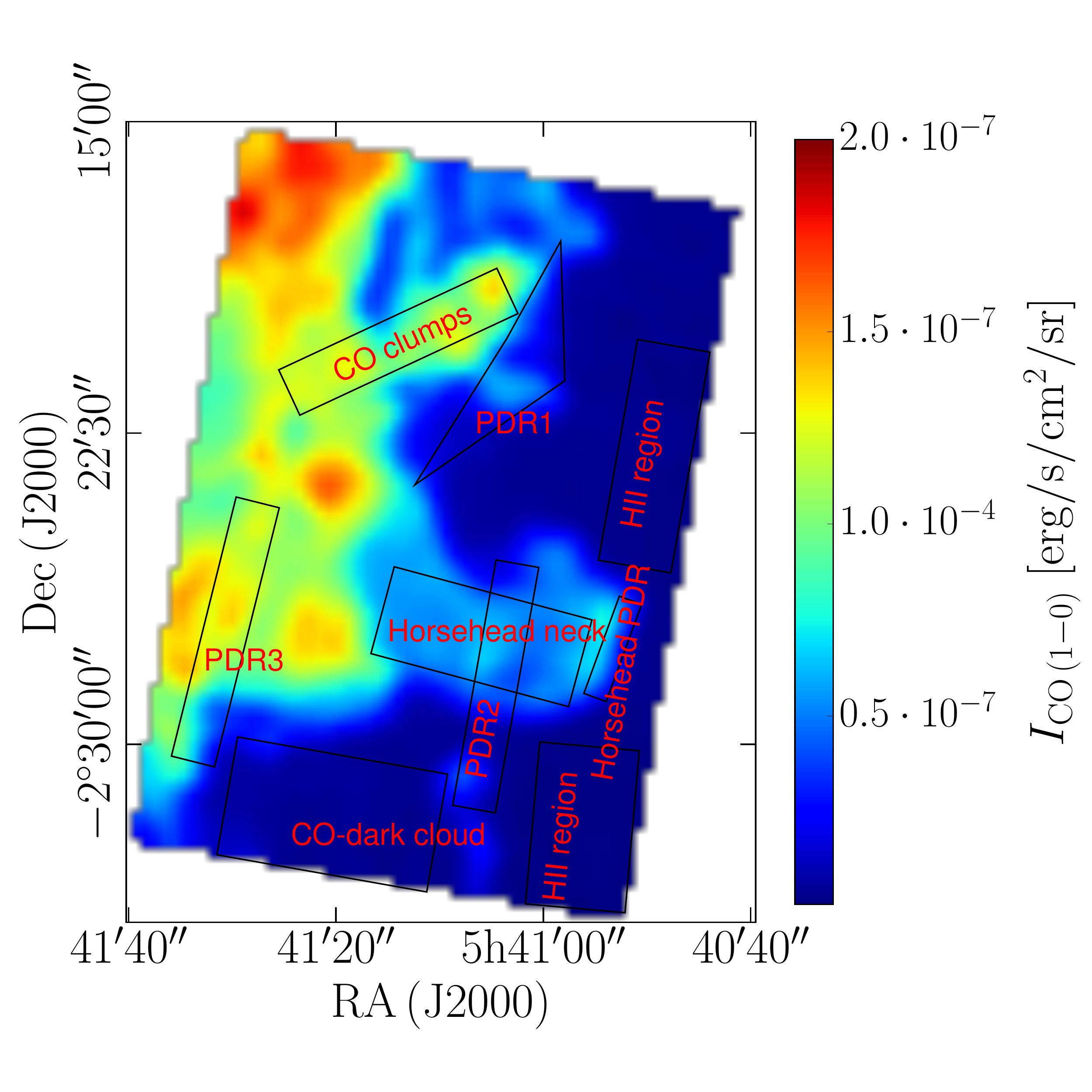}
\caption{CO$\,(1\mhyphen 0)$ line-integrated intensity convolved to $36\arcsec$ resolution with selected regions (see Sec. \ref{sec.GlobMorph}) indicated.}
\label{Fig.CO}
\end{minipage}
\end{figure*}

\subsection{Dust SED Analysis}
\label{sec.SED}
In this study we make use of the dust temperature and dust optical depth maps released by \cite{Lombardi}. \cite{Lombardi} fit a spectral energy distribution (SED) to {\it Herschel}/PACS and SPIRE observations of the Orion molecular cloud complex in the PACS $100\,\mu\mathrm{m}$ and $160\,\mu\mathrm{m}$, and SPIRE $250\,\mu\mathrm{m}$, $350\,\mu\mathrm{m,}$ and $500\,\mu\mathrm{m}$ bands. The photometric data, convolved to the SPIRE $500\,\mu\mathrm{m}$ $36\arcsec$ resolution, are modeled as a modified blackbody,
\begin{align}
I(\lambda) = B(\lambda,T_{\rm d})\, \tau_0\left(\frac{\lambda_0}{\lambda}\right)^{\beta} \label{eq.I},
\end{align}
with $T_{\rm d}$ the effective dust temperature, $\tau_0$ the dust optical depth at the reference wavelength $\lambda_0$, and $\beta$ the grain-emissivity index. \cite{Lombardi} use the all-sky $\beta$ map with $35\arcmin$ resolution by the Planck collaboration, interpolated to the grid on which the SED is performed; only the effective dust temperature and $\tau_0$ are free parameters in this fit. The $\beta$ map shows a smooth increase of about 3\% from the north-east to the south-west in the area surveyed in [C\,{\sc ii}], with a mean of 1.56. Lombardi et al. present their dust optical depth map at $\lambda_0=850\,\mu\mathrm{m}$, following the Planck standard, but for our analysis we convert $\tau_{850}$ to $\tau_{160}$ using the $\beta$ data. We integrate Eq. (\ref{eq.I}) from $\lambda_{\mathrm{min}}=20\,\mu\mathrm{m}$ to $\lambda_{\mathrm{max}}=1000\,\mu\mathrm{m}$ to obtain the far-infrared intensity $I_{\mathrm{FIR}}$.

We notice that the Horsehead PDR has comparatively low dust temperature in the SED fit, $T_\mathrm{d}\simeq 20\mhyphen22 \,\mathrm{K}$. This could be due to beam dilution. In the models of \cite{Habart2005}, the dust temperature is $T_{\mathrm{d}}\simeq 30\,\mathrm{K}$ at the edge, dropping to $T_{\mathrm{d}}\simeq 22\,\mathrm{K}$ for a hydrogen nucleus gas density of $n_{\mathrm{H}}=2\cdot 10^4\,\mathrm{cm}^{-3}$ within $12\arcsec$, and to $T_{\mathrm{d}}\simeq 13.5\,\mathrm{K}$ for $n_{\mathrm{H}}=2\cdot 10^5\,\mathrm{cm}^{-3}$. Throughout this paper,  by "gas density" we mean the hydrogen nucleus gas density: $n_{\mathrm{H}}=n_{\mathrm{H\,\textsc{i}}}+2\,n_{\mathrm{H_2}}$.

The derived effective dust temperature and dust optical depth can depend significantly on the choice of $\beta$: The temperature can be up to $3\mhyphen4\,\mathrm{K}$ lower if $\beta=2$ instead of $\beta=1.5$; $\tau_{160}$ then increases by a factor of two. The FIR intensity is less sensitive to $\beta$: it only decreases by 10\% for $\beta=2$.

Furthermore, we employ {\it Spitzer}/IRAC observations in the $8\,\mu\mathrm{m}$ band, which is dominated by PAHs but which can be influenced by very small grains. We use a super mosaic image retrieved from the Spitzer Heritage Archive, created October 22, 2012. We also make use of the $850\,\mu\mathrm{m}$ observations from the Submillimetre Common-User Bolometer Array 2 (SCUBA-2) around NGC 2023/2024 first presented by \cite{Kirk} as part of the JCMT Gould Belt Survey (GBS). These trace dense regions within the molecular cloud. However, we do not use the map reduced by the GBS group, but we retrieved the data from the Canadian Astronomy Data Centre (CADC) archive, processed on October 1, 2015.

\subsection{CO$\,(1\mhyphen 0)$ Observations}
In this work we make use of part of the $^{12}$CO$\,(1\mhyphen 0)$ large-scale map at $115.271\,\mathrm{GHz}$ obtained by \cite{Pety2017} with the Eight Mixer Receiver (EMIR) 090 at the IRAM $30\,\mathrm{m}$ telescope. The fully sampled on-the-fly line maps were taken with a channel spacing of $195\,\mathrm{kHz}$ (a velocity resolution of $\sim 0.5\,\mathrm{km\,s}^{-1}$). CO-emission contamination from the reference position was eliminated by adding dedicated frequency-switched line observations of the reference position itself (see \cite{Pety2017} for details). The median noise levels range from 100 to 180\,mK (in the $T_{\rm mb}$ scale) per resolution channel. Here we use the CO$\,(1\mhyphen 0)$ line-integrated intensity map in the $v_{\rm LSR}=9\mhyphen16\,\mathrm{km\,s}^{-1}$ range\footnote{The integration range is truncated at $v_{\rm LSR}=9\,\mathrm{km\,s}^{-1}$ to avoid contamination from a second CO component at $v_{\rm LSR}\sim 5\,\mathrm{km\,s}^{-1}$ \citep[cf.][]{Pety2017}.}, convolved to the $36\arcsec$ angular resolution of SPIRE $500\,\mu\mathrm{m}$. The resulting map is shown in Fig. \ref{Fig.CO}.

\subsection{H$\alpha$ Observations}
In this study we use the H$\alpha$ image of the Horsehead Nebula and its environs in L1630 and the H\,{\sc ii} region IC 434 taken by the Mosaic 1 wide field imager on Kitt Peak National Observatory (KPNO). For calibration of the KPNO image, we use H$\alpha$ data of the Horsehead Nebula collected by the Hubble Space Telescope as part of the Hubble Heritage program. We obtained the image from the archive of the National Optical Astronomy Observatory (NOAO), but it was taken as part of the program presented in \cite{Reipurth1998}.

The bright star at $05\mathrm{h}41\arcmin02.70\arcsec, -02^{\circ}18\arcmin17.77\arcsec$ in the H$\alpha$ image is a foreground star; it is visible in the IRAC $8\,\mu\mathrm{m}$ image, as well. We masked it before convolution, such that it does not show in the convolved images.

\section{Analysis}

\subsection{Kinematics: velocity channel maps}
\label{sec.channels}

\begin{figure*}[htb]
\centering
\includegraphics[width=1\textwidth, height=0.67\textwidth]{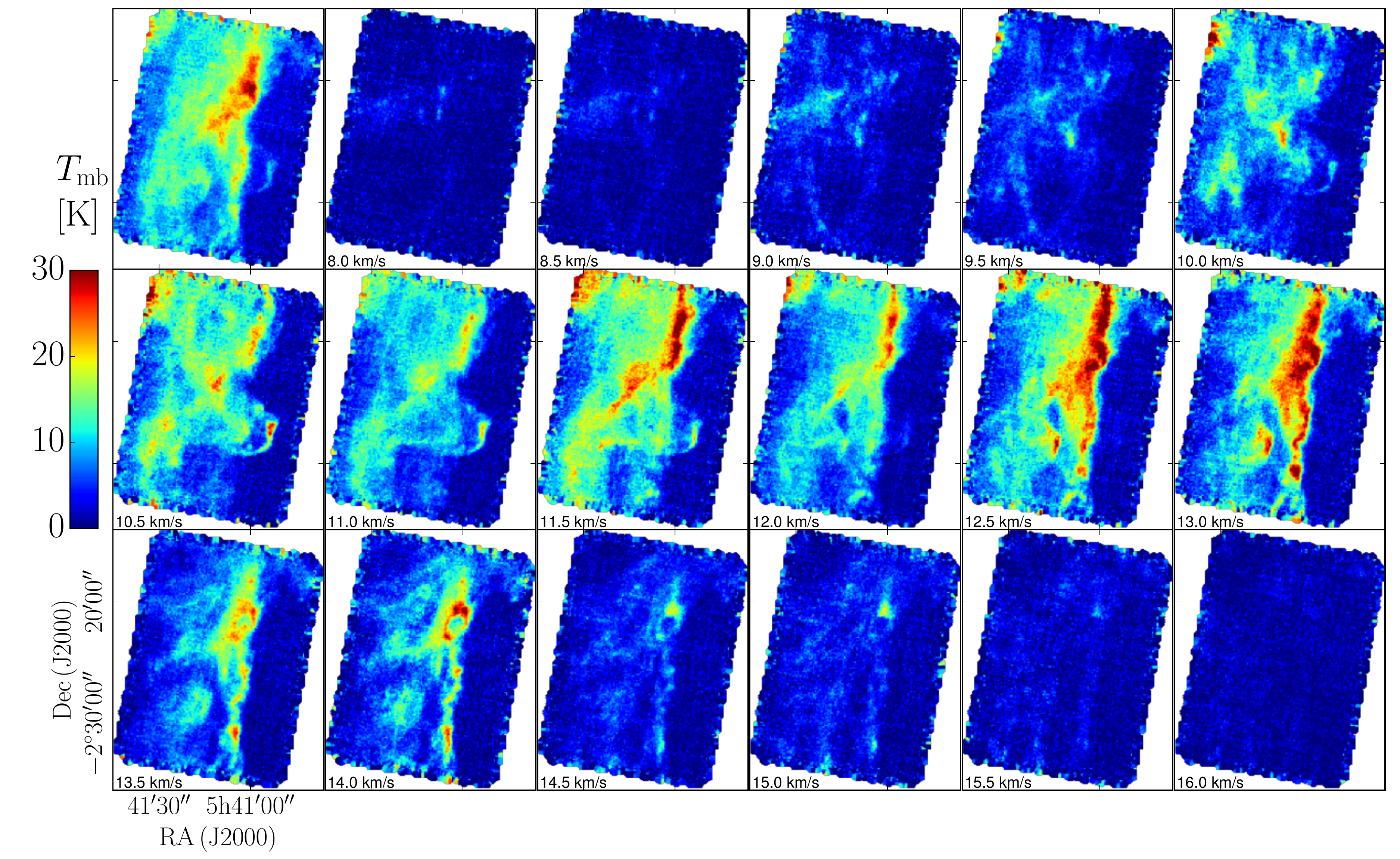}
\caption{[C\,{\sc ii}] channel maps from $8.0\,\mathrm{km\,s^{-1}}$ to $16.0\,\mathrm{km\,s^{-1}}$ in steps of $\mathrm{d}v=0.5\,\mathrm{km\,s^{-1}}$ at $15.9\arcsec$ resolution. The main-beam temperature $T_{\mathrm{mb}}$ is averaged over the step size $\mathrm{d}v$. The first panel shows the line-integrated intensity.}
\label{Fig.channels}
\end{figure*}

Perusing the [C\,{\sc ii}] channel maps from $8.0\,\mathrm{km\,s^{-1}}$ to $16.0\,\mathrm{km\,s^{-1}}$ shown in Fig. \ref{Fig.channels}, we recognize several continuous structures in space-velocity.
From $10.5\mhyphen11.5\,\mathrm{km\,s^{-1}}$, we observe a [C\,{\sc ii}] front that runs from the south-east to the north-west of the map. From $12.5\mhyphen14.0\,\mathrm{km\,s^{-1}}$, a front runs from the north to the south. The Horsehead mane is visible from $10.0\mhyphen11.5\,\mathrm{km\,s^{-1}}$. From $12.5\mhyphen13.0\,\mathrm{km\,s^{-1}}$, an intermediate [C\,{\sc ii}] front lights up. Based on the kinematic behavior and assuming that [C\,{\sc ii}] emission is related to PDR surfaces, we divide the [C\,{\sc ii}] fronts into four groups: PDR1 in the north-west of the molecular cloud, PDR2 in the south-west, PDR3 in the south-east, and the Horsehead PDR. The intermediate PDR front we do not discuss in detail.

In the luminous north, an almost circular cavity forms in the center of the region of highest intensity. Its boundary lights up in the $14.0\,\mathrm{km\,s^{-1}}$ map. In the $12.5\,\mathrm{km\,s^{-1}}$ and $13.0\,\mathrm{km\,s^{-1}}$ channels, we see bright emission where the rim of the cavity is. This cavity appears quite clearly in the unconvolved IRAC $8\,\mu\mathrm{m}$ image (see Fig. \ref{Fig.cuts}). Comparison with the $8\,\mu\mathrm{m}$ map reveals a (proto-)star at the northern edge of the bubble. This star is visible in the H$\alpha$ image as well, but it is not identified as a pre-main-sequence (PMS) object in \cite{Mookerjea}. Here it is listed as MIR-29, a more evolved star in the vicinity of NGC2023, identified by the Two Micron All Sky Survey (2MASS).
In addition to the main emission in the velocity range $v_{\mathrm{LSR}} = 6\mhyphen 20\,\mathrm{km\,s^{-1}}$, we see a faint [C\,{\sc ii}] component at $v_{\mathrm{LSR}} \simeq 5\,\mathrm{km\,s^{-1}}$, that has also been detected in CO observations by \cite{Pety2017}. Due to its faintness, however, we will ignore it in our analysis.

\subsection{Global morphology}
\label{sec.GlobMorph}

\begin{table*}[htb]
\centering
\begin{tabular}{lccccc}
 & $\bar{\eta}$ & $\bar{I}_{\mathrm{[C\,\textsc{ii}]}}$ & $\bar{I}_{\mathrm{[C0\,(1\mhyphen 0)]}}$ & $\bar{I}_{\mathrm{FIR}}$ & $\bar{\tau}_{160}$ \\
Region & [$10^{-2}$] & [$\mathrm{erg\,s^{-1}\,cm^{-2}\,sr^{-1}}$] & [$\mathrm{erg\,s^{-1}\,cm^{-2}\,sr^{-1}}$] & [$\mathrm{erg\,s^{-1}\,cm^{-2}\,sr^{-1}}$] & [$10^{-3}$] \\ \hline \rule{0pt}{3ex}\noindent
L1630 &$1.3(0.5)$& $2.8(1.6)\cdot 10^{-4}$ & $6.5(4.9)\cdot 10^{-8}$ & $2.5(1.2)\cdot 10^{-2}$ & 4.9(6.3)\\ 
Horsehead PDR & $1.0(0.3)$ & $1.5(0.3)\cdot 10^{-4}$ & $4.3(1.5)\cdot 10^{-8}$ & $1.5(0.5)\cdot 10^{-2}$ & 5.1(2.6)\\
PDR1 & $1.1(0.3)$ & $5.5(0.5)\cdot 10^{-4}$ & $3.2(1.6)\cdot 10^{-8}$ & $5.2(0.9)\cdot 10^{-2}$ & 2.2(0.4)\\ 
PDR2 & $2.2(0.4)$ & $3.5(0.6)\cdot 10^{-4}$ & $3.5(1.9)\cdot 10^{-8}$ & $1.7(0.5)\cdot 10^{-2}$ & 2.5(2.0)\\ 
PDR3 & $1.1(0.2)$ & $2.9(0.2)\cdot 10^{-4}$ & $10(3.0)\cdot 10^{-8}$ & $2.9(0.6)\cdot 10^{-2}$ & 6.9(3.7)\\
CO-dark cloud & $1.3(0.1)$ & $1.9(0.4)\cdot 10^{-4}$ & $0.7(0.4)\cdot 10^{-8}$ & $1.7(0.4)\cdot 10^{-2}$ & 0.9(0.2)\\
CO clumps & $1.1(0.1)$ & $3.5(0.4)\cdot 10^{-4}$ & $11(1.6)\cdot 10^{-8}$ & $3.1(0.6)\cdot 10^{-2}$ & 4.5(1.3)\\
H\,{\sc ii} region & $1.9(0.5)$ & $0.6(0.3)\cdot 10^{-4}$ & $0.4(0.2)\cdot 10^{-8}$ & $0.3(0.1)\cdot 10^{-2}$ & 0.7(0.2)\\
\end{tabular}
\vspace{0.2cm}
\caption{Mean values (standard deviation between brackets) of several quantities in the several regions ($\eta = I_{\mathrm{[C\,\textsc{ii}]}}/I_{\mathrm{FIR}}$). L1630 is the entire molecular cloud (without H\,{\sc ii} region) in the mapped area. Face-on values are values calculated from integration along the depth into the cloud from the surface with respect to the incident FUV radiation (see Sec. \ref{sec.face-on}).}
\label{tab:2}
\end{table*}

\begin{figure*}[ht]
\includegraphics[width=\textwidth, height=0.67\textwidth]{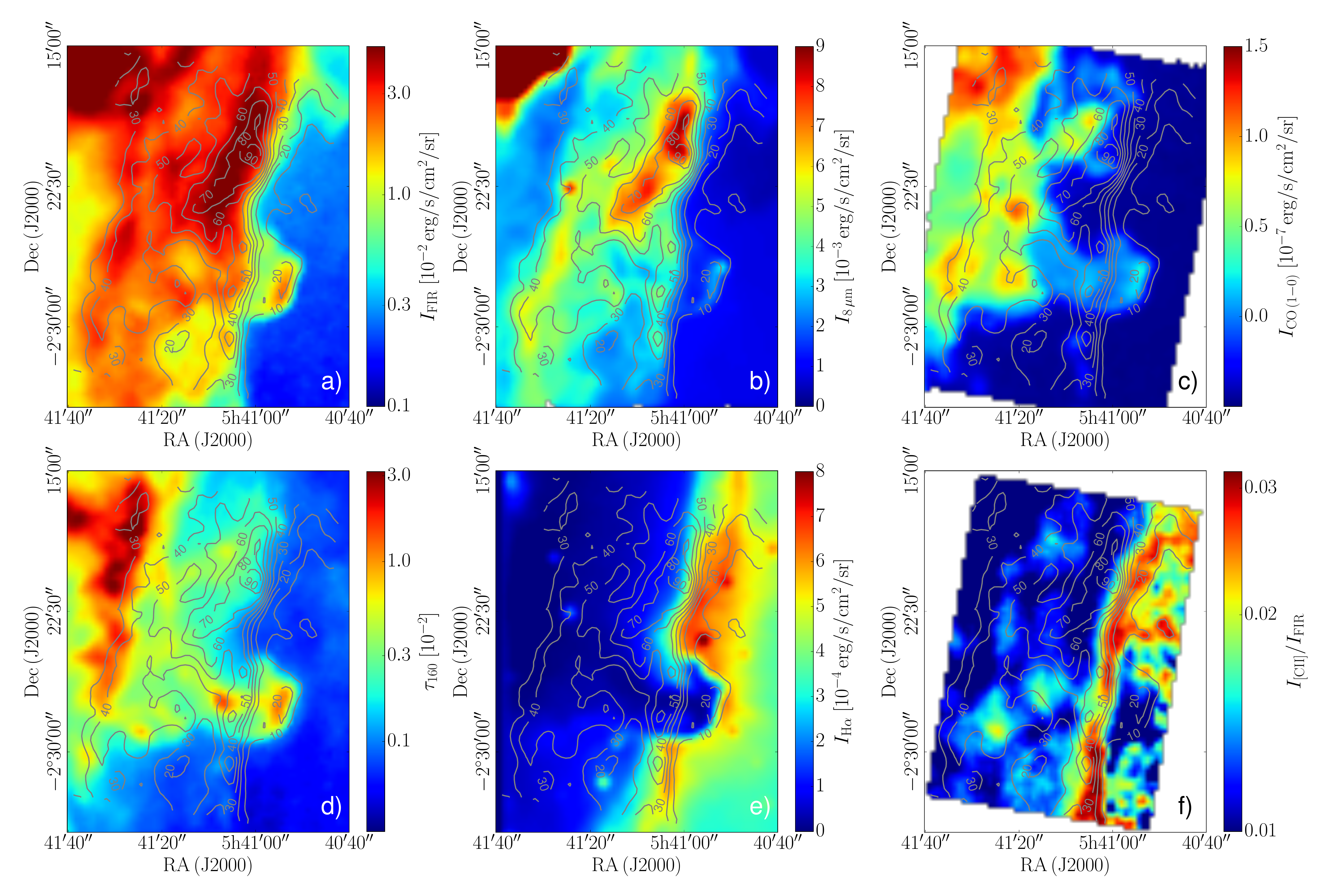}
\caption{Different quantities with [C\,{\sc ii}] emission in units of $\mathrm{K\,km\,s^{-1}}$ in contours: $I_{\mathrm{FIR}}$ tracing the UV radiation field re-radiated in the FIR by dust particles, $I_{8\,\mu\mathrm{m}}$ tracing the UV radiation field by fluorescence of PAHs, $I_{\mathrm{CO}\,(1\mhyphen 0)}$ tracing the molecular gas, $\tau_{160}$ tracing the dust column, $I_{\mathrm{H}\alpha}$ emitted by ionized gas, and, finally, the ratio $I_{\mathrm{[C\,\textsc{ii}]}}/I_{\mathrm{FIR}}$. All maps are convolved to $36\arcsec$ spatial resolution and re-gridded to a pixel size of $14\arcsec$, that of the SPIRE $500\,\mu\mathrm{m}$ map.}
\label{Fig.contours}
\end{figure*}

Apart from the four PDR surfaces discussed in Sec. \ref{sec.channels}, we singled out other specific regions that stand out in their morphology in the respective quantities [C\,{\sc ii}], CO, and IRAC $8\,\mu\mathrm{m}$ emission (see Fig. \ref{Fig.contours}). The $8\,\mu\mathrm{m}$ emission is a tracer of UV-pumped polycyclic aromatic hydrocarbons (PAHs) and therefore of PDR surfaces. Ionized gas is traced by H$\alpha$ emission, and CO traces the molecular hydrogen gas. The regions are indicated in Figs. \ref{Fig.Cplusrepro} ([C\,{\sc ii}] map) and \ref{Fig.CO} (CO$\,(1\mhyphen 0)$ map). We outline the boundary between the H\,{\sc ii} region IC 434 and the molecular cloud L1630 by the onset of significant [C\,{\sc ii}] emission at the molecular cloud surface where we also have been guided by the H$\alpha$ contour of highest emission (see Fig. \ref{Fig.Ha_contours}).

The four PDR regions, among them the Horsehead PDR\footnote{What we call PDR here really is the PDR surface.}, are distinct in the IRAC $8\,\mu\mathrm{m}$ map. The Horsehead PDR is not the most luminous part of the region in all maps. The brightest part is region PDR1. We define the neck of the Horsehead Nebula that is traced by the CO$\,(1\mhyphen 0)$ line; in itself, it has little [C\,{\sc ii}] emission, but part of it is covered by the [C\,{\sc ii}]-emitting molecular cloud. Another part of the cloud, where there is little CO emission, we call CO-dark cloud. Deeper inside the molecular cloud, CO emission is high and we define a region of CO cores or clumps. The Horsehead PDR is likely to suffer from beam dilution in all images, since its scale length is found to be less than $10\arcsec$ \citep{Habart2005}, which is smaller than the beam sizes in question. To the north-east of the map, we recognize the reflection nebula NGC 2023, which has been studied with SOFIA/GREAT in [C\,{\sc ii}] emission by \cite{Sandell2015}. This region will not be discussed here.

Figure \ref{Fig.contours}a shows the FIR intensity with [C\,{\sc ii}] contours in the mapped area. The FIR intensity peaks close to [C\,{\sc ii}] in the most luminous part (PDR1), but slightly deeper into the cloud. The Horsehead mane is bright in both [C\,{\sc ii}] and FIR; the emission overlaps very well. PDR2 is more pronounced in [C\,{\sc ii}] emission than in the FIR. PDR3 can only be surmised in $I_{\mathrm{FIR}}$, but it cannot be distinguished very clearly in the integrated [C\,{\sc ii}] map either.

In Fig. \ref{Fig.contours}b, we compare the $8\,\mu\mathrm{m}$ emission with [C\,{\sc ii}] in contours. The $8\,\mu\mathrm{m}$ emission behaves in a similar way as $I_{\mathrm{FIR}}$, but structures stand out more decidedly. The $8\,\mu\mathrm{m}$ emission, too, peaks slightly deeper into the cloud than [C\,{\sc ii}]. The bright regions in the Horsehead mane overlap; in both maps it is a thin filament. PDR2 is more pronounced in [C\,{\sc ii}], but is distinguishable in $I_{8\,\mu\mathrm{m}}$ as well. PDR3 is more distinct in $I_{8\,\mu\mathrm{m}}$.

The CO$\,(1\mhyphen 0)$ emission in the mapped area does not resemble the pattern of [C\,{\sc ii}] emission (Fig. \ref{Fig.contours}c). In CO, the entire Horsehead and its neck light up with nearly equal intensity while the surroundings remain dark. PDR1 and 2 are not very bright in CO. Interestingly, there is a CO spot in PDR1, right where the cavity is observed in (unconvolved) [C\,{\sc ii}] and $8\,\mu\mathrm{m}$ emission (see Figs. \ref{Fig.channels} and \ref{Fig.cuts}, respectively). A "finger" of CO emission, the "CO clumps", points towards PDR1. PDR3 can be inferred as shadow in CO emission, that is, a ridge of low CO emission.

The $\tau_{160}$ map (Fig. \ref{Fig.contours}d) resembles the CO$\,(1\mhyphen 0)$ map. The Horsehead and its neck have higher dust optical depth than their surroundings; the material directly behind the mane is a peak in $\tau_{160}$, that overlaps partially with the [C\,{\sc ii}] peak, but it peaks slightly deeper into the Horsehead. The [C\,{\sc ii}] peak in PDR1 does not correspond to a peak in dust optical depth, although the onset of the molecular cloud is traced by an increase in $\tau_{160}$. PDR3 corresponds to an optically thin region compared to its environment. The region of highest dust optical depth at the eastern border of the map (not containing NGC 2023) corresponds to a region with little [C\,{\sc ii}] emission, but high CO emission. The CO "finger" relates to enhanced dust optical depth.

PDR1 and PDR2 border on the H\,{\sc ii} region, as can be seen from Fig. \ref{Fig.contours}e. PDR2 overlaps with a region of significant H$\alpha$ emission, tracing the ionized gas at the surface of the molecular cloud. The H$\alpha$ map and the logarithmic [C\,{\sc ii}] cooling efficiency $I_{\mathrm{[C\,\textsc{ii}]}}/I_{\mathrm{FIR}}$ map (Fig. \ref{Fig.contours}f) resemble each other. High [C\,{\sc ii}] over FIR intensity ratios are found near the boundary with high H$\alpha$ emission. However, $I_{\mathrm{[C\,\textsc{ii}]}}/I_{\mathrm{FIR}}$ is a misleading measure in the H\,{\sc ii} region, since $I_{\mathrm{FIR}}$ does not trace the radiation field well here and $I_{\mathrm{[C\,\textsc{ii}]}}$ is sufficiently low to be significantly affected by noise. Variations in $I_{\mathrm{[C\,\textsc{ii}]}}/I_{\mathrm{FIR}}$ across the map span a range from $3\cdot 10^{-3}$ up to $3\cdot 10^{-2}$.
Table \ref{tab:2} lists the mean values of the [C\,{\sc ii}] cooling efficiency $\eta=I_{\mathrm{[C\,\textsc{ii}]}}/I_{\mathrm{FIR}}$, [C\,{\sc ii}] intensity $I_{\mathrm{[C\,\textsc{ii}]}}$, CO$(1\mhyphen 0)$ intensity $I_{\mathrm{[C0\,(1\mhyphen 0)]}}$, FIR intensity $I_{\mathrm{FIR}}$, and dust optical depth $\tau_{160}$ for the several regions defined above.

\subsection{Kinematics: velocity-resolved line spectra}
\label{sec.spectra}

\begin{figure}[ht]
\centering
\includegraphics[width=0.5\textwidth, height=0.5\textwidth]{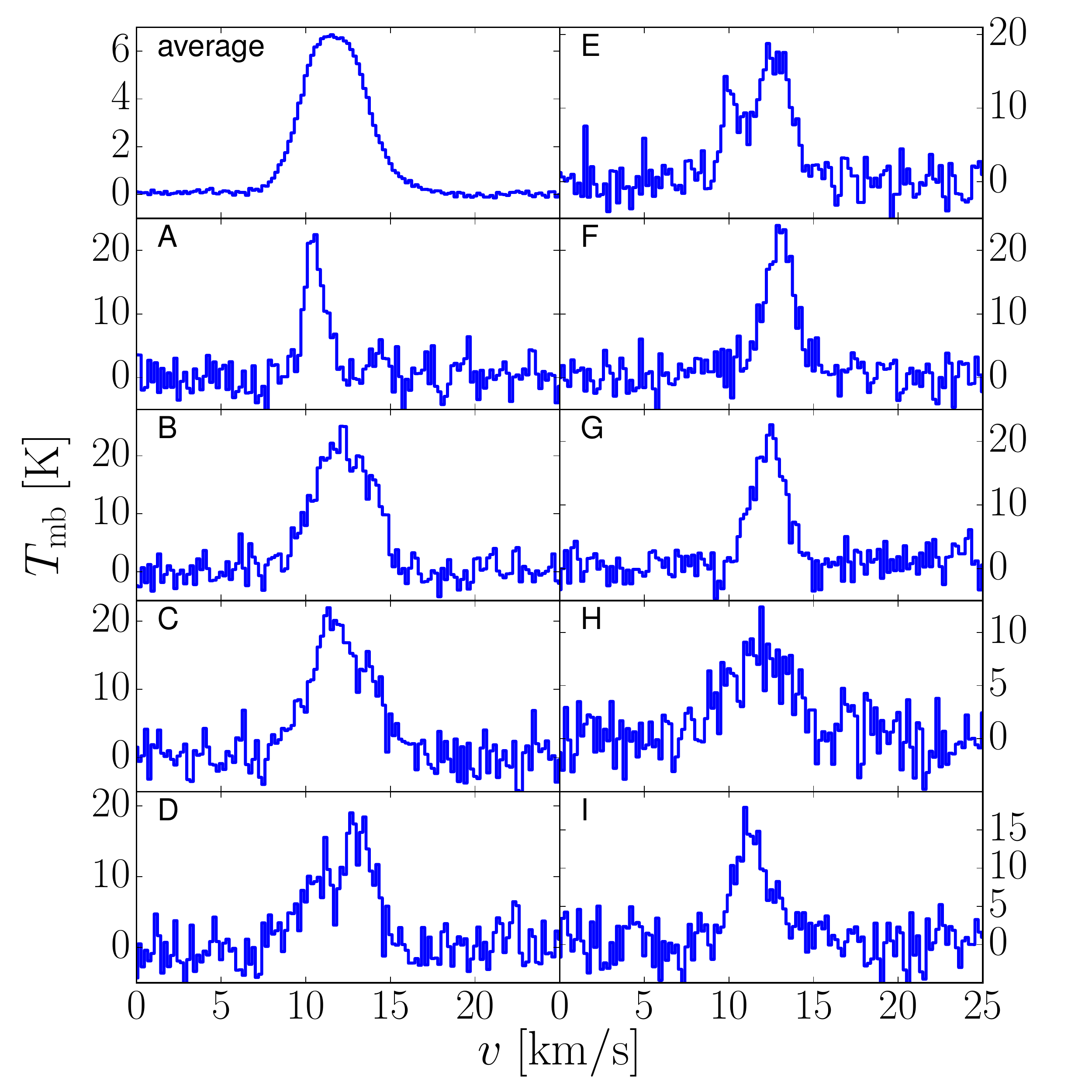}
\caption{Line spectra towards points A--I with average spectrum over the entire map (including H\,{\sc ii} region) in the top left panel. Point A corresponds to the Horsehead PDR, points B to D are located in PDR1, points E and F lie in PDR2, point G represents the intermediate PDR front, point H is located in a region of little [C\,{\sc ii}] emission, and point I represents PDR3.}
\label{Fig.spectra}
\end{figure}

Figure \ref{Fig.spectra} displays spectra extracted towards different positions in the map, as indicated in Fig. \ref{Fig.Cplus}. The positions are chosen as representative single points for the different regions we identified earlier. Details on the spectra, that is, peak position, peak temperature, and line width, are given in Table \ref{tab:1}. Point A represents the Horsehead mane, B lies in the most luminous part of the map, C just north of it, whereas D is displaced to the east; all three represent PDR1. We chose additional points in PDR1, because B might be affected by the bubble structure discussed later. Point E lies in PDR2 behind the Horsehead and F in the southern part of PDR2. Point G is located in the intermediate PDR front which we do not discuss in detail. Point I represents PDR3, whereas H is chosen in the CO-dark cloud, where there is little [C\,{\sc ii}] emission (and little emission in other tracers).

\begin{table}[ht]
\centering
\begin{tabular}{cccccc}
 &  RA & Dec & v & $T_{\mathrm{P}}$ & FWHM \\
pos. & (J2000)& (J2000)& [$\mathrm{km\,s^{-1}}$] & [K] & [$\mathrm{km\,s^{-1}}$] \\ \hline \rule{0pt}{3ex}\noindent
A & $5\mbox{h}40\arcmin 53\arcsec$ & $-2^{\circ}27\arcmin 37\arcsec$ &10.5&21.1&1.4 \\ 
B & $5\mbox{h}41\arcmin 00\arcsec$ & $-2^{\circ}20\arcmin 27\arcsec$ &12.0&22.5&3.6 \\
 & & &14.2&6.9&1.2 \\ 
C & $5\mbox{h}41\arcmin 00\arcsec$ & $-2^{\circ}19\arcmin 34\arcsec$ &11.3&6.1&1.2 \\ 
 & & &12.2&16.2&4.6 \\
D & $5\mbox{h}41\arcmin 05\arcsec$ & $-2^{\circ}20\arcmin 42\arcsec$ &10.7&9.5&2.7 \\
 & & &13.2&16.1&1.9\\
E & $5\mbox{h}41\arcmin 02\arcsec$ & $-2^{\circ}26\arcmin 44\arcsec$ &10.0&12.4&1.1 \\
 & & &12.6&17.4&2.5 \\
F & $5\mbox{h}41\arcmin 06\arcsec$ & $-2^{\circ}30\arcmin 46\arcsec$ &13.0&21.9&2.1 \\
G & $5\mbox{h}41\arcmin 14\arcsec$ & $-2^{\circ}28\arcmin 38\arcsec$ &12.4 &20.3&2.3\\
H & $5\mbox{h}41\arcmin 33\arcsec$ & $-2^{\circ}30\arcmin 53\arcsec$ &11.8&8.1&4.5 \\ 
I & $5\mbox{h}41\arcmin 24\arcsec$ & $-2^{\circ}31\arcmin 16\arcsec$ &11.3&13.9&2.6
\end{tabular}
\vspace{0.2cm}
\caption{Results from Gaussian fit (points B--E with two components) to individual spectra with a spatial resolution of $15.9\arcsec$ sampled at $7.55\arcsec$. Positions are indicated in Fig. \ref{Fig.Cplus}. Listed are the velocity of the peak, the peak temperature, and the full width at half maximum (FWHM) of the peak. Note to spectrum I: From a Lorentzian fit we obtain $T_{\mathrm{P}}=15.6\,\mathrm{K}$, which fits the spectrum better by eye; velocity and FWHM are similar.}
\label{tab:1}
\end{table}

The spectrum taken towards the Horsehead PDR (point A) shows a narrow line. Opposed to this is the line width of the spectrum extracted towards the most luminous part of the molecular cloud, point B: Here, the line is broadened. It peaks at a slightly higher velocity than the Horsehead PDR. From comparison with the dust optical depth, we conclude that the broadening of the line is not due to a high column density (if dust density and gas density are related). The same holds for point C in PDR1. Here there appears a small side peak at higher velocity, which could also be inferred for point B (as a shoulder). Point D evidently has a spectrum with two peaks. From the distinctly different morphology of the channel maps at the two peak velocities (cf. Fig. \ref{Fig.channels} at $10.5\,\mathrm{km\,s^{-1}}$ and $13.0\,\mathrm{km\,s^{-1}}$), we surmise that the two peaks correspond to two distinct emitting components, rather than to one emission component with foreground absorption. The same goes for point E in PDR2, which also has two peaks. The southern part of PDR2, point F, has only one rather narrow peak. The intermediate PDR, point G, exhibits a strong narrow line, as well. The spectrum taken in the western PDR, point I, shows a somewhat broader line with somewhat lower intensity. At point H, where the intensity is low in all tracers, the [C\,{\sc ii}] line is also broader.

Strikingly, the peak velocity of point D is shifted towards lower velocity by $1\,\mathrm{km\,s^{-1}}$ with respect to points B and C (all PDR1). However, one component of this spectrum lies at about $11\,\mathrm{km\,s^{-1}}$, which is also the velocity of PDR3 (point I). This is further evidence that PDR3 and a part of PDR1 are spatially connected, as concluded from the channel maps. Point D in PDR1 has a component at about $13\,\mathrm{km\,s^{-1}}$, which is the velocity of PDR2. In B and C (PDR1) this component might be hidden beneath the strong side peak at $14\,\mathrm{km\,s^{-1}}$. The affiliation of the second component at point E in PDR2 is unclear; there might be another layer of gas behind or in front of the main component, or it could originate from the gas of PDR1 and PDR3 at $11\,\mathrm{km\,s^{-1}}$.

\subsection{Edge-on PDR models}
\label{sec.model_description}
We supplement the correlation plots in the following section with model runs that are based on the PDR models of \cite{TielensHollenbach1985}, with updates like those found in \cite{Wolfire2010} and \cite{Hollenbach2012}. We include the most recent computations on fine-structure excitations of C$^+$ by collisions with H by \cite{Barinovs2005} and with H$_2$ by \cite{WiesenfeldGoldsmith2014}, and adopt a fractional gas-phase carbon abundance of $1.6\cdot 10^{-4}$ \citep{SofiaApril2004}. The line intensities are calculated for an edge-on case by storing the run of level populations with molecular cloud depth for the excited level of CO and C$^+$ as calculated in the face-on model. For each line of sight, the intensity is found from integrating eq. (B14) in \cite{TielensHollenbach1985} through the layer of length $z = N_{\mathrm{H}}/n_{\mathrm{H}}$ with the gas column density $N_{\mathrm{H}}$ and the gas density $n_{\mathrm{H}}$, where we replace the factor $(2 \pi)^{-1}$ for a semi-infinite slab with $(4 \pi)^{-1}$. The cooling rate is given by eq. (B1) in \cite{TielensHollenbach1985} in the limit of no background radiation. For the escape probability we take the line-of-sight formulation
\begin{align}
\beta(\tau) = \frac{1-e^{-\tau}}{\tau},
\end{align}
where $\tau$ is the line optical depth calculated as in eq. (B8) in \cite{TielensHollenbach1985}.

The FIR continuum intensity in the edge-on case is calculated from the run of dust temperature with depth into the molecular cloud. We find the dust temperature, $T_{\mathrm{d}}$, from the prescription given in \cite{HollenbachTT1991}. We integrate the dust absorption efficiency, $Q_{\mathrm{abs}}$, through the layer of length $z$
\begin{align}
I_{\mathrm{FIR}} = \frac{1}{4\pi} \int 4 \pi a^2\,\frac{n_{\mathrm{d}}}{n_{\mathrm{H}}} Q_{\mathrm{abs}}\, \sigma T_{\mathrm{d}}^4\, n_{\mathrm{H}}\, \mathrm{d}z,
\end{align}
where we take the grain size $a = 0.1\mathrm{\,\mu m}$, and $n_{\mathrm{d}}/n_{\mathrm{H}} = 6.36\cdot 10^{-12}$, which gives a grain cross section per hydrogen atom of $2.0\cdot 10^{-21}\,\mathrm{cm}^2$. For $Q_{\mathrm{abs}}$ we use the average silicate and graphite value $Q_{\mathrm{abs}} = 1.0\cdot 10^{-6} (a/0.1\mathrm{\,\mu m}) (T_{\mathrm{d}}/\mathrm{K})^2$ from \cite{Draine2011}.

\begin{figure*}[!ht]
\includegraphics[width=\textwidth, height=0.6\textwidth]{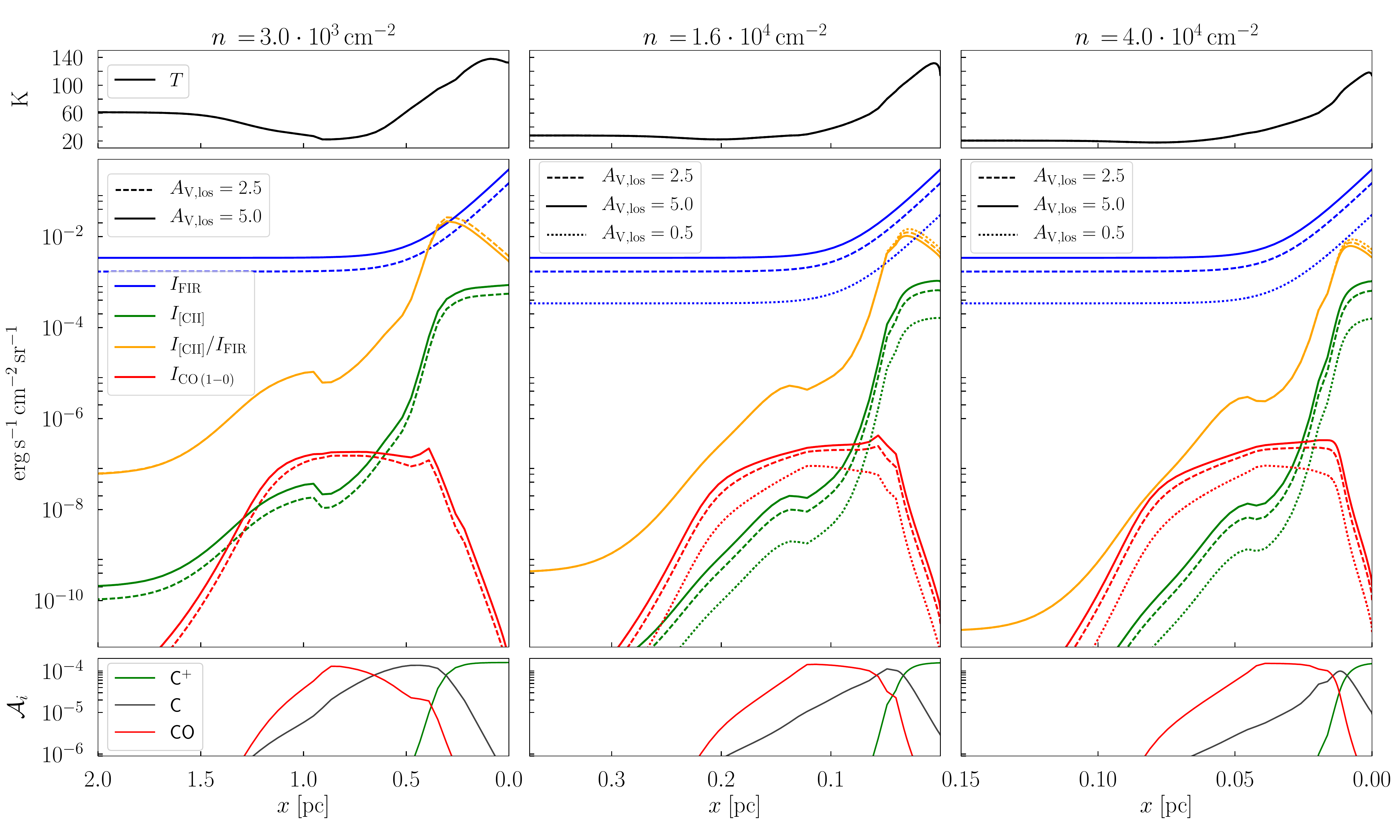}
\caption{Results of our edge-on models described in Sec. \ref{sec.model_description}. The panels show the gas temperature $T$ (upper panels), $I_{\mathrm{FIR}}$, $I_{\mathrm{[C\,\textsc{ii}]}}$, $I_{\mathrm{[C\,\textsc{ii}]}}/I_{\mathrm{FIR,}}$ and $I_{\mathrm{CO\,(1\mhyphen 0)}}$ (middle panels) and C$^+$, C, and CO fractional abundances (lower panel) versus physical scale, for the gas densities $n_{\mathrm{H}}=3.0\cdot 10^3\,\mathrm{cm^{-3}},\;1.6\cdot 10^4\,\mathrm{cm^{-3}},\;4.0\cdot 10^4\,\mathrm{cm^{-3}}$ (left to right panels), and $A_\mathrm{V,los}=0.5,\,2.5,\,\mbox{ and }5.0$. We note that the gas temperature does not vary with $A_\mathrm{V,los}$.}
\label{Fig.model}
\end{figure*}

In Fig. \ref{Fig.model}, we present the results of the models with an incident FUV intensity of $G_0 = 100$ appropriate for $\sigma$ Ori (\cite{Abergel2003} and references therein) and a Doppler line width of $\Delta v = 1.5\,\mathrm{km\,s}^{-1}$ for different densities on a physical scale. The $x$-axes share the same range of visual extinction, $A_\mathrm{V}=0.0\mhyphen9.3$. We computed models for gas densities $n_{\mathrm{H}}=3.0\cdot 10^3\,\mathrm{cm^{-3}},\;1.6\cdot 10^4\,\mathrm{cm^{-3}}$, and $4.0\cdot 10^4\,\mathrm{cm^{-3}}$; those densities we estimate from the line cuts in Sec. \ref{sec.crosscuts} for different parts of the molecular cloud. We integrate along a length $A_\mathrm{V,los}$ of the line of sight estimated in Sec. \ref{sec.column} for each density, where we assumed $N_{\mathrm{H}}=2.0\cdot 10^{21}\,\mathrm{cm}^{-2} A_{\mathrm{V}}$. $A_\mathrm{V,los}=2.5\mbox{ and }5.0$ with $n_{\mathrm{H}}=3.0\cdot 10^3\,\mathrm{cm^{-3}}$ correspond to PDR1 and PDR2, respectively; $A_\mathrm{V,los}=0.5\mbox{ and }2.5$ with $n_{\mathrm{H}}=1.6\cdot 10^4\,\mathrm{cm^{-3}}\mbox{ and }4.0\cdot 10^4\,\mathrm{cm^{-3}}$ correspond to potentially dense cloud surfaces in PDR1 and PDR2. The Horsehead PDR should be matched with $A_\mathrm{V,los}=2.5\mbox{ or }5.0$ with $n_{\mathrm{H}}=4.0\cdot 10^4\,\mathrm{cm^{-3}}$.

The three models substantially show the same result, with a luke-warm surface layer where the gas cools through the [C\,{\sc ii}] line. The colder gas deeper in the cloud emits mainly through CO. Not surprisingly, the FIR dust emission also peaks at the surface. The line-of-sight depth of the molecular cloud ($A_\mathrm{V,los}=0.5,2.5,\mbox{ or }5.0$) only slightly affects the ratios of FIR, [C\,{\sc ii}]-, and CO-line intensities.

\subsection{Correlation diagrams}
\label{sec.corr}

\begin{figure*}[!ht]
\begin{minipage}{0.49\textwidth}
\includegraphics[width=\textwidth, height=0.67\textwidth]{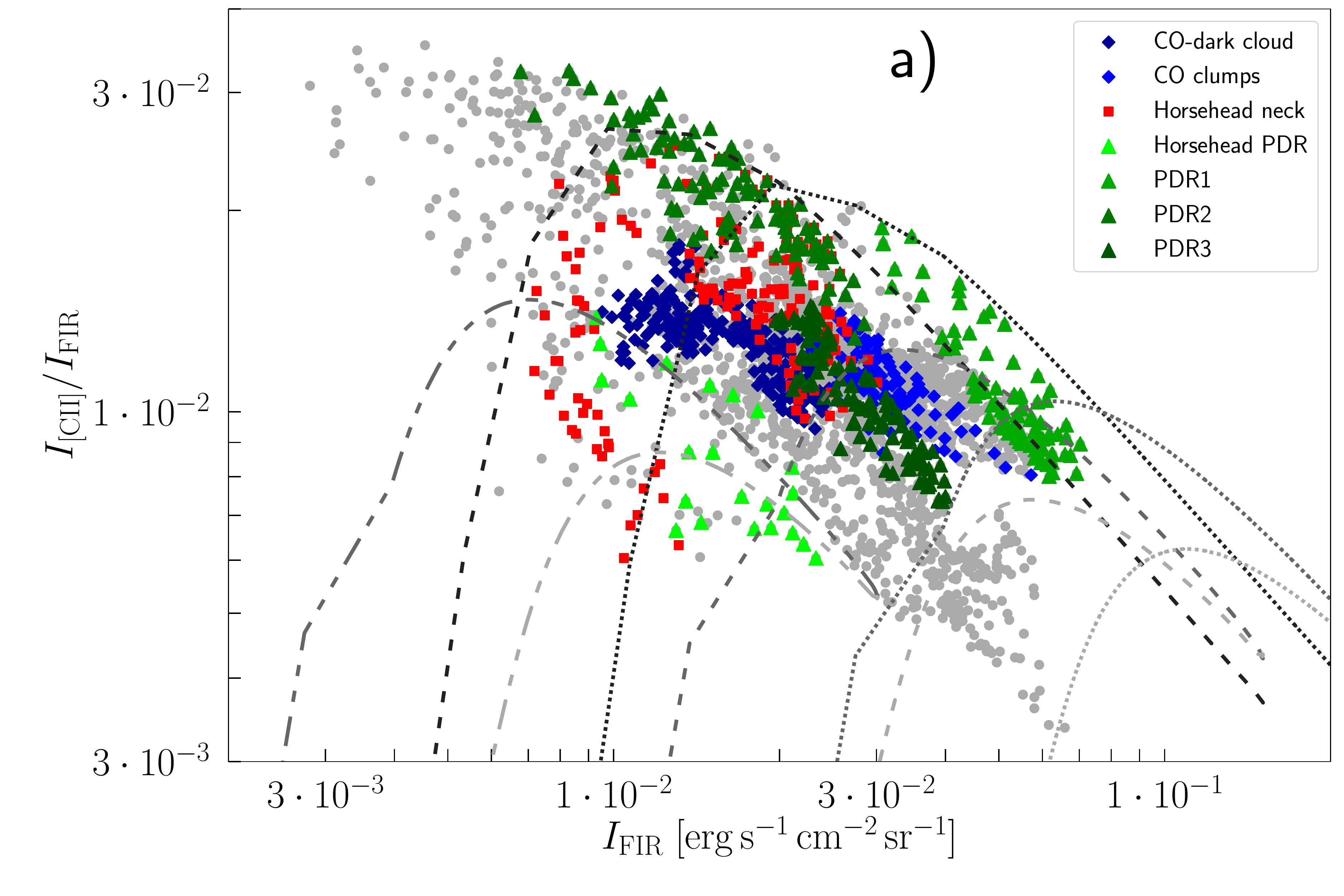}
\end{minipage}
\begin{minipage}{0.49\textwidth}
\includegraphics[width=\textwidth, height=0.67\textwidth]{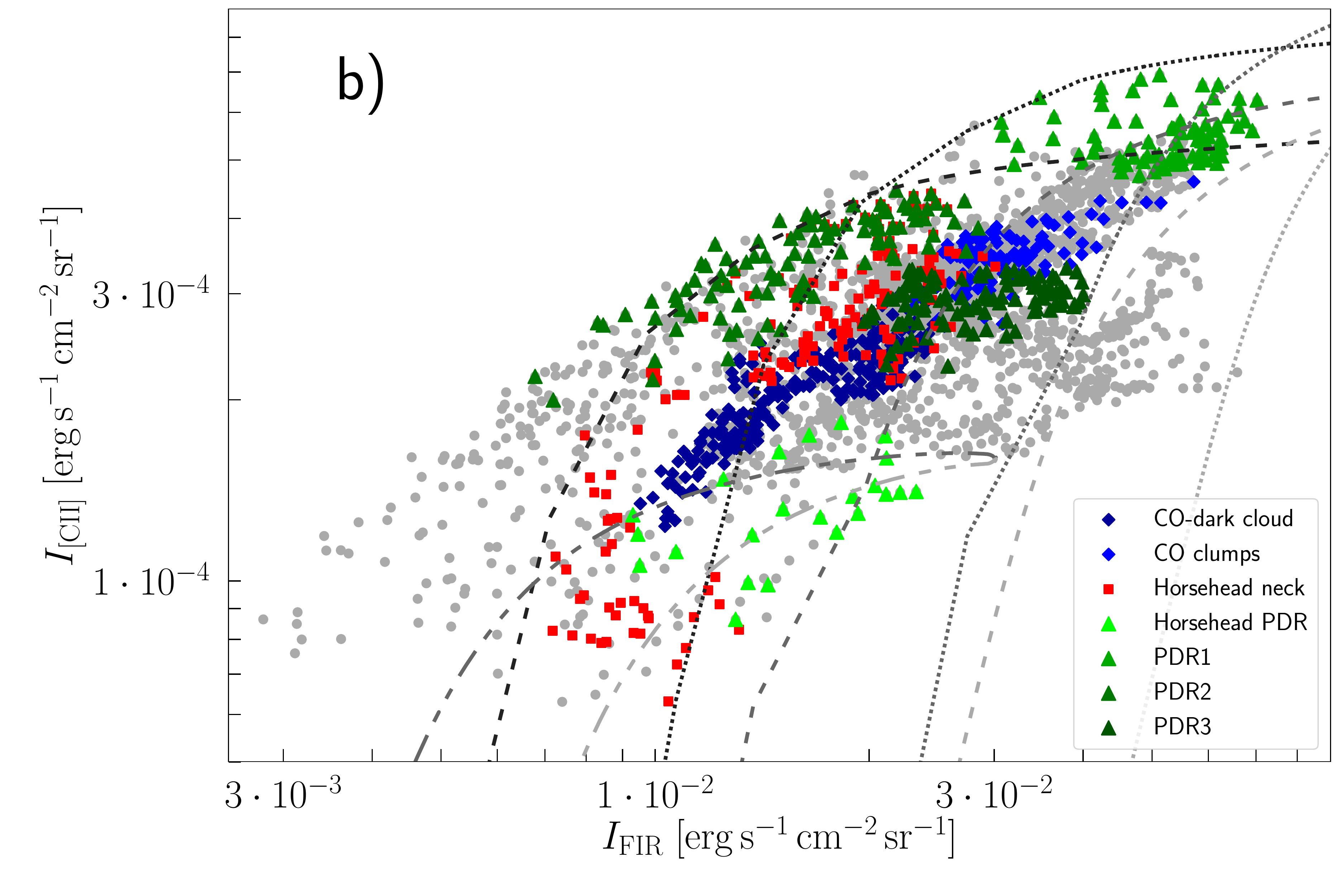}
\end{minipage}

\begin{minipage}{0.49\textwidth}
\includegraphics[width=\textwidth, height=0.67\textwidth]{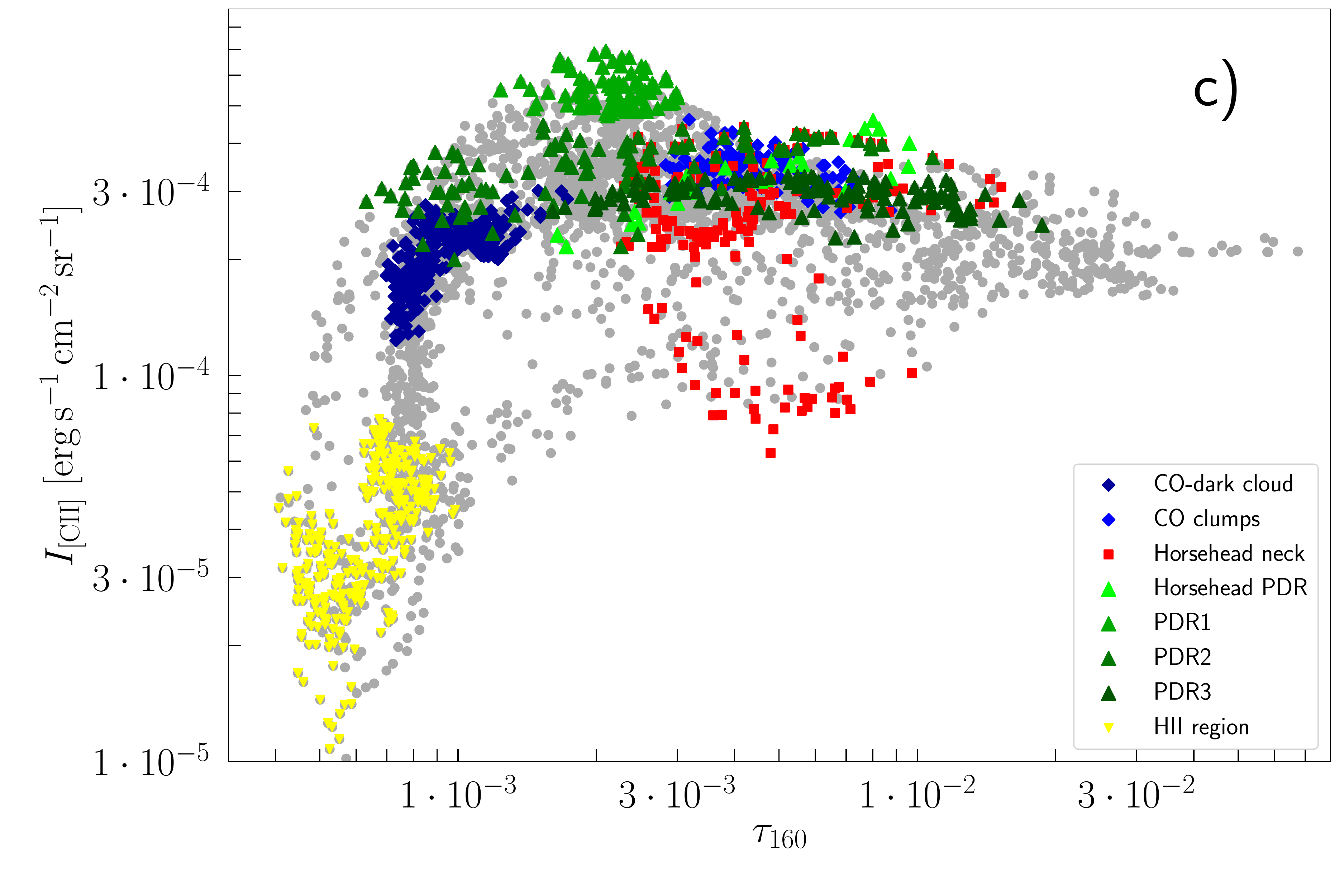}
\end{minipage}
\begin{minipage}{0.49\textwidth}
\includegraphics[width=\textwidth, height=0.67\textwidth]{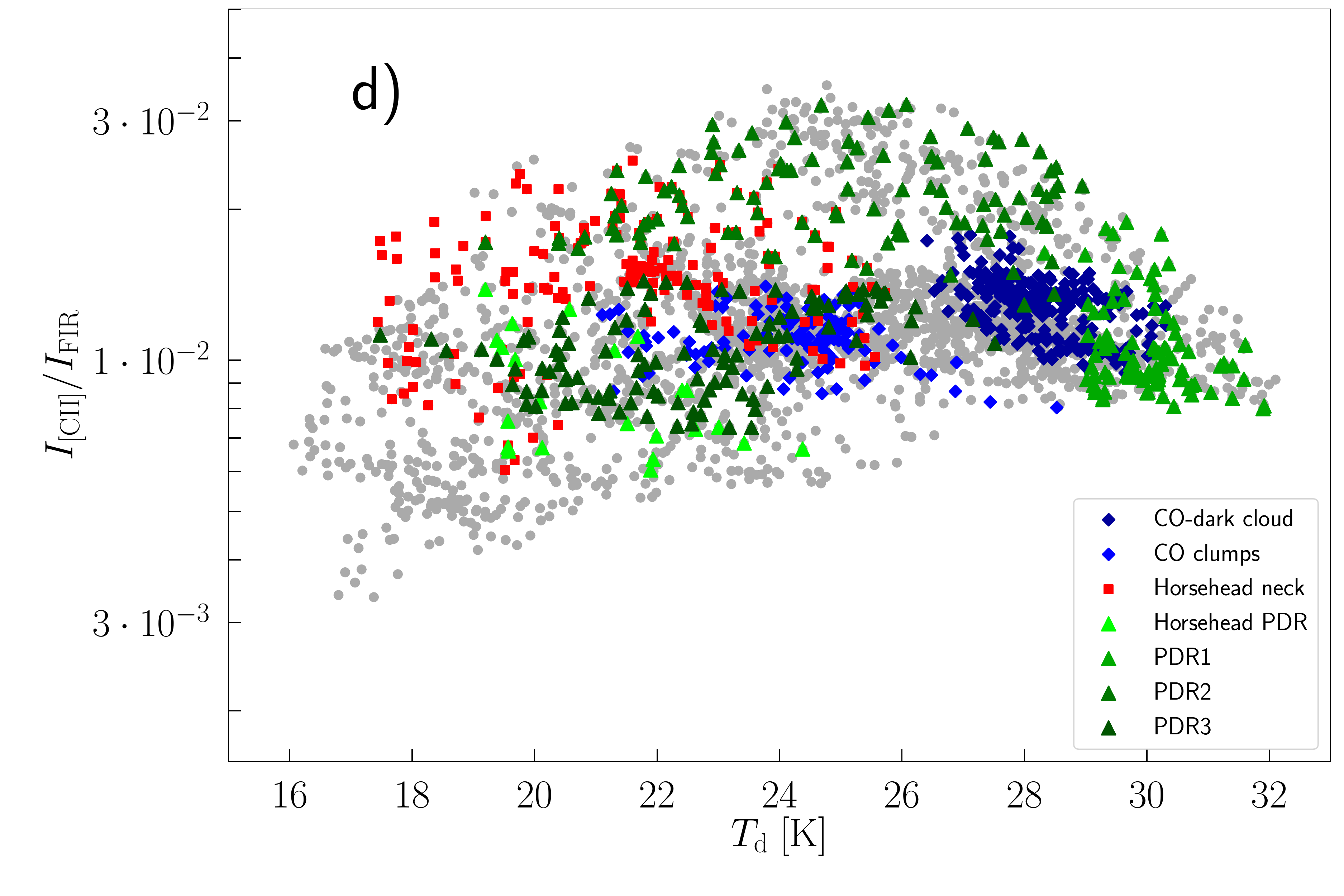}
\end{minipage}

\begin{minipage}{0.49\textwidth}
\includegraphics[width=\textwidth, height=0.67\textwidth]{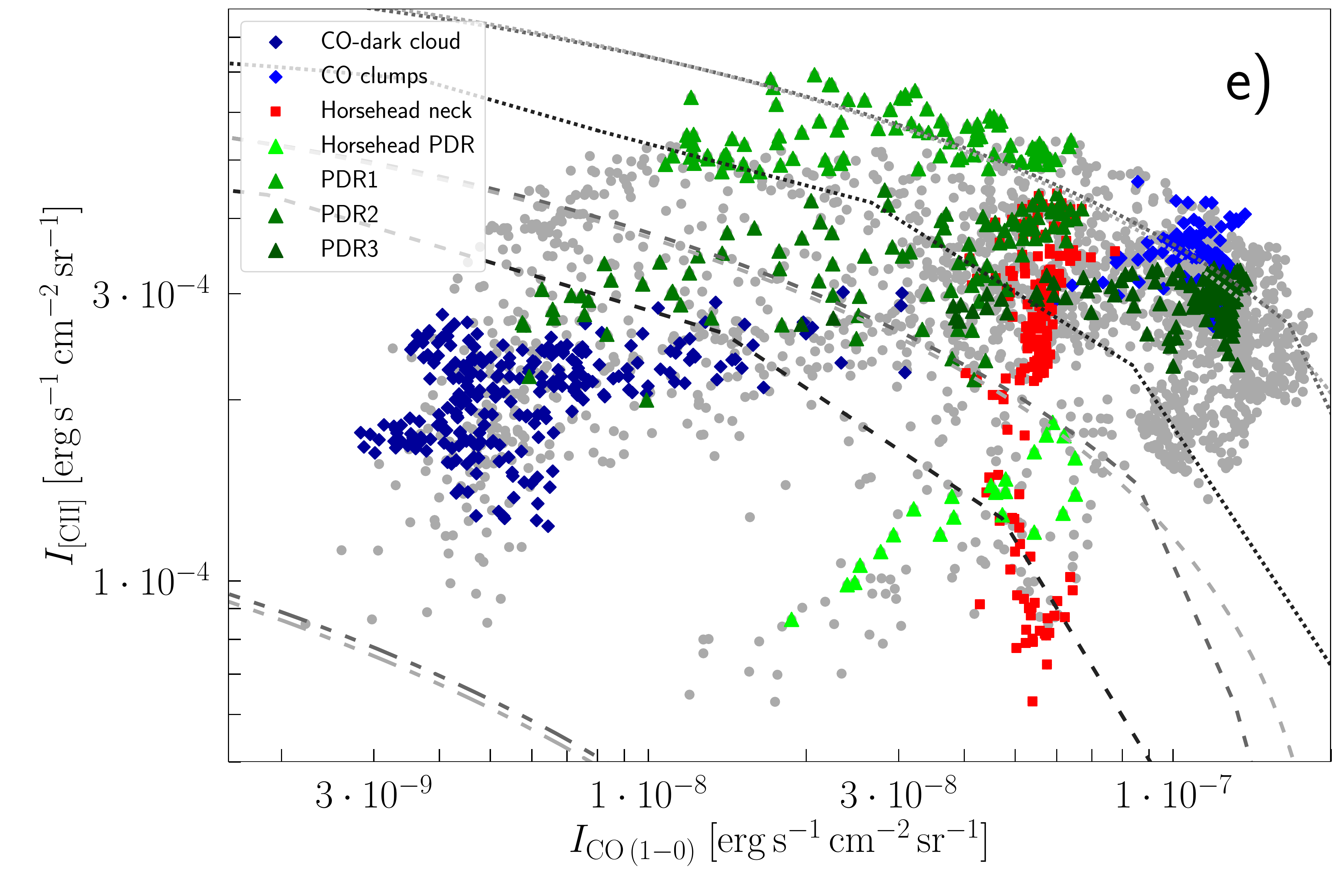}
\end{minipage}
\begin{minipage}{0.49\textwidth}
\includegraphics[width=\textwidth, height=0.67\textwidth]{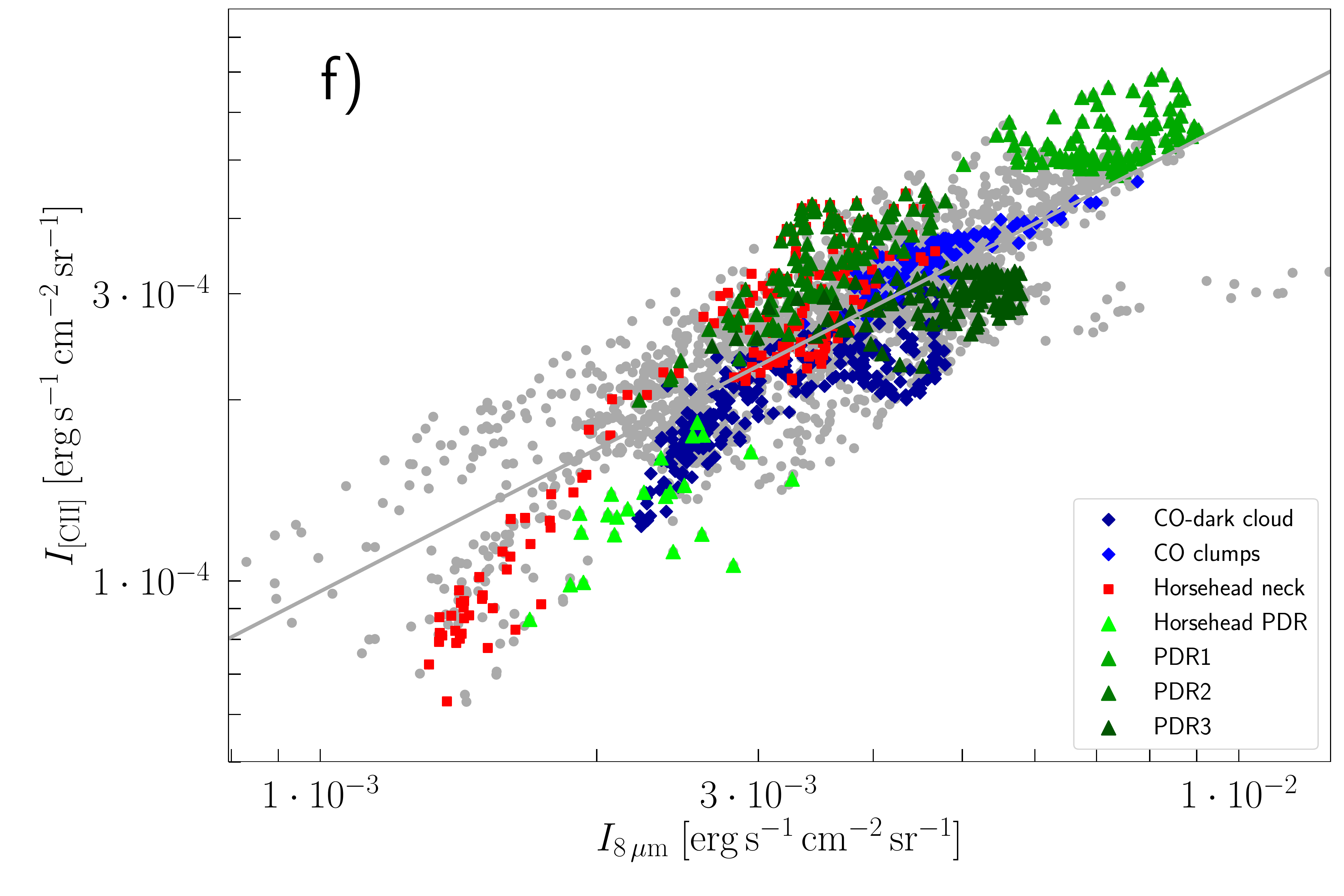}
\end{minipage}
\caption{Correlation plots extracted from the Orion B maps, convolved to a uniform resolution of $36\arcsec$ and pixel size of $15\arcsec$. Dark blue diamonds represent the CO-dark cloud, blue diamonds represent the CO clumps, red squares represent the Horsehead neck, triangles in different shades of green represent the PDRs (bright green is the Horsehead PDR), and yellow triangles represent the H\,{\sc ii} region. Edge-on model predictions for selected gas densities are plotted as lines. Dashed lines are for $A_{\mathrm{V,los}}=2.5,$ with dark gray corresponding to a gas density of $n_{\mathrm{H}}=3.0\cdot 10^3\,\mathrm{cm}^{-3}$, medium gray to $n_{\mathrm{H}}=1.6\cdot 10^4\,\mathrm{cm}^{-3}$, and light gray to $n_{\mathrm{H}}=4.0\cdot 10^4\,\mathrm{cm}^{-3}$; dotted lines are the same but for $A_{\mathrm{V,los}}=5.0$; the medium and light gray dash-dotted lines are for $A_{\mathrm{V,los}}=0.5,$ with $n_{\mathrm{H}}=1.6\cdot 10^4\,\mathrm{cm}^{-3}$ and $n_{\mathrm{H}}=4.0\cdot 10^4\,\mathrm{cm}^{-3}$, respectively. In panel f, the best fit is plotted as a line.}
\label{Fig.corr}
\end{figure*}

Figure \ref{Fig.corr} shows correlation diagrams between several quantities. The different colors indicate the selected regions assigned in Sec. \ref{sec.GlobMorph} and shown in Figs. \ref{Fig.Cplus} and \ref{Fig.CO}. Gray points represent points that do not lie in either of the defined regions.

Figure \ref{Fig.corr}a shows that the [C\,{\sc ii}] cooling efficiency $I_{\mathrm{[C\,\textsc{ii}]}}/I_{\mathrm{FIR}}$ decreases with increasing $I_{\mathrm{FIR}}$. The different PDRs lie on distinct curves, with similar slopes. Figure \ref{Fig.corr}b shows increasing $I_{\mathrm{[C\,\textsc{ii}]}}$ with increasing $I_{\mathrm{FIR}}$, but again in distinct branches for the different regions. The relation between $I_{\mathrm{[C\,\textsc{ii}]}}$ and $I_{8\,\mu\mathrm{m}}$ resembles that, as can be seen from Fig. \ref{Fig.corr}f, but simple fits reveal a tighter relation of $I_{\mathrm{[C\,\textsc{ii}]}}$ with $I_{8\,\mu\mathrm{m}}$ than with $I_{\mathrm{FIR}}$. Over-plotted is a least-square fit $I_{\mathrm{[C\,\textsc{ii}]}} \simeq 2.2\cdot 10^{-2} I_{8\,\mu\mathrm{m}}^{0.79}$ ($\rho=0.85$).

\begin{figure}[ht]
\includegraphics[width=0.5\textwidth, height=0.33\textwidth]{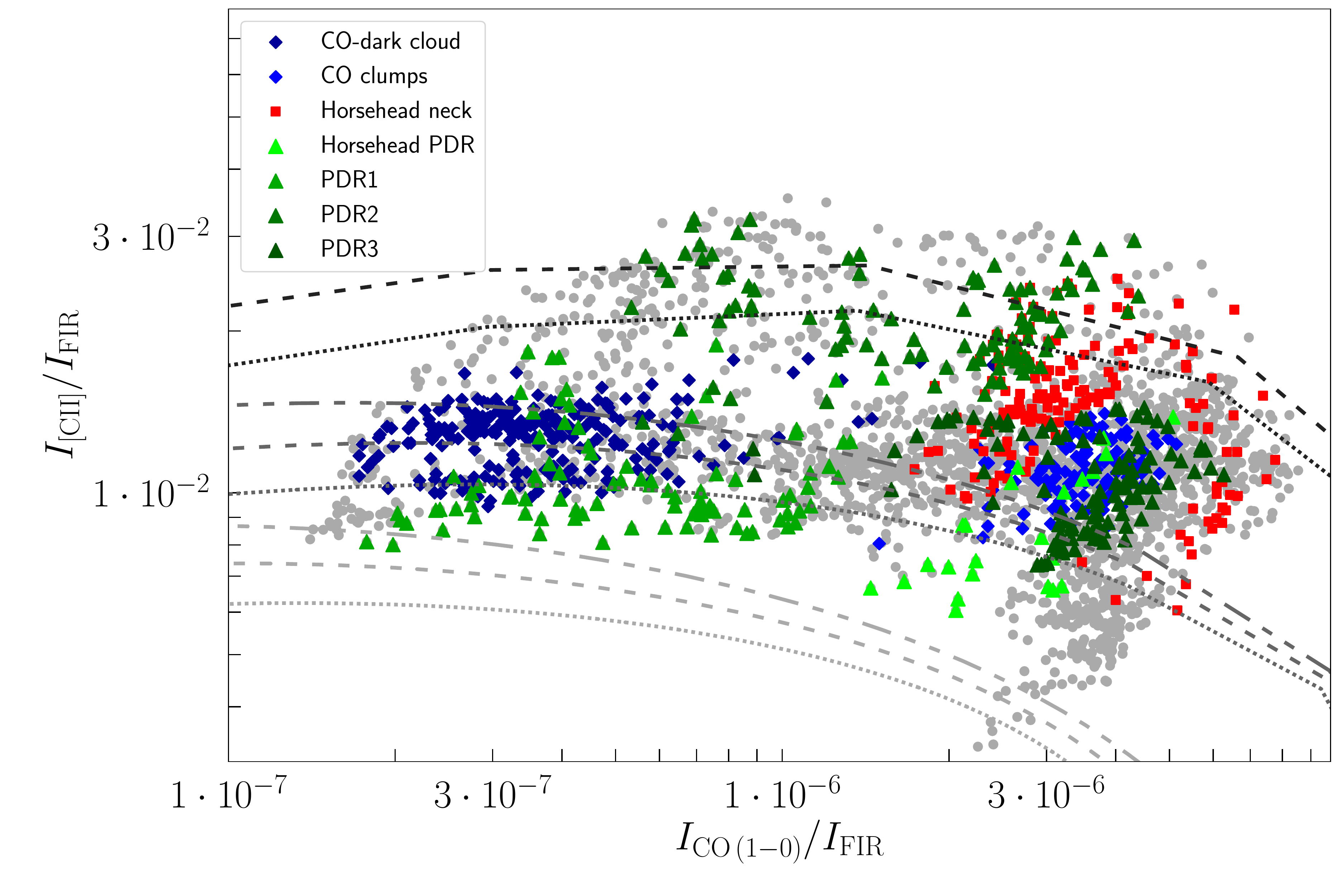}
\caption{Correlation plot of $I_{\mathrm{[C\,\textsc{ii}]}}/I_{\mathrm{FIR}}$ versus $I_{\mathrm{CO}\, (1\mhyphen 0)}/I_{\mathrm{FIR}}$. Edge-on model predictions for selected gas densities are plotted as lines. Dashed lines are for $A_{\mathrm{V,los}}=2.5,$ with dark gray corresponding to a gas density of $n_{\mathrm{H}}=3.0\cdot 10^3\,\mathrm{cm}^{-3}$, medium gray to $n_{\mathrm{H}}=1.6\cdot 10^4\,\mathrm{cm}^{-3}$, and light gray to $n_{\mathrm{H}}=4.0\cdot 10^4\,\mathrm{cm}^{-3}$; dotted lines are the same but for $A_{\mathrm{V,los}}=5.0$; the medium and light gray dash-dotted lines are for $A_{\mathrm{V,los}}=0.5,$ with $n_{\mathrm{H}}=1.6\cdot 10^4\,\mathrm{cm}^{-3}$ and $n_{\mathrm{H}}=4.0\cdot 10^4\,\mathrm{cm}^{-3}$, respectively.}
\label{Fig.CplusCO}
\end{figure}

As Fig. \ref{Fig.corr}c shows, $I_{\mathrm{[C\,\textsc{ii}]}}$ is roughly constant for higher $\tau_{160}$, that is, in PDR3 and the Horsehead PDR. This might reflect the fact that there is colder, non-PDR material located behind the PDR surfaces. PDR1 and PDR2 at the onset of the molecular cloud, where we do not expect a huge amount of colder material along the line of sight, show a nice correlation: PDR1 lies at twice as high $\tau_{160}$ and has twice as high $I_{\mathrm{[C\,\textsc{ii}]}}$. For small $\tau_{160}$ (i.e., in the H\,{\sc ii} region and parts of the CO-dark cloud), the data show a steep rising slope. In Fig. \ref{Fig.corr}d, there is no obvious relation between $I_{\mathrm{[C\,\textsc{ii}]}}/I_{\mathrm{FIR}}$ and the dust temperature $T_{\mathrm{d}}$. Figure \ref{Fig.corr}e shows $I_{\mathrm{[C\,\textsc{ii}]}}$ versus $I_{\mathrm{CO}\,(1\mhyphen 0)}$. Here we notice that the various regions populate distinct areas in the plot. In the diagram of $I_{\mathrm{[C\,\textsc{ii}]}}/I_{\mathrm{FIR}}$ versus $I_{\mathrm{CO}\, (1\mhyphen 0)}/I_{\mathrm{FIR}}$ (Fig. \ref{Fig.CplusCO}), we observe no obvious correlation, only the Horsehead PDR exhibits a significant slope.

The relation between $I_{\mathrm{[C\,\textsc{ii}]}}$ and $I_{8\,\mu\mathrm{m}}$ resembles the relation of $I_{\mathrm{[C\,\textsc{ii}]}}$ with $I_{\mathrm{FIR}}$, as can be seen from Fig. \ref{Fig.corr}f, but simple fits reveal a tighter relation of $I_{\mathrm{[C\,\textsc{ii}]}}$ with $I_{8\,\mu\mathrm{m}}$ than with $I_{\mathrm{FIR}}$. Over-plotted is a least-square fit, $I_{\mathrm{[C\,\textsc{ii}]}} \simeq 2.2\cdot 10^{-2}\, (I_{8\,\mu\mathrm{m}}[\,\mathrm{erg\,s^{-1}\,cm^{-2}\,sr^{-1}}])^{0.79}\,\mathrm{erg\,s^{-1}\,cm^{-2}\,sr^{-1}}$ ($\rho=0.85$).

\begin{figure}[ht]
\includegraphics[width=0.5\textwidth, height=0.33\textwidth]{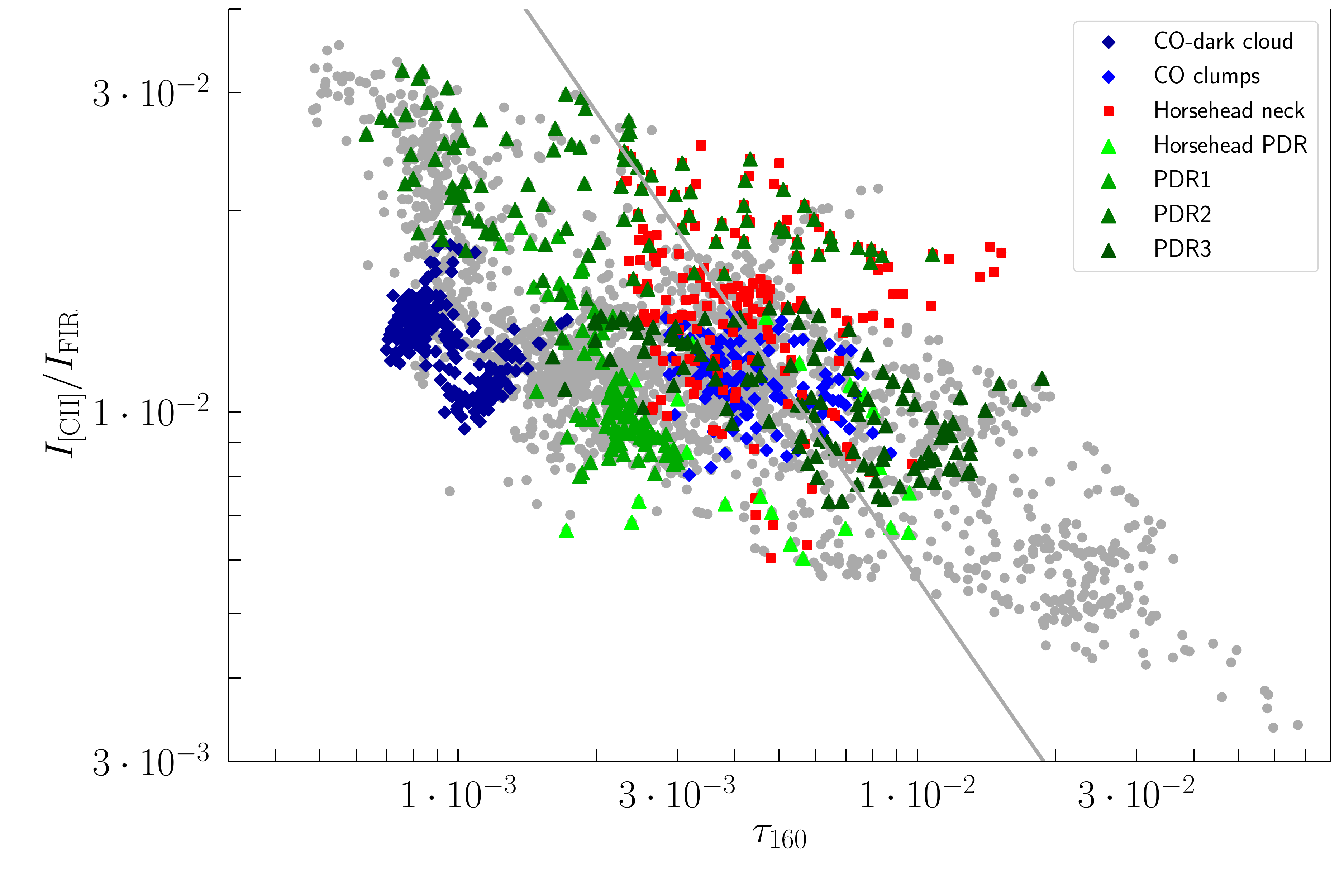}
\caption{Correlation plot of $I_{\mathrm{[C\,\textsc{ii}]}}/I_{\mathrm{FIR}}$ versus $\tau_{160}$. The plotted line corresponds to the relation expected from a simple face-on slab model, $C/(1-\exp(-\tau_{160}))$; it is drawn such that it runs through the mean of $I_{\mathrm{[C\,\textsc{ii}]}}/I_{\mathrm{FIR}}$ and $\tau_{160}$.}
\label{Fig.Cplustau}
\end{figure}

Figure \ref{Fig.Cplustau} shows that the Horsehead PDR lies at the high end of the $\tau_{160}$ distribution. The general trend does not exactly fit a slope of $\simeq -1,$ as does OMC1 in a first approximation (from $I_{\mathrm{[C\,\textsc{ii}]}}/I_{\mathrm{FIR}} \simeq C/(1-e^{-\tau_{160}})$, \cite{Goico}), indicating that the emission cannot be modeled by a homogeneous face-on slab of dust with [C\,{\sc ii}] foreground emission. Of course, dust temperature differences should be taken into account, yet here we assume a constant pre-factor. Moreover, this simple model is derived from a face-on geometry, whereas here we are likely to deal with PDRs viewed edge-on.

\begin{figure}[ht]
\includegraphics[width=0.5\textwidth, height=0.33\textwidth]{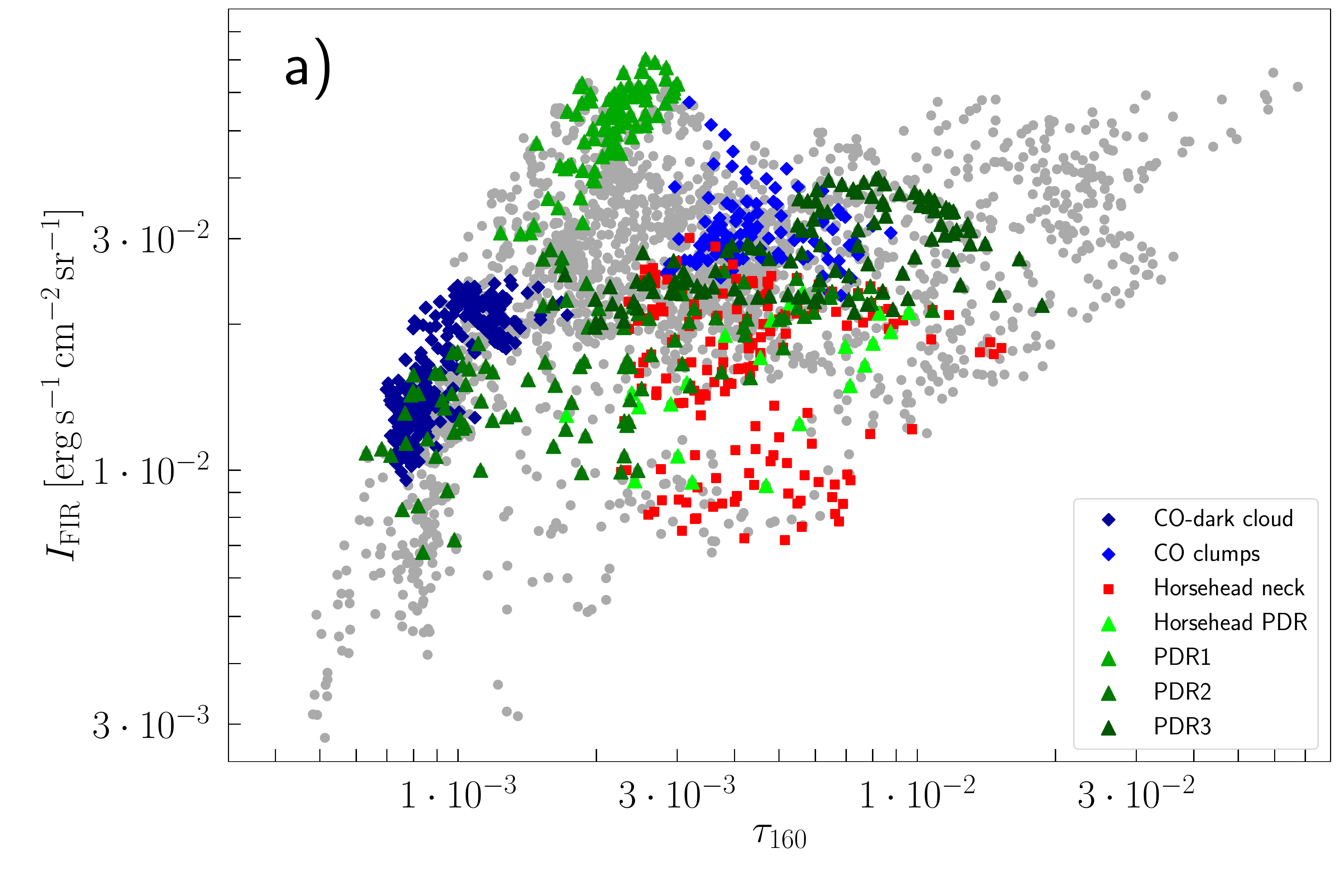}
\includegraphics[width=0.5\textwidth, height=0.33\textwidth]{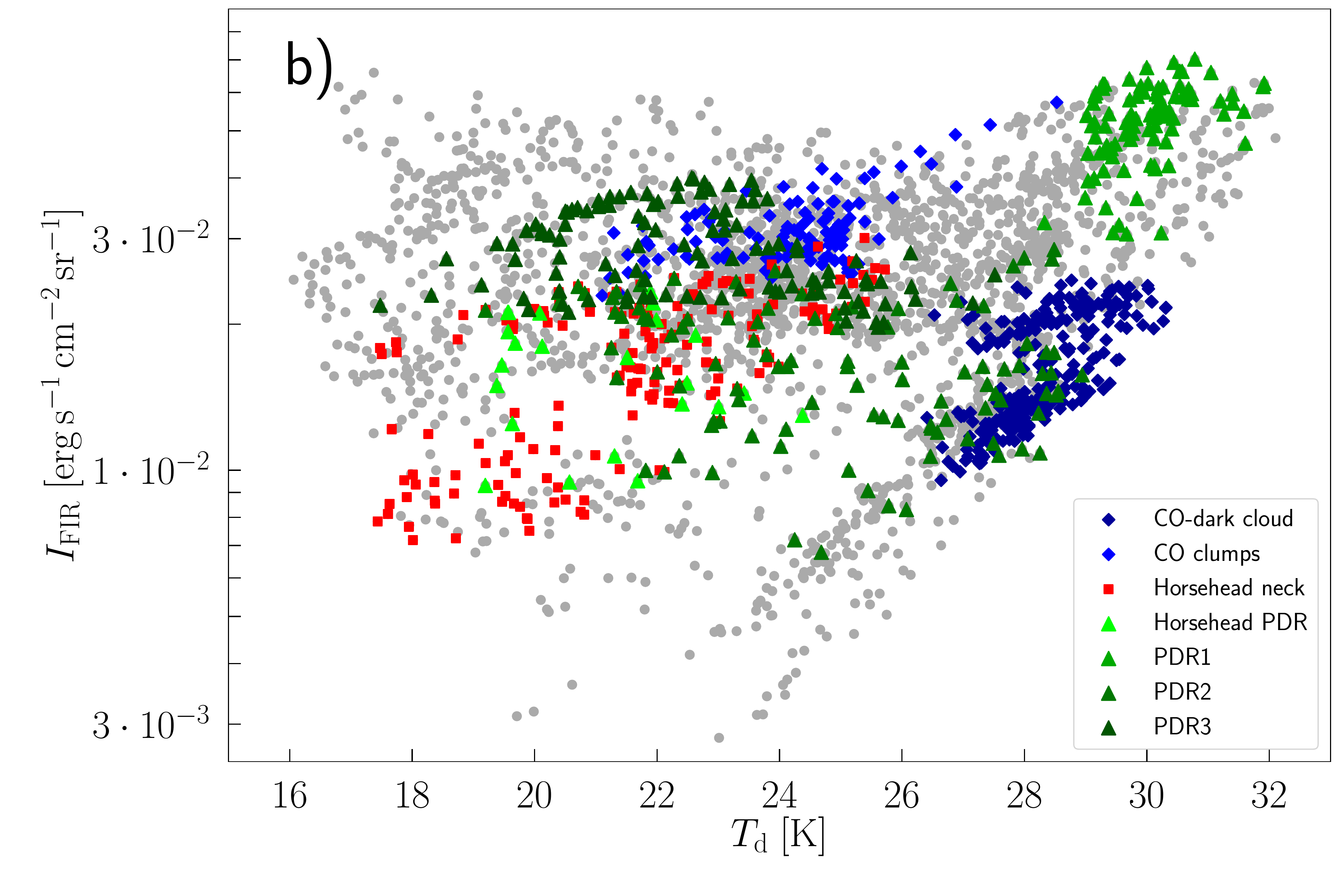}
\includegraphics[width=0.5\textwidth, height=0.33\textwidth]{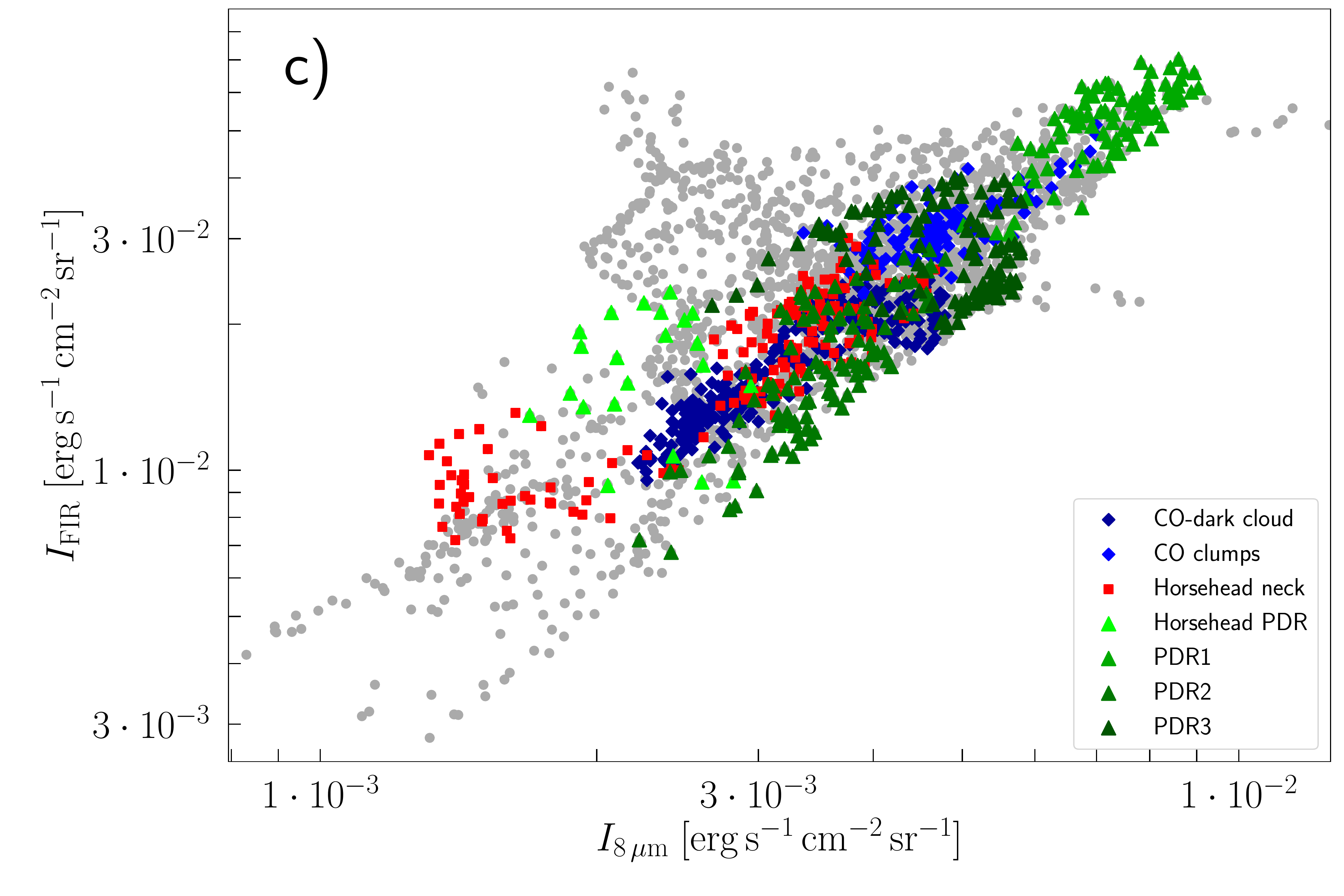}
\caption{Correlation plots of $I_{\mathrm{FIR}}$ versus $\tau_{160}$, $T_{\mathrm{d}}$, and $I_{8\,\mu\mathrm{m}}$, respectively.}
\label{Fig.FIR}
\end{figure}

FIR intensity increases with increasing $\tau_{160}$, as Fig. \ref{Fig.FIR}a shows, but temperature differences play a role. The dust is comparatively hotter in PDR1 and PDR2, and in the CO-dark cloud. The FIR intensity scatters a lot when related to $T_{\mathrm{d}}$, as can be seen from Fig. \ref{Fig.FIR}b. Opposed to that, $I_{\mathrm{FIR}}$ seems to be well-correlated to $I_{8\,\mu\mathrm{m}}$ (Fig. \ref{Fig.FIR}c).

\section{Discussion}

\subsection{[C\,{\sc ii}] Emission from the PDR}
The total [C\,{\sc ii}] luminosity from the mapped area of $\simeq 210$ square arcmin is $L_{\mathrm{total}}\simeq 14\,L_{\sun}$. The luminosity stemming from the molecular cloud is $L_{\mathrm{cloud}}\simeq 13\,L_{\sun}$ and that from the H\,{\sc ii} region is $L_{\mathrm{H\,\textsc{ii}}} \simeq 1\,L_{\sun}$. Thus, about 5\% of the total [C\,{\sc ii}] luminosity of the surveyed area originates from the H\,{\sc ii} region; 95\% stems from the irradiated molecular cloud. This compares to 20\% and 80\%, respectively, of the area. For comparison, the total FIR luminosity from the mapped area is $L_{\mathrm{FIR}}\simeq 1245\,L_{\sun}$, of which $1210\,L_{\sun}$ belong to the molecular cloud and $35\,L_{\sun}$ to the H\,{\sc ii} regio. However, a small part of the luminosity may be attributed to NGC 2023: $0.2\,L_{\sun}$ in [C\,{\sc ii}] and $35\,L_{\sun}$ in FIR luminosity. Since the $8\,\mu\mathrm{m}$ intensity as a cloud surface tracer is very well correlated to the [C\,{\sc ii}] intensity, we may conclude that in the studied region of the Orion molecular cloud complex most of the [C\,{\sc ii}] emission originates from PDR surfaces. This is in agreement with the distribution of [C\,{\sc ii}] emission in OMC1: Here, \cite{Goico} find that 85\% of the [C\,{\sc ii}] emission arises from the irradiated surface of the molecular cloud. On Galactic scales, however, \cite{Pineda} find that ionized gas contributes about 20\% and dense PDRs about 30\% to the total [C\,{\sc ii}] luminosity (the remainder stemming in equal amounts from cold H\,{\sc i} gas and CO-dark H$_2$ gas).

The [C\,{\sc ii}] line-integrated intensity $I_{\mathrm{[C\,\textsc{ii}]}}$ ranges from $10^{-5}\,\mathrm{erg\,s^{-1}\,cm^{-2}\,sr^{-1}}$ in the H\,{\sc ii} region up to a maximum of $7\cdot 10^{-4}\,\mathrm{erg\,s^{-1}\,cm^{-2}\,sr^{-1}}$ in PDR1, with an average over the mapped area of $\bar{I}=2.4\cdot10^{-4}\,\mathrm{erg\,s^{-1}\,cm^{-2}\,sr^{-1}}$. The [C\,{\sc ii}] cooling efficiency $\eta=I_{\mathrm{[C\,\textsc{ii}]}}/I_{\mathrm{FIR}}$ takes its highest values at the edge of the molecular cloud, bordering on the H$\alpha$ emitting region. Its peak value is about $3\cdot 10^{-2}$, ranging down to $3\cdot 10^{-3}$. The separation of molecular cloud and H\,{\sc ii} region emission is not trivial, since we think that the surface of the cloud is not straight, but warped. However, [C\,{\sc ii}] emission from the region exclusively associated with the ionized gas in IC 434 is very weak and has a much wider line profile (cf. Figs. \ref{Fig.spectra} and \ref{Fig.H2spectra}; see also Sec. \ref{sec.HII}). Hence, we assume that at the edge of the molecular cloud the [C\,{\sc ii}] and FIR emission from ionized gas is minor compared to emission stemming from the molecular cloud itself.

Considering the average [C\,{\sc ii}] cooling efficiencies, where beam-dilution and column-length effects should divide out, we note that PDR2 has twice as high [C\,{\sc ii}] cooling efficiency as the Horsehead PDR and PDR1 have (see Table \ref{tab:2}). We remark that PDR2 lies in a region where there still is significant H$\alpha$ emission, indicating a corrugated edge structure. Since the average [C\,{\sc ii}] emission in the H\,{\sc ii} region is quite low, we do not expect [C\,{\sc ii}] emission from the ionized gas to be responsible for the enhanced [C\,{\sc ii}] cooling efficiency in PDR2. However, $I_{\mathrm{FIR}}$ is unexpectedly low in PDR2, which may account for the mismatch in $I_{\mathrm{[C\,\textsc{ii}]}}/I_{\mathrm{FIR}}$.

\subsection{[C\,{\sc ii}] Emission from the H\,{\sc ii} region}
\label{sec.HII}
From the H$\alpha$ emission in the studied region, originating from the ionized gas to the west of the molecular cloud, we can estimate the density of the H\,{\sc ii} region \citep{Ochsendorf}. The radiated intensity of the H$\alpha$ line can be calculated by
\begin{align}
I_{\mathrm{H}\alpha} = \int\limits_0^d j_{\mathrm{H}\alpha}\,\mathrm{d}z = \int\limits_0^d \frac{4\pi j_{\mathrm{H}\beta}}{n_pn_e}\frac{j_{\mathrm{H}\alpha}}{j_{\mathrm{H}\beta}}\frac{n_pn_e}{4\pi}\,\mathrm{d}z.
\end{align}
Assuming a gas temperature of $T\simeq 10^4\,\mathrm{K}$, we use $4\pi j_{\mathrm{H}\beta}/n_pn_e = 8.30\cdot 10^{26}\,\mathrm{erg\,cm^3 s^{-1}}$ and $j_{\mathrm{H}\alpha}/j_{\mathrm{H}\beta}=2.86$ \citep{Osterbrock1989}. Further, we assume a homogeneous gas distribution along the line of sight, which we take to be $d\sim 1\,\mathrm{pc}$. Hence, we obtain
\begin{align}
I_{\mathrm{H}\alpha} \simeq 7.0 \cdot 10^{-8}\, n_{\mathrm{e}}^2 \;\mathrm{erg\,cm^{4}\,s^{-1}\,sr^{-1}},
\end{align}
where $n_p=n_e$. When a molecular cloud is photoevaporated into a cavity, as the surface of L1630 is, we expect an exponential density profile as a function of distance from the surface. Fitting an exponential to the observed H$\alpha$ emission along a line cut, we obtain a density law with $n_{e,0}=95\,\mathrm{cm}^{-2}$ at the ionization front and a scale length of $1.2\,\mathrm{pc}$, which is in good agreement with \cite{Ochsendorf}. In the surveyed area, the density varies between 60 and $100\,\mathrm{cm}^{-3}$.

Applying again $T\simeq 10^4\,\mathrm{K}$ for the gas temperature to the cooling law of [C\,{\sc ii}] (eq. (2.36) in \cite{Tielens}), we obtain
\begin{align}
n^2\Lambda \simeq 2.7\cdot 10^{-24} \frac{n_e}{1+\frac{n_{\mathrm{cr}}}{3n_e}}\;\mathrm{erg\,s^{-1}\,cm^{-3}},
\end{align}
where we assumed an ionization fraction of $x=1$ and, hence, consider collisions with electrons only; the critical density scales with $T$ and is, at $T=10^4\,\mathrm{K}$, $n_{\mathrm{cr}}\simeq 50\,\mathrm{cm}^{-3}$ \citep{Goldsmith2012}. Neglecting $\frac{n_{\mathrm{cr}}}{3n_e}$ and assuming again $d\sim 1\,\mathrm{pc}$ for the length of the line of sight, the above yields
\begin{align}
I_{\mathrm{[C\,\textsc{ii}]}} \simeq 7\cdot 10^{-5} \left(\frac{n_e}{10^2\,\mathrm{cm}^{-2}}\right)\;\mathrm{erg\,s^{-1}\,cm^{-2}\,sr^{-1}}.
\end{align}
The observed intensity values lie in the range of $10^{-5}\mhyphen 10^{-4}\,\mathrm{erg\,s^{-1}\,cm^{-2}\,sr^{-1}}$, which is, given the range of densities, in good agreement with the values derived from H$\alpha$ emission. However, it is difficult to recognize a declining trend in the [C\,{\sc ii}] intensity away from the molecular cloud, since the signal is very noisy in the H\,{\sc ii} region due to the low intensity.

\begin{figure}[htb]
\centering
\includegraphics[width=0.5\textwidth, height=0.17\textwidth]{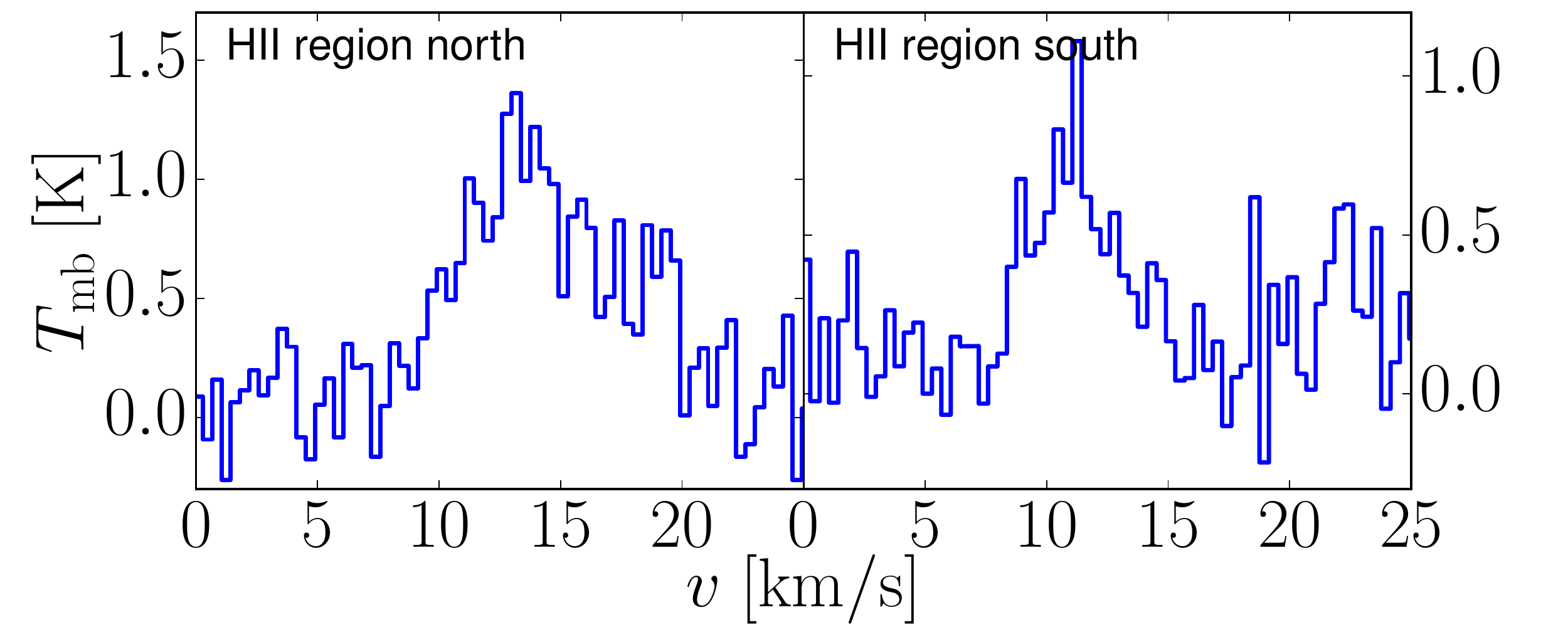}
\caption{[C\,{\sc ii}] spectra towards the H\,{\sc ii} region, averaged over 156 (left) and 187 (right) pixels. The left panel represents the H\,{\sc ii} region north of the Horsehead Nebula, the right panel represents the part south of the Horsehead Nebula. For the northern part, we obtain $T_{\mathrm{P}}=1.0\,\mathrm{K}$, $\mbox{FWHM}=8.7\,\mathrm{km\,s^{-1}}$ , and $v_{\mathrm{P}}=14.1\,\mathrm{km\,s^{-1}}$; for the southern part, a Gaussian fit yields  $T_{\mathrm{P}}=0.7\,\mathrm{K}$, $\mbox{FWHM}=5.2\,\mathrm{km\,s^{-1}}$ , and $v_{\mathrm{P}}=11.2\,\mathrm{km\,s^{-1}}$.}
\label{Fig.H2spectra}
\end{figure}

The [C\,{\sc ii}] spectra extracted from the H\,{\sc ii} region show a very weak and very broad feature (Fig. \ref{Fig.H2spectra}) with a peak main-beam temperature of $\sim 1\,\mathrm{K}$ and an FWHM of $5\mhyphen10\,\mathrm{km\,s^{-1}}$, as compared to $10\mhyphen20\,\mathrm{K}$ and $2\mhyphen4\,\mathrm{km\,s^{-1}}$ for the PDR regions in the molecular cloud. This is distinct from the spectra taken towards the cloud. We cannot distinguish a broad feature in the spectra taken towards the molecular cloud, although there is some H$\alpha$ emission and we should expect ionized gas in front of the multiple PDR surfaces. Most likely the intensity is simply too low, approximately ten times lower than the intensity towards the H\,{\sc ii} region, assuming that the H\,{\sc ii} column in front of the molecular cloud as seen along the line of sight is $\sim 0.1\,\mathrm{pc}$, which renders the signal undetectable.

\subsection{FIR emission and beam-dilution effects}
\label{sec.FIR}
We expect that beam dilution affects all maps to some extent when convolved to the SPIRE $500\,\mu\mathrm{m}$ $36\arcsec$ resolution, since the unconvolved IRAC $8\,\mu\mathrm{m}$ map at $1.98\arcsec$ resolution reveals features and delicate structures (see Fig. \ref{Fig.cuts}) that disappear upon convolution. The upGREAT beam has an FWHM of $15.9\arcsec$, thus beam dilution might be noticeable in [C\,{\sc ii}] observations towards thin filaments. From the IRAC $8\,\mu\mathrm{m}$ emission, we infer a dilution factor of $\sim 2.5$ for the Horsehead PDR going from the native resolution of the $8\,\mu\mathrm{m}$ image to $36\arcsec$ resolution, but only for the narrowest (densest) part of the PDR. The average $8\,\mu\mathrm{m}$ emission is not significantly beam-diluted when convolved to $36\arcsec$. PDR3 possibly suffers significantly from beam dilution as well, since it shows as a rather sharp filament in $8\,\mu\mathrm{m}$, where it reaches high peak values (higher than the Horsehead PDR peak values). Values of quantities we observe and compute from that are taken to be beam-averaged values in the respective resolution.

Due to the edge-on geometry of the PDRs in L1630 with respect to the illuminating star system $\sigma$ Ori (and the low dust optical depth), we expect that $I_{\mathrm{FIR}}$ depends on the re-radiating column along the line of sight, which might explain the excess intensity in $8\,\mu\mathrm{m}$, [C\,{\sc ii}], and FIR in PDR1 as compared to other PDR surfaces here. The commonly expected value of incident FUV radiation, re-emitted in $I_{\mathrm{FIR}}$, is $G_0\simeq 100$, calculated from properties of $\sigma$ Ori (\cite{Abergel2003} and references therein). Given the edge-on geometry of the cloud-star complex, the formula by \cite{HollenbachTielens},
\begin{align}
G_0=\frac{1}{2}\frac{I_{\mathrm{FIR}}}{1.3\cdot 10^{-4}\, \mathrm{erg\,s^{-1}\,cm^{-2}\,sr^{-1}}} \label{eq.G0},
\end{align}
cannot be used to infer the intensity of the incident UV radiation. The FIR intensity varies substantially across the mapped area; for a face-on geometry with a single UV-illuminating source, one would expect less divergent values. This realization corroborates the assumption of an edge-on geometry and has been the rationale for building edge-on models with different molecular-cloud depths along the line of sight.

The dust optical depth $\tau_{160}$ does not trace the PDR surface column, but the total gas and dust column. The FIR intensity does not increase with increasing $\tau_{160}$ for PDR1, PDR2, and PDR3 (in comparison to $\tau_{160}\simeq 2\cdot 10^{-3},\,1\cdot 10^{-3},\,3\cdot 10^{-3}$, respectively\footnote{PDR2 partially overlaps with the dense region of the neck of the Horsehead Nebula; hence we take $\tau_{160}$ for PDR2 from the region where it does not overlap.}). Especially in the case of PDR3, $\tau_{160}$ likely traces not only the PDR but the molecular cloud interior, as well. A fraction of $I_{\mathrm{FIR}}$ might stem from deeper, cooler parts of the molecular cloud and not from the PDR surface, although $I_{\mathrm{FIR}}$ is biased towards the hot PDR surface, as is $\tau_{160}$. In PDR2, $I_{\mathrm{FIR}}$ is lower than we would expect, leading to significantly higher $I_{\mathrm{[C\,\textsc{ii}]}}/I_{\mathrm{FIR}}$ than in PDR1 and PDR3. The dust temperature $T_{\mathrm{d}}$ in PDR2 is determined to be considerably lower than in PDR1, although the environments seem similar.

\subsection{Column densities, gas temperature, and mass}
\label{sec.column}
Since we do not detect the [$^{13}$C\,{\sc ii}] line in single spectra, we cannot determine the [C\,{\sc ii}] optical depth by means of it. The noise rms of the spectra is too high to put a significant constraint on $\tau_{\mathrm{[C\,\textsc{ii}]}}$. In averaged spectra, we can detect the [$^{13}$C\,{\sc ii}] $F=2\mhyphen 1$ line just above the noise level (see Sec. \ref{sec.13CII}). However, knowing the $\mathrm{C}^+$ column density and the intrinsic line width, we can estimate the [C\,{\sc ii}] optical depth and the excitation temperature (see Appendix \ref{app.1}) from single spectra. We compute the $\mathrm{C}^+$ column density of the PDR surface from the dust optical depth, assuming standard dust properties and that all carbon in the PDR surface is singly ionized. Additionally, we expect beam dilution to be insignificant for the [C\,{\sc ii}] observations. From the native IRAC $8\,\mu\mathrm{m}$ map, we infer a dilution factor of 1.5 when going to $15.9\arcsec$ resolution, but only towards the thinnest filament in the Horsehead mane. This equally yields a peak temperature of $T_{\mathrm{P}} \simeq 20\,\mathrm{K}$ there, as does the brightest part of the Horsehead PDR.

The gas column density can be computed from the dust optical depth $\tau_{160}$, assuming a theoretical absorption coefficient. This yields
\begin{align}
N_{\mathrm{H}} \simeq \frac{100\,\tau_{160}}{\kappa_{160}m_{\mathrm{H}}}\simeq 5\cdot 10^{24}\,\mathrm{cm}^{-2}\;\tau_{160},
\end{align}
where we have used a gas-to-dust mass ratio of 100 and assumed $\kappa_{\mathrm{abs}}=2.92\cdot 10^{5}(\lambda\,[\mu\mbox{m}])^{-2}\,\mathrm{cm}^2\mbox{/g}$ \citep{Draine2001}. With the fractional gas-phase carbon abundance $[\mathrm{C}/\mathrm{H}] = 1.6\cdot 10^{-4}$ \citep{SofiaApril2004}, we can estimate the $\mathrm{C}^+$ column density in the PDR surface from the dust optical depth, under the assumption that all carbon in the line of sight is ionized:
\begin{align}
N_{\mathrm{C}^+} \simeq 8\cdot 10^{20}\,\mathrm{cm}^{-2}\;\tau_{160}.
\end{align}
Later studies have reported somewhat differing values for $[\mathrm{C}/\mathrm{H}]$, varying by a factor of two for different sight lines (see e.g. \cite{Sofia2009,Sofia2011}). However, the average is not found to deviate substantially from the earlier value of $[\mathrm{C}/\mathrm{H}] = 1.6\cdot 10^{-4}$; the general uncertainty seems to be quite large. We discuss the effect of the uncertainty in the column density on the derived gas properties in the following. For PDR1 and PDR2 we have $\tau_{160}\simeq 2\cdot 10^{-3}$ and $\tau_{160}\simeq 10^{-3}$, respectively, from the $\tau_{160}$ map, where we assume that all the dust actually is in the PDR surface. However, these values for the dust optical depth may be affected by significant uncertainties, up to a factor of two, since $\tau_{160}$ depends on the assumed dust properties in the SED fit (see discussion in Sec. \ref{sec.SED}). For PDR3, we suppose that there is a significant amount of cold material located along the line of sight, which renders $\tau_{160}$ an inaccurate measure for the depth of the PDR along the line of sight here.

Inferring the PDR dust optical depth of the Horsehead PDR requires further effort. According to \cite{Habart2005}, there is a large density gradient from the surface to the bulk of the Horsehead PDR. In the surface the gas density might be as low as $n_{\mathrm{H}}\sim 10^4\,\mathrm{cm}^{-3}$, whereas in the bulk it assumes $n_{\mathrm{H}}\sim 2\cdot 10^5\,\mathrm{cm}^{-3}$. \cite{Abergel2003} find $n_{\mathrm{H}}\sim 2\cdot 10^4\,\mathrm{cm}^{-3}$ as a lower limit for the density of the gas directly behind the illuminated filament. The dust optical depth is likely to be beam diluted in the SPIRE $500\,\mu\mathrm{m}$ resolution; the filament has an extent of only $5\mhyphen 10\arcsec$. Assuming that the maximum $\tau_{160}\simeq 10^{-2}$ occurs in the densest (inner) part of the PDR and that the length along the line of sight remains approximately the same, we conclude that the dust optical depth in the Horsehead PDR surface must be significantly lower than the maximum value, by a factor of approximately ten, due to the decreased density.

From deep integration of the Horsehead PDR with SOFIA/upGREAT, however, we are able to infer a [C\,{\sc ii}] optical depth of $\tau_{\mathrm{[C\,\textsc{ii}]}}\simeq 2$ from the brightest [$^{13}$C\,{\sc ii}] line, which can be detected in these data (C. Guevara, priv. comm.; paper in prep. C. Guevara, J. Stutzki et al.). According to Eq. \ref{eq.tau2}, this translates into a $\mathrm{C}^+$ column density of $N_{\mathrm{C}^+}\simeq 7\cdot 10^{17}\,\mathrm{cm}^{-2}$, that is $\tau_{160}\simeq 10^{-3}$, which is ten times lower than the maximum value in the Horsehead bulk. However, from our models, this [C\,{\sc ii}] optical depth corresponds to a twice as large $\mathrm{C}^+$ column density of $N_{\mathrm{C}^+}\simeq 1.6\cdot 10^{18}\,\mathrm{cm}^{-2}$, if we assume that all carbon is ionized within our beam, which might not be the case.

We calculate the [C\,{\sc ii}] optical depth $\tau_{\mathrm{[C\,\textsc{ii}]}}$ and excitation temperature $T_{\mathrm{ex}}$ for PDR1 and PDR2, and $T_{\mathrm{ex}}$ for the Horsehead PDR, using the formulas of Appendix \ref{app.1}. \footnote{We have checked that including raditative excitation by the dust FIR background is insignificant.} The results are shown in Table \ref{tab:3}. In principle, the values we infer for $N_{\mathrm{C}^+}$ are upper limits, but the general uncertainty in the $\mathrm{C}^+$ column density is potentially larger than the deviation from the upper limit. The dust optical depth, from which we calculate the $\mathrm{C}^+$ column density, is not well-determined (cf. Sec. \ref{sec.SED}) and the carbon fractional abundance may deviate. If we assume an error margin of the $\mathrm{C}^+$ column density of $\pm$50\%, this results in ranges of [C\,{\sc ii}] optical depth and excitation temperature of $\tau_{\mathrm{[C\,\textsc{ii}]}}\simeq 0.6\mhyphen 2.7$ and $T_{\mathrm{ex}}\simeq 60\mhyphen 90\,\mathrm{K}$, respectively, for PDR1, and $\tau_{\mathrm{[C\,\textsc{ii}]}}\simeq 0.4\mhyphen 2.2$ and $T_{\mathrm{ex}}\simeq 60\mhyphen 100\,\mathrm{K}$, respectively, for PDR2. Certainly, these values are subject to uncertainties in the inferred gas density and the spectral parameters, as well; however, the uncertainty in $N_{\mathrm{C}^+}$ seems to be the most significant, leading to considerable deviations, so we will not discuss the influence of the other uncertainties here. In the Horsehead PDR, the precise value of $\tau_{\mathrm{[C\,\textsc{ii}]}}$ does not overly influence the excitation temperature which we calculate: it is $T_{\mathrm{ex}}\simeq 60\pm 2\,\mathrm{K}$.\\

\begin{table}[ht]
\centering
\begin{tabular}{cccccc}
 & RA & Dec & $N_{\mathrm{C}^+}$ & $\tau_{\mathrm{[C\,\textsc{ii}]}}$ & $T_{\mathrm{ex}}$ \\
pos. & (J2000) & (J2000) & [$\mathrm{cm}^{-2}$] & & [K] \\ \hline \rule{0pt}{3ex}\noindent
A & $5\mbox{h}40\arcmin 53\arcsec$ & $-2^{\circ}27\arcmin 37\arcsec$ & $7\cdot 10^{17}$ & 2.0 & 58 \\
B & $5\mbox{h}41\arcmin 00\arcsec$ & $-2^{\circ}20\arcmin 27\arcsec$ & $1.6\cdot 10^{18}$ & 1.7 & 63 \\
F & $5\mbox{h}41\arcmin 06\arcsec$ & $-2^{\circ}30\arcmin 46\arcsec$& $8\cdot 10^{17}$ & 1.4 & 65 \\
\end{tabular}
\vspace{0.2cm}
\caption{Results from solving Eq. (\ref{eq.tau2}). Position A corresponds to the Horsehead PDR, B is PDR1, F lies in PDR2. The spectrum at point B consists of two components, but here we consider only the main component (the second component is much weaker). Table \ref{tab:1} shows the spectral parameters $T_{\mathrm{P}}$ and $\Delta v_{\mathrm{FWHM}}$.}
\label{tab:3}
\end{table}

If the density of the gas is known, one can compute the gas temperature from the excitation temperature:
\begin{align}
T=\frac{T_{\mathrm{ex}}}{1-\frac{T_{\mathrm{ex}}}{91.2\mathrm{ K}}\ln(1+\frac{n_{\mathrm{cr}}}{n})},
\end{align}
with the critical density $n_{\mathrm{cr}}=\beta(\tau_{\mathrm{[C\,\textsc{ii}]}})\frac{A}{\gamma_{\mathrm{ul}}}$, where $\beta(\tau_{\mathrm{[C\,\textsc{ii}]}})$ is the [C\,{\sc ii}] $158\,\mu\mathrm{m}$ photon escape probabilty, $A\simeq 2.3\cdot 10^{-6}\,\mathrm{s^{-1}}$ is the Einstein coefficient for spontaneous radiative de-excitation of C$^+$, and $\gamma_{\mathrm{ul}}$ is the collisional de-excitation rate coefficient, which is $\gamma_{\mathrm{ul}}\simeq 7.6\cdot 10^{-10}\,\mathrm{cm^3\,s^{-1}}$ for C$^+$--H collisions \citep{Goldsmith2012} and $\gamma_{\mathrm{ul}}\simeq 5.1\cdot 10^{-10}\,\mathrm{cm^3\,s^{-1}}$ for C$^+$--H$_2$ collisions \citep{WiesenfeldGoldsmith2014} at gas temperatures of $\simeq 100\,\mathrm{K}$; $n$ is the collision partner density. At $T\sim 100\,\mathrm{K,}$ $\gamma_{\mathrm{ul}}$, and thereby $n_{\mathrm{cr}}$, is only weakly dependent on temperature; for C$^+$--H collisions, $n_{\mathrm{cr}}\simeq \beta(\tau_{\mathrm{[C\,\textsc{ii}]}})\cdot 3.0\cdot 10^{3}\,\mathrm{cm}^{-3}$, and for C$^+$--H$_2$ collisions, $n_{\mathrm{cr}}\simeq \beta(\tau_{\mathrm{[C\,\textsc{ii}]}})\cdot 4.5\cdot 10^{3}\,\mathrm{cm}^{-3}$. At the cloud edge, collisions with H dominate the excitation of C$^+$, while deeper into the cloud H$_2$ excitation dominates. We expect that excitation caused by collisions involving both H and H$_2$ contribute within our beam. Since the points we chose in PDR1 and PDR2 lie close to the surface, we consider collisions with H; in the Horsehead PDR the choice of collision partner does not affect the derived gas temperature significantly due to the higher gas density. We take the photon escape probability to be $\beta(\tau)=(1-e^{-\tau})/\tau$, as used in our edge-on PDR models. The densities are discussed in Sec. \ref{sec.crosscuts}.

Using $n_{\mathrm{H}}\simeq 3\cdot 10^{3} \,\mathrm{cm}^{-3}$ for PDR1 and PDR2, we obtain $T\simeq 86\,\mathrm{K}$ and $T\simeq 93\,\mathrm{K}$, respectively, for C$^+$--H collisions. In the Horsehead PDR, we compute $T\simeq 60\,\mathrm{K}$. From models (see Secs. \ref{sec.model_description} and \ref{sec.models}) we compute a gas temperature of about $T\simeq 100\mhyphen140\,\mathrm{K}$ in the top layers of a PDR (cf. Fig. \ref{Fig.model}). This is in reasonable agreement with the values derived from our observations in PDR1 and PDR2. In the Horsehead PDR, however, the results likely are affected by beam dilution, since the gas temperature drops quickly on the physical scale (within the beam size of $15.9\arcsec$ of the [C\,{\sc ii}] observations). Disregarding this, with a low $\mathrm{C}^+$ column density of $N_{\mathrm{C^+}}=1.6\cdot 10^{17}\,\mathrm{cm}^{-2}$ we can match the gas temperature measured by \cite{Habart2011} from H$_2$ observations, $T\simeq 264\,\mathrm{K}$; this results in a [C\,{\sc ii}] optical depth of $\tau_{\mathrm{[C\,\textsc{ii}]}}\simeq 0.1$. We can fit a gas temperature of $T\simeq 120\,\mathrm{K}$, as predicted by our models for conditions appropriate for the Horsehead PDR, with a column density of $N_{\mathrm{C^+}}\simeq 2.0\cdot 10^{17}\,\mathrm{cm}^{-2}$. This would yield a [C\,{\sc ii}] optical depth of $\tau_{\mathrm{[C\,\textsc{ii}]}}\simeq 0.3$. Both values are significantly lower than what is observed in [$^{13}$C\,{\sc ii}].

According to \cite{Goldsmith2012}, the [C\,{\sc ii}] line is effectively optically thin, meaning the peak temperature is linearly proportional to the C$^+$ column density, if $T_{\mathrm{P}} < J(T)/3$, where $J(T)$ is the brightness temperature of the gas. Hence, even though our derived [C\,{\sc ii}] optical depth for PDR1 and PDR2 is $>1$, the line is still effectively optically thin in these regions. It is optically thick in the Horsehead PDR.

From the H column densities, that is, from the dust optical depth, we can estimate the mass of the gas: $M_{\mathrm{gas}}=N_{\mathrm{H}} m_{\mathrm{H}} A$, where $A$ is the surface that is integrated over. The total gas mass of the molecular cloud (not including the H\,{\sc ii} region IC 434 and the north-eastern corner of gas and dust associated with NGC 2023) in the [C\,{\sc ii}] mapped area is $M_{\mathrm{gas,total}}\simeq 280\,M_{\sun}$. The Horsehead Nebula and its shadow add $M_{\mathrm{gas}}\simeq 33\,M_{\sun}$ to the total mass, the Horsehead PDR surface contributing $M_{\mathrm{gas}}\simeq 3\,M_{\sun}$. The mass of CO-emitting gas (also computed from $N_{\mathrm{H}}$) is $M_{\mathrm{gas}}\simeq 250\,M_{\sun}$. The assumption that CO-emitting gas contributes the bulk of the mass and that [C\,{\sc ii}] traces PDR surfaces yields a gas mass of PDR surfaces of $M_{\mathrm{gas}}\simeq 20\,M_{\sun}$. Additionally, $\simeq 12\,M_{\sun}$ are located in gas that emits neither strongly in [C\,{\sc ii}] nor in CO. The H\,{\sc ii} region comprises $M_{\mathrm{gas}}\simeq 10\,M_{\sun}$ as derived from the dust optical depth; from H$\alpha$ emission, we estimate $M_{\mathrm{gas}}\simeq 5\,M_{\sun}$.

Table \ref{tab:4} compares the masses computed from the dust optical depth and from CO$\,(1\mhyphen 0$) emission, respectively, for the several regions we defined. Since the uncertainties (in the dust optical depth, but also in the conversion factors, especially in the $X_{\mathrm{CO}}$ factor \citep[cf.][]{Bolatto2013}) are substantial, we cannot draw clear-cut conclusions. We note, moreover, that PDR regions might overlap along the line of sight with CO-emitting regions whose emission stems from a different layer of the molecular cloud, as certainly is the case in PDR3. It seems that there is a fair amount of gas mass unaccounted for by CO emission, but within the error margin of 30\%, this mass can be between 10 and $100\,M_{\sun}$. In a larger area mapped in CO, comprising the area mapped in [C\,{\sc ii}], \cite{Pety2017} find that the CO-traced mass is in fact greater than the dust-traced mass; this especially influences the deeper layers of the Orion B molecular cloud (see Table 4 in \cite{Pety2017}), thus these findings might not apply to our study field in L1630. However, they conclude that CO tends to overestimate the gas mass, whereas the dust optical depth underestimates it. In this way, with an area of $\sim 2\,\mathrm{pc}^2$, we find a gas mass surface density of $100\mhyphen 150\,M_{\sun}\,\mathrm{pc}^{-2}$ in the [C\,{\sc ii}] mapped area.

\begin{table}[ht]
\centering
\begin{tabular}{lcc}
Region & $M_{\mathrm{dust}}$ [$M_{\sun}$] & $M_{\mathrm{CO}}$ [$M_{\sun}$] \\ \hline \rule{0pt}{3ex}\noindent
entire map & 294 & 200 \\
L1630 & 283 & 197 \\ 
Horsehead PDR & 3 & 1 \\
PDR1 & 5 & 3 \\ 
PDR2 & 6 & 4 \\ 
PDR3 & 17 & 13 \\
CO-dark cloud & 5 & 2 \\
CO clumps & 12 & 14 \\
H\,{\sc ii} region & 11 & 3 \\
CO-emitting gas & 251 & 175 
\end{tabular}
\vspace{0.2cm}
\caption{Masses within the several regions defined in Sec. \ref{sec.GlobMorph} calculated from the dust optical depth $\tau_{160}$ and from the $\mathrm{CO}\,(1\mhyphen 0)$ intensity $I_{\mathrm{CO\,(1\mhyphen 0)}}$. For the latter we use an $X_{\mathrm{CO}}$ of $2\cdot 10^{20}\,\mathrm{cm^{-2}\,(K\,km\,s^{-1})^{-1}}$ (\cite{Bolatto2013} and references therein). "CO-emitting gas" refers to the part of the map with extended CO emission, that is, the deeper layers of L1630.}
\label{tab:4}
\end{table}

\subsection{Excitation properties from [$^{13}$C\,{\sc ii}]}
\label{sec.13CII}

\begin{figure}[htb]
\centering
\includegraphics[width=0.5\textwidth, height=0.17\textwidth]{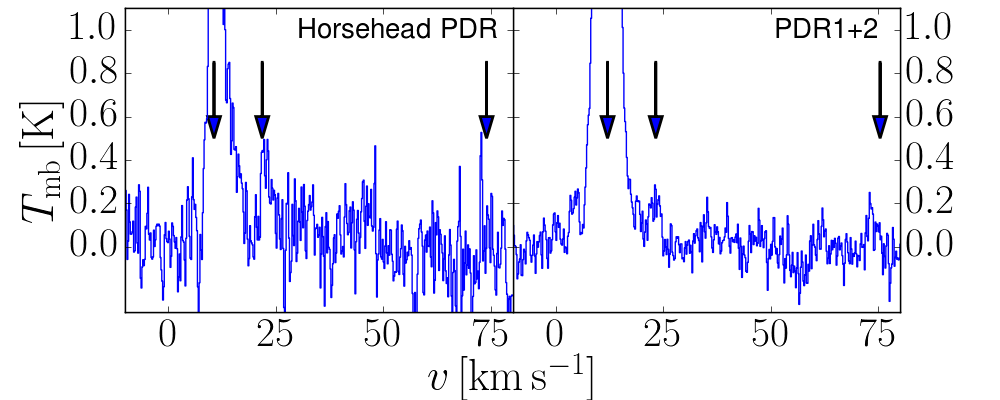}
\caption{[C\,{\sc ii}] spectra towards the Horsehead PDR and PDR1+2, averaged over 180 (left) and 3140 (right) pixels. Arrows indicate the positions of the [$^{12}$C\,{\sc ii}] line and the [$^{13}$C\,{\sc ii}] $F=2\mhyphen 1$ and $F=1\mhyphen 1$ lines (from left to right); the [$^{13}$C\,{\sc ii}] $F=1\mhyphen 0$ line falls outside the spectral range of our map.}
\label{Fig.13Cspectra}
\end{figure}

If we average over a large number of spectra in [C\,{\sc ii}]-bright areas, we are able to identify the [$^{13}$C\,{\sc ii}] $F=2\mhyphen 1$ line above the $5\sigma$ level (see Fig. \ref{Fig.13Cspectra}); we cannot detect the other two (weaker) [$^{13}$C\,{\sc ii}] lines. From the [$^{13}$C\,{\sc ii}] $F=2\mhyphen 1$ line, we can compute average values of the [C\,{\sc ii}] optical depth, the excitation temperature, and the $\mathrm{C}^+$ column density. From an average spectrum of the [C\,{\sc ii}]-bright regions of PDR1 and PDR2 (an area of 50 square arcmin, which is a third of the cloud area and corresponds to 3140 spectra), we obtain $\tau_{\mathrm{[C\,\textsc{ii}]}}\simeq 1.5$ (with an uncertainty of 20\%). To obtain this result, we used [$^{13}$C\,{\sc ii}] line parameters established by \cite{Ossenkopf} and the $^{12}$C/$^{13}$C isotopic ratio of 67 for Orion \citep{Langer1990,Milam2005}. This yields a $\mathrm{C}^+$ column density of $N_{\mathrm{C}^+}\simeq 3\cdot 10^{17}\,\mathrm{cm}^{-2}$. From an average over the Horsehead PDR (an area of 3 square arcmin, corresponding to 180 spectra), we obtain $\tau_{\mathrm{[C\,\textsc{ii}]}}\simeq 5$ (also with 20\% uncertainty) and $N_{\mathrm{C}^+}\simeq 1\cdot 10^{18}\,\mathrm{cm}^{-2}$. We note that $\tau_{\mathrm{[C\,\textsc{ii}]}}$ in PDR1 and PDR2 does match the value calculated in Sec. \ref{sec.column} from the dust optical depth and single spectra, and $N_{\mathrm{C}^+}$ does not, whereas in the Horsehead PDR $\tau_{\mathrm{[C\,\textsc{ii}]}}$ does not agree, but $N_{\mathrm{C}^+}$ does. Excitation temperatures from the averaged spectra are $T_{\mathrm{ex}}\simeq 40\,\mathrm{K}$, which is lower than what we infer from single spectra. Regions with lower-excitation [C\,{\sc ii}] contribute to the averaged spectra, but we have to include them to obtain a sufficient signal-to-noise ratio. We stress that this spectral averaging over a large area with varying conditions will weigh the emission differently for the [$^{12}$C\,{\sc ii}] and [$^{13}$C\,{\sc ii}] lines in accordance with the excitation temperatures and optical depths involved. Hence, the averaged spectrum will not be the same as the spectrum of the average. In our analysis, we have elected to rely on the analysis based on the dust column density rather than this somewhat ill-defined average.

\subsection{Photoelectric heating and energy balance}
\label{sec.heating}
\begin{figure}[ht]
\includegraphics[width=0.5\textwidth, height=0.33\textwidth]{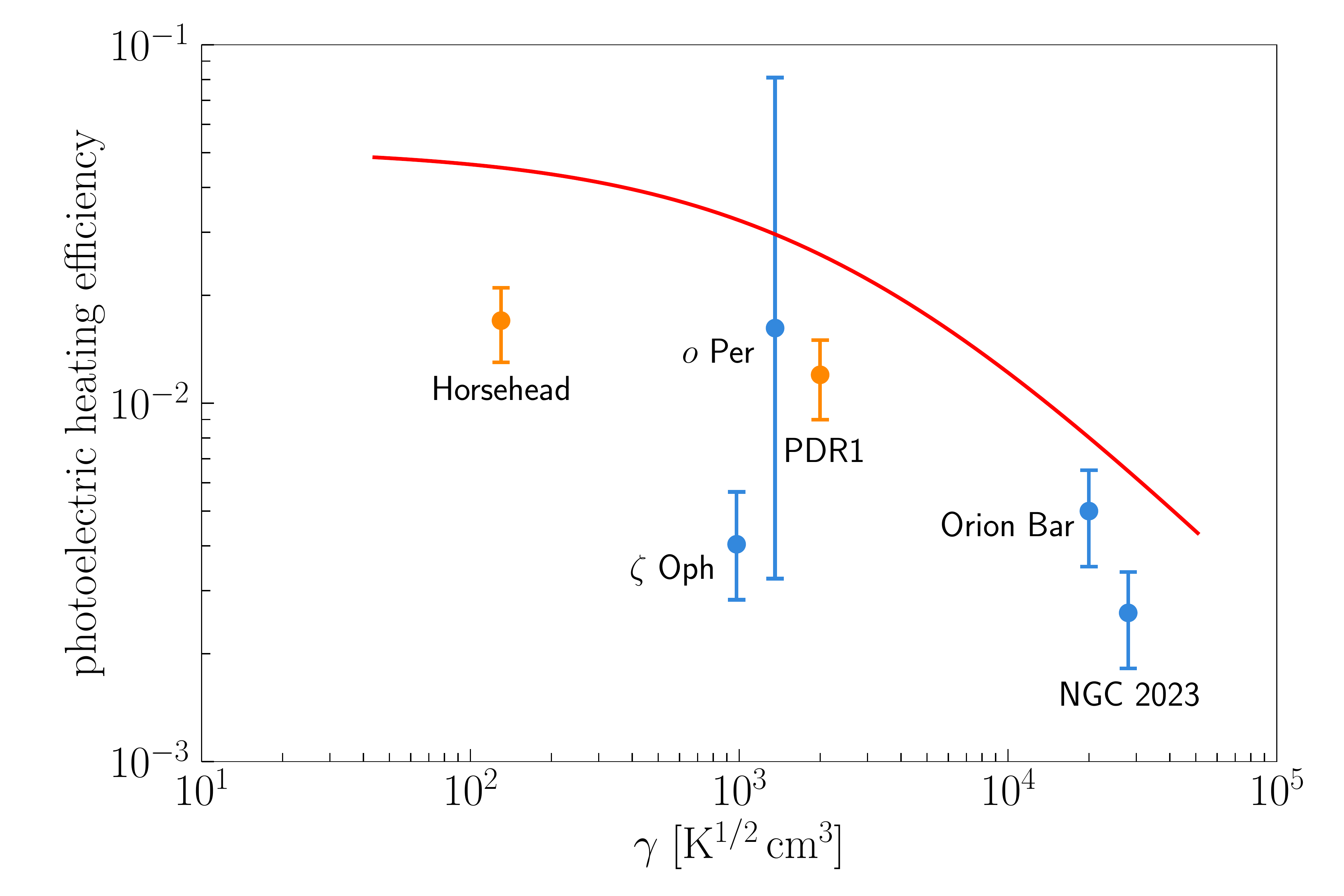}
\caption{Theoretical photoelectric heating efficiency of PAHs, clusters of PAHs, and very small grains is plotted against the ionization parameter $\gamma=G_0T^{0.5}/n_e$ \citep{BakesTielens1994}. We added the orange data points for the Horsehead PDR and PDR1. Blue data points are for the diffuse-ISM sight lines $\zeta$ Oph and $o$ Per, and the PDRs Orion Bar and NGC 2023. Figure adapted from \cite{Tielens2008}.}
\label{Fig.pe}
\end{figure}

The most substantial heating source of PDRs is photoelectric heating by PAHs, clusters of PAHs, and very small grains. The photoelectric heating rate is deeply built into PDR models and controls the detailed structure and emission characteristics to a large extent. The heating rate drops with increasing ionization of these species. It can be parametrized by the ionization parameter $\gamma=G_0T^{0.5}/n_e$, where $G_0$ is the incident radiation, $T$ is the gas temperature, and $n_e$ is the electron density; the corresponding theoretical curve as derived by \cite{BakesTielens1994} is shown in Fig. \ref{Fig.pe}. \cite{Okada2013} confirm in a study of six PDRs, which represent a variety of environments, the dependence of the photoelectric heating rate on PAH ionization and conclude on the dominance of photoelectric heating by PAHs.

The [C\,{\sc ii}] $158\,\mu\mathrm{m}$ cooling rate increases with gas density and temperature. In the high-density limit ($n_{\mathrm{H}} \gg n_{\mathrm{cr}}$), it scales with $n_{\mathrm{H}}$, whereas for low densities it scales with $n_{\mathrm{H}}^2$; the temperature dependence is largely captured in a factor $\exp(-\Delta E/k_BT)$, where $\Delta E$ is the energy level spacing.\footnote{The same is true of other cooling lines, for example, the [O\,{\sc i}] $63\,\mu\mathrm{m}$ line, with differing critical densities, however.} The gas density of the Horsehead PDR, $n_{\mathrm{H}}\simeq 4\cdot 10^{4} \,\mathrm{cm}^{-3}$, lies above the critical density for C$^+$, $n_{\mathrm{cr}}=\beta(\tau_{\mathrm{[C\,\textsc{ii}]}})\cdot 3.0\cdot 10^{3}\,\mathrm{cm}^{-3}$; the densities of PDR1 and PDR2 are close to $n_{\mathrm{cr}}$, that is, in the intermediate density regime. The similar gas temperatures and densities of PDR1 and PDR2, however, do not reflect the difference in the [C\,{\sc ii}] cooling efficiencies $I_{\mathrm{[C\,\textsc{ii}]}}/I_{\mathrm{FIR}}$ calculated in Table \ref{tab:2} and visualized in Fig. \ref{Fig.corr}a (cf. Sec. \ref{sec.FIR}). In the case of the Horsehead PDR, gas cooling through the [O\,{\sc i}] $63\,\mu\mathrm{m}$ line becomes important; the [O\,{\sc i}] surface brightness is comparable to the surface brightness of the [C\,{\sc ii}] line \citep{Goico09}.

We emphasize that we can directly measure the temperature and the density of the emitting gas in the PDR from our observations (see Secs. \ref{sec.column} and \ref{sec.crosscuts}, respectively). Hence, we can test the theory in a rather direct way. Specifically, we assume that all electrons come from C ionization, hence $n_e=1.6\cdot 10^{-4}\,n_{\mathrm{H}}$, where we have adopted the gas-phase abundance of carbon estimated by \cite{SofiaApril2004}. For the Horsehead PDR with $n_{\mathrm{H}}\simeq 4\cdot 10^4\,\mathrm{cm}^{-3}$ and $T\simeq 60\,\mathrm{K,}$ we compute an ionization parameter of $\gamma \simeq 1 \cdot 10^2\,\mathrm{K^{1/2}\,cm^{3}}$; for PDR1 and PDR2 with $n_{\mathrm{H}}\simeq 3\cdot 10^3\,\mathrm{cm}^{-3}$ and $T\simeq 100\,\mathrm{K,}$ we obtain $\gamma \simeq 2 \cdot 10^3\,\mathrm{K^{1/2}\,cm^{3}}$. From the cooling lines, that is, assuming that all the heating is converted into [C\,{\sc ii}] and [O\,{\sc i}] emission, we arrive at a heating efficiency of $1.7\pm 0.4\cdot 10^{-2}$ (with $I_{\mathrm{[O\,\textsc{i}]}}\simeq 1.04\pm 0.14\,\mathrm{erg\,s^{-1}\,cm^{-2}\,sr^{-1}}$ from \cite{Goico09} at similar spatial resolution) for the Horsehead PDR, and $1.1\pm 0.3\cdot 10^{-2}$ for PDR1. From their [O\,{\sc i}] study of the Horsehead Nebula, \cite{Goico09} find a heating efficiency of $1\mhyphen 2\cdot 10^{-2}$, which is consistent with our findings. For PDR2, we calculate an average [C\,{\sc ii}] cooling efficiency of $2.2\pm 0.4\cdot 10^{-2}$. However, $I_{\mathrm{FIR}}$ is unexpectedly low in PDR2, so we wonder whether the mismatch in $I_{\mathrm{[C\,\textsc{ii}]}}/I_{\mathrm{FIR}}$ between PDR1 and PDR2 really is due to an erroneous determination of $I_{\mathrm{FIR}}$ and is thereby deceptive.

The general behavior of the observationally obtained heating efficiency is indeed very similar to the theoretical curve for photoelectric heating by PAHs, clusters of PAHs, and very small grains, as shown in Fig. \ref{Fig.pe}, except that theoretical values seems to be offset to higher efficiency by about a factor of two. Such a shift might reflect a somewhat different abundance of PAHs and related species in the studied regions. We note that these differences between theory and observations can lead to considerable differences in the derived physical conditions. For example, adopting the theoretical relationship and solving the energy balance for the gas density and FUV field appropriate for the Horsehead PDR would result in a derived gas temperature of $T\simeq 100\,\mathrm{K}$, while the temperature as measured from the pure rotational H$_2$ lines by \cite{Habart2011} is $T\simeq 264\,\mathrm{K}$, and we determine it to be $T\simeq 60\,\mathrm{K}$, which is beam-averaged. For PDR1 and PDR2, the theoretical relationship would imply a temperature of $T\simeq 125\,\mathrm{K}$, while we measure $T\simeq 90\,\mathrm{K}$ from the peak [C\,{\sc ii}] intensity. From a theoretical perspective, we expect the temperature to decrease with increasing density (cf. Fig. 9.4 in Tielens (2010)). This is what we see in [C\,{\sc ii}] observations. However, from studies by \cite{Habart2011} and \cite{Habart2005}, the observational temperature lies in the regime $T\simeq 200\mhyphen 300\,\mathrm{K}$, albeit in a very narrow surface gas layer. Here, the observational temperature in the Horsehead PDR appears to be higher than the theoretical temperature, despite the heating efficiency being underestimated. This may seem inconsistent; it certainly suggests that we have to be very careful with the assumptions we make. However, the discrepancy may be due to various reasons, not all of them implying inconsistency, that we will not discuss in this paper. Calculating the ionization parameter with either differing observational temperatures or the theoretical temperature does not shift the data points in Fig. \ref{Fig.pe} significantly towards the theoretical curve. Clearly, further validation of the theoretical relationship in a variety of environments is important as photoelectric heating is at the core of all PDR research, including studies on the phase structure of the ISM \citep{Wolfire1995, Wolfire2003, HollenbachTielens}.
The significance of PAH photoelectric heating is reflected in the tight correlation between PAH and [C\,{\sc ii}] emission from the PDRs (see Figs. \ref{Fig.corr}f and \ref{Fig.corrAB}d). In a future study, we intend to return to this issue of the importance of PAHs to the heating of interstellar gas.

\subsection{Line cuts}
\label{sec.crosscuts}
\begin{figure}[ht]
\centering
\includegraphics[width=0.5\textwidth, height=0.45\textwidth]{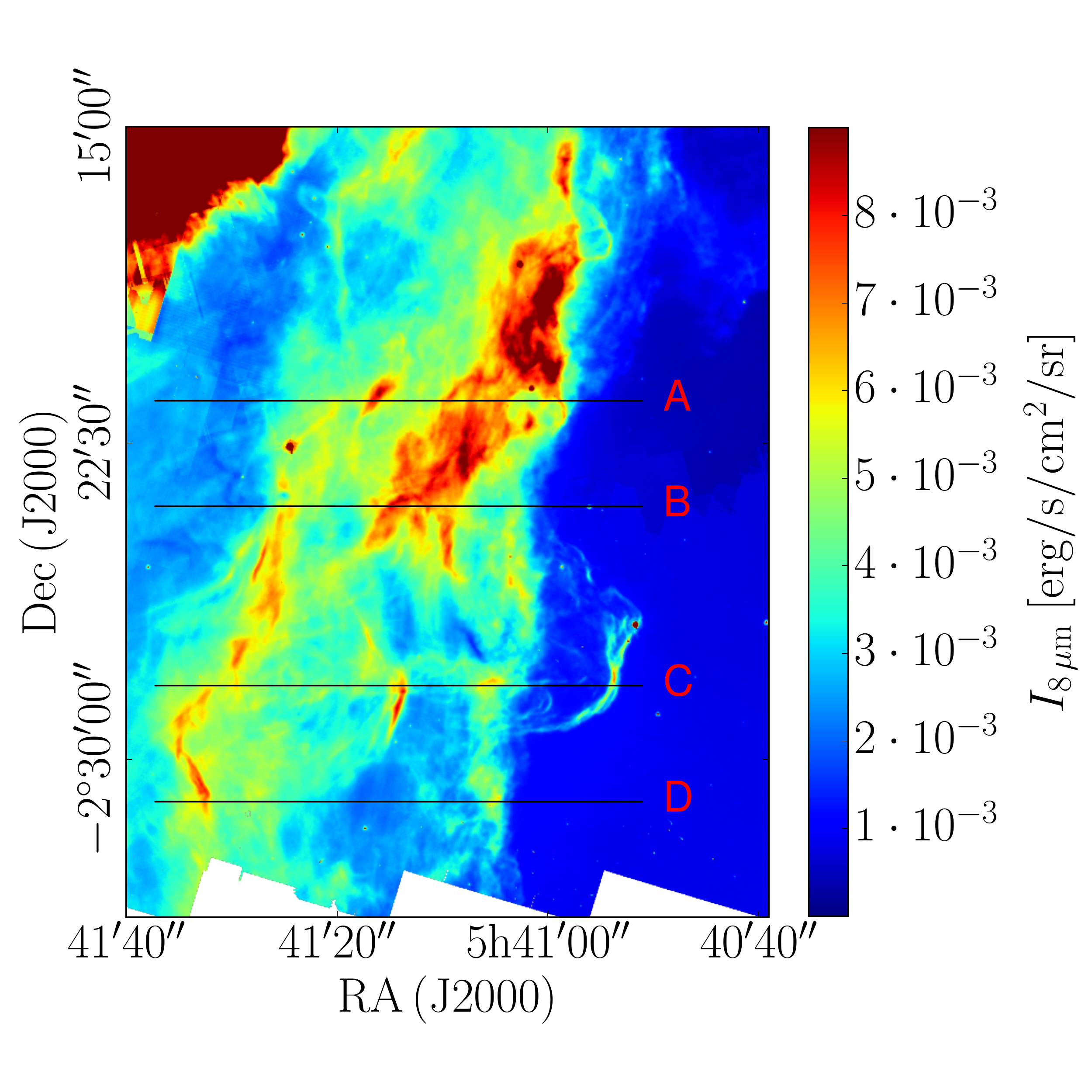}
\caption{IRAC $8\,\mu\mathrm{m}$ image with lines A, B, C, and D indicated.}
\label{Fig.cuts}
\end{figure}

\begin{figure*}[ht]
\centering
\includegraphics[width=\textwidth, height=0.67\textwidth]{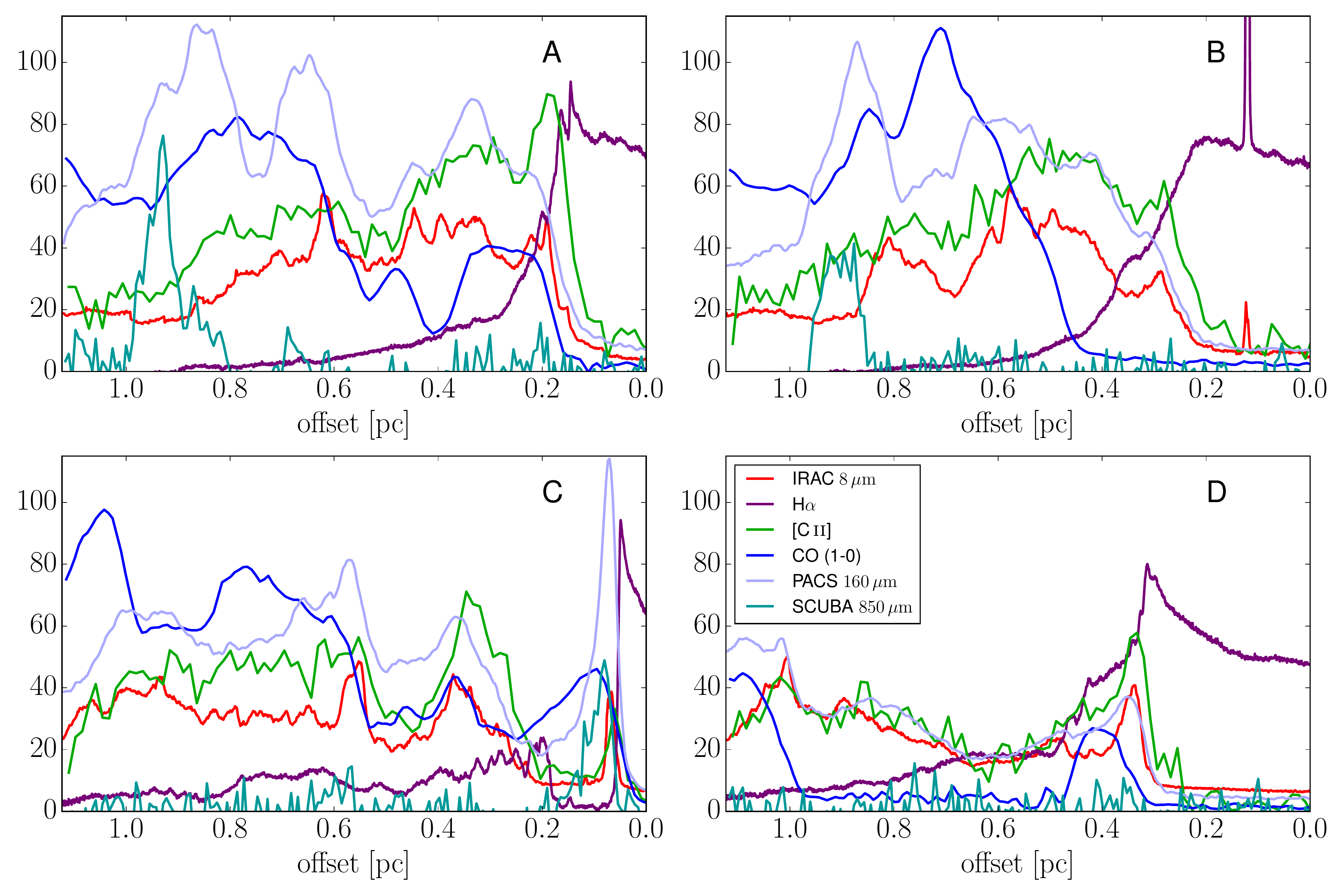}
\caption{IRAC $8\,\mu\mathrm{m}$ (PAH) intensity, H$\alpha$ intensity, [C\,{\sc ii}] and $\mathrm{CO\,(1\mhyphen 0)}$ line-integrated intensity, PACS $160\,\mu\mathrm{m}$ intensity, and SCUBA $850\,\mu\mathrm{m}$ intensity plotted in their respective native resolution along lines A, B, C, and D of Fig. \ref{Fig.cuts} in their respective original spatial resolution. Multiply plotted values by $1.4\cdot 10^{-4}$, $2.9\cdot 10^{-3}$, $7.0\cdot 10^{-6}$, $1.6\cdot 10^{-9}$, $3,9\cdot 10^{-5}$, and $2.2\cdot 10^{-7}$, respectively, for $I_{8\,\mu\mathrm{m}}$, $I_{\mathrm{H\alpha}}$, $I_{\mathrm{[C\,\textsc{ii}]}}$, $I_{\mathrm{CO\,(1\mhyphen 0)}}$, $I_{\mathrm{160\,\mu m}}$, and $I_{\mathrm{850\,\mu m}}$, respectively, in $\mathrm{erg\,s^{-1}\,cm^{-2}\,sr^{-1}}$.}
\label{Fig.cross-cuts}
\end{figure*}

The edge-on nature of the PDR in the Orion B molecular cloud (L1630) is well illustrated by line cuts taken from the surface of the molecular cloud into the bulk (cf. Figs. \ref{Fig.cuts} and \ref{Fig.cross-cuts}). In addition to previously employed tracers ($8\,\mu\mathrm{m}$, H$\alpha$, [C\,{\sc ii}], CO$\,(1\mhyphen 0)$), we compare with SCUBA 850$\mu$m observations, that trace dense clumps, and compare PACS 160$\mu$m data as a measure for $I_{\mathrm{FIR}}$.

Along line cut C, the Horsehead PDR is clearly distinguishable; H$\alpha$ drops immediately and the other four tracers peak, which indicates high density. Assuming $A_{\mathrm{V}}$=2 for the transition from $\mathrm{C}^+$/C to CO and $N_{\mathrm{H}}=2\cdot 10^{21}\,\mathrm{cm}^{-2}\,A_{\mathrm{V}}$, which is consistent with our models (see Fig. \ref{Fig.model}), we obtain from the physical position of the transition, $d\simeq 0.03\,\mathrm{pc}$, $n_{\mathrm{H}}\simeq 4\cdot 10^{4}\,\mathrm{cm}^{-3}$. Since there are no further indications of dense clumps in $850\,\mu\mathrm{m}$ emission, we assume that the rest of the gas located along cut C is relatively diffuse. Having established that, we tentatively assign the CO peak between 0.6 and $0.8\,\mathrm{pc}$ to the PDR surface at $0.3\,\mathrm{pc}$. We infer a gas density of $n_{\mathrm{H}}\simeq 3\cdot 10^{3}\,\mathrm{cm}^{-3}$. Identifying the CO peak at $1.1\,\mathrm{pc}$ with the PDR surface at $0.6\,\mathrm{pc}$ yields about the same density.

If we perform the same procedure for line cut A, we derive about the same densities, $n_{\mathrm{H}}\simeq 3\cdot 10^{3} \,\mathrm{cm}^{-3}$. Here, we assign the broad prominent CO feature at $0.8\,\mathrm{pc}$ to the broad PDR feature at $0.4\,\mathrm{pc}$. The small CO peak at $0.5\,\mathrm{pc}$ possibly relates to the PDR feature at $0.4\,\mathrm{pc}$, hence $n_{\mathrm{H}}\simeq 10^{4} \,\mathrm{cm}^{-3}$. The broad CO feature between 0.2 and $0.4\,\mathrm{pc}$ could originate from a dense surface, located at $0.2\,\mathrm{pc}$.

Line cut B is difficult to interpret. Due to their respective shapes, the PAH peaks around $0.6\,\mathrm{pc}$ might correspond to the CO peaks at 0.7 and $0.85\,\mathrm{pc}$, yielding a density of $n_{\mathrm{H}}\simeq 5\cdot 10^{3}\mhyphen 10^4 \,\mathrm{cm}^{-3}$. There is no indication of dense clumps in the SCUBA map at this point. The gas at the surface is likely to be relatively diffuse, since there is no distinct CO peak that could be related.

The gas at the surface cut by line D is denser again. Here, we estimate $n_{\mathrm{H}}\simeq 3\cdot 10^{4} \,\mathrm{cm}^{-3}$. PDR3 at $1.0\,\mathrm{pc}$ might be relatively dense too, since there is CO emission peaking directly behind the PDR front. However, this CO emission could certainly originate from deeper, less dense layers of the molecular cloud, as well. There is no $850\,\mu\mathrm{m}$ peak corresponding to the CO peak, rendering the latter hypothesis more plausible.

There are a number of CO peaks we cannot relate to a specific structure, for example, the one at $0.4\,\mathrm{pc}$ in line cut C. This might indicate that there are layers of gas with higher density stacked along the line of sight. Overall, this analysis suffers from numerous uncertainties and unknowns. Nevertheless, the calculated densities seem to be reasonable for a molecular cloud.

\subsection{Geometry of the L1630 molecular cloud}

\begin{figure}[ht]
\includegraphics[width=0.5\textwidth, height=0.375\textwidth]{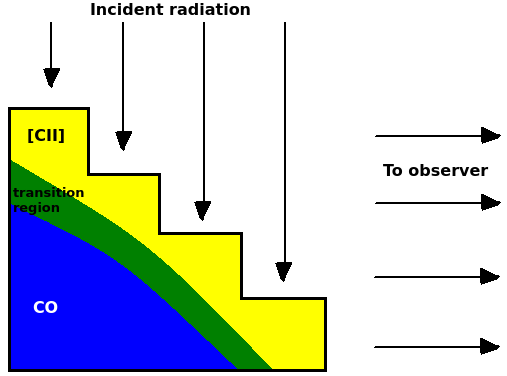}
\caption{Geometry of the L1630 molecular cloud surface.}
\label{Fig.drawing}
\end{figure}

Our data suggest that the L1630 molecular cloud as we see it consists of stacked layers of PDRs along the line of sight, which are offset against each other. Figure \ref{Fig.drawing} shows a schematic illustration, which we inferred from our data, of the edge-on geometry of the studied region of the Orion B molecular cloud. The surface of the molecular cloud is inclined with respect to the incident radiation and the observer, but with steps where strong PDR surfaces can be discerned in the IRAC $8\,\mu\mathrm{m}$ image, for instance. This implies that we generally cannot compare our data with face-on PDRs; in the correlation plots, we need to correlate along line cuts and not along the line of sight. Also, the incident radiation is not easily estimated (see discussion in Sec. \ref{sec.FIR}).

The CO emission in the Horsehead neck stems from a different distance along the line of sight than the overlapping [C\,{\sc ii}] emission. The CO emission stems from the shadow of the Horsehead Nebula, whereas $I_{\mathrm{FIR}}$ and $I_{\mathrm{[C\,\textsc{ii}]}}$ originate from the surface of the bulk molecular cloud. They are spatially not coincident and are likely to correspond to different densities. Hence, the [C\,{\sc ii}] and the CO emission from this part cannot be correlated.

\begin{figure}[ht]
\includegraphics[width=0.5\textwidth, height=0.475\textwidth]{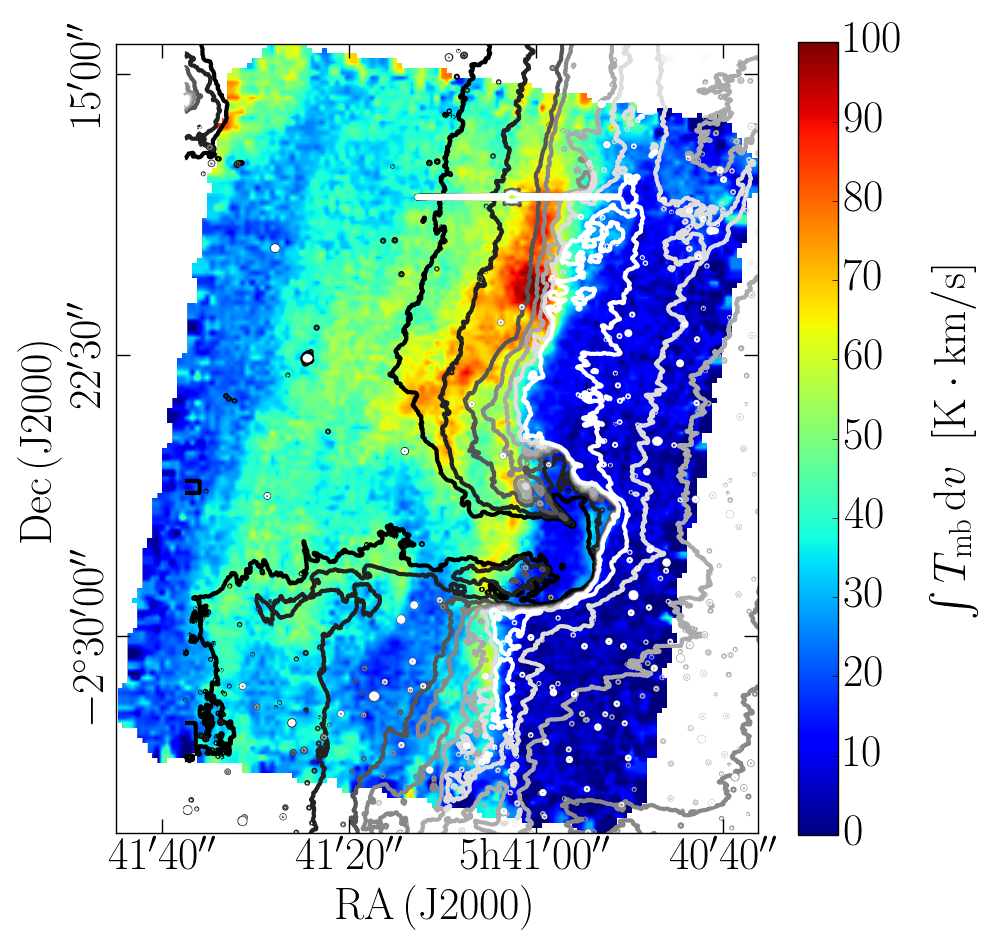}
\caption{Original [C\,{\sc ii}] image (line-integrated intensity) with H$\alpha$ emission in contours (slightly smoothed). Contours from black to white: $I_{\mathrm{H\alpha}}=0.5,1.0,2.0,3.0,4.0,5.0,6.0\;\cdot 10^{-3}\,\mathrm{erg\,s^{-1}\,cm^{-2}\,sr^{-1}}$.}
\label{Fig.Ha_contours}
\end{figure}

We do not see sharp H$\alpha$ edges along the line cuts, as we would expect for multiple PDR fronts, but only on the primary surface of the molecular cloud. We clearly notice the onset of the bulk cloud behind the Horsehead Nebula in line cut C, and in line cut D there is a distinct shoulder to the primary H$\alpha$ peak. Apart from that, the H$\alpha$ emission across the molecular cloud is somewhat diffuse. We do notice coincidence, however, between the H$\alpha$ emission contours and the PDR surfaces as traced by [C\,{\sc ii}] emission. In Fig. \ref{Fig.Ha_contours}, PDR1 and PDR2 are lined-out by contours of high $I_{\mathrm{H\alpha}}$, whereas PDR3 and the intermediate PDR fronts are traced by contours of lower $I_{\mathrm{H\alpha}}$.

From the dust optical depth and the inferred densities, we can estimate the length of the PDR along the line of sight. At the surface of the Horsehead Nebula, $\tau_{160} \simeq 10^{-3}$ together with $n_{\mathrm{H}}\simeq 4\cdot 10^4\,\mathrm{cm}^{-3}$ yields an estimate of about $l_{\mathrm{PDR}}\simeq 0.05\,\mathrm{pc}$. In the bulk, however, we would expect something closer to the projected extent, about $\simeq 0.2\,\mathrm{pc}$. Assuming a density of $n_{\mathrm{H}}\simeq 2\cdot 10^4\,\mathrm{cm}^{-3}$, \cite{Habart2005} arrive at $l_{\mathrm{PDR}}\simeq 0.5\,\mathrm{pc}$, but comment that this seems implausibly high; they conclude that the density must actually be higher. If we adopt $n_{\mathrm{H}}\sim 2\cdot 10^{5}\,\mathrm{cm}^{-3}$ \citep{Habart2005} for the bulk and a dust optical depth of $\tau_{160}\simeq 10^{-2}$, we obtain $l_{\mathrm{Horsehead}}\simeq 0.17\,\mathrm{pc}$.

For PDR1 and PDR2, we estimate $l_{\mathrm{PDR1}}\simeq 1\,\mathrm{pc}$ and $l_{\mathrm{PDR2}}\simeq 0.5\,\mathrm{pc}$, respectively. PDR3 yields $l_{\mathrm{PDR3}}\simeq 1.5\,\mathrm{pc}$, but this is most likely not the length of the PDR but of the whole PDR+molecular cloud interior column. From the $G_0$ values of Sec. \ref{sec.FIR}, assuming that $G_0 \propto l_{\mathrm{PDR}}$, we infer that $l_{\mathrm{PDR3}}$ lies in between $l_{\mathrm{PDR1}}$ and $l_{\mathrm{PDR2}}$.

\subsection{Comparison with models}
\label{sec.models}
As described in Sec. \ref{sec.model_description}, we ran PDR models for an edge-on geometry with varying length along the line of sight and varying density (see also Fig. \ref{Fig.model}) at $G_0=100$. The respective model relations between tracers are plotted in Figs. \ref{Fig.corr}a, b, e, and \ref{Fig.CplusCO}: $I_{\mathrm{[C\,\textsc{ii}]}}/I_{\mathrm{FIR}}$ versus $I_{\mathrm{FIR}}$, $I_{\mathrm{[C\,\textsc{ii}]}}$ versus $I_{\mathrm{FIR}}$, $I_{\mathrm{[C\,\textsc{ii}]}}$ versus $I_{\mathrm{CO\,(1\mhyphen 0)}}$, and $I_{\mathrm{[C\,\textsc{ii}]}}/I_{\mathrm{FIR}}$ versus $I_{\mathrm{CO\,(1\mhyphen 0)}}/I_{\mathrm{FIR}}$, respectively.

We have constructed models for selected depths of the molecular cloud along the line of sight $A_{\mathrm{V,los}}$, which we calculated from the gas column density using $N_{\mathrm{H}}=2.0\cdot 10^{21}\,\mathrm{cm}^{-2}\,A_{\mathrm{V}}$, and gas densities $n_{\mathrm{H}}$, that we inferred from $\tau_{160}$ in Sec. \ref{sec.column} and the line cuts in Sec. \ref{sec.crosscuts}, respectively. For PDR1, we estimated $A_{\mathrm{V,los}}\simeq 5.0$, whereas for PDR2 we got $A_{\mathrm{V,los}}\simeq 2.5$. The Horsehead PDR ranges from $A_{\mathrm{V,los}}\simeq 2.5$ at the edge to $A_{\mathrm{V,los}}\gtrsim 5.0$ deeper in the Horsehead Nebula. For the gas in PDR1 and PDR2, we assume $n_{\mathrm{H}}=3.0\cdot 10^3\,\mathrm{cm}^{-3}$; for the Horsehead PDR we take $n_{\mathrm{H}}=4.0\cdot 10^4\,\mathrm{cm}^{-3}$, but we note that the density increases with depth into the PDR according to a number of studies on that region \citep[cf., e.g.,][]{Habart2005}. Further, we ran models for $n_{\mathrm{H}}=1.6\cdot 10^4\,\mathrm{cm}^{-3}$ to probe a density region that might either be occupied by the Horsehead PDR or by a denser (surface) structure overlaid on PDR1 and PDR2.

Figure \ref{Fig.corr}a shows that the model predictions for the Horsehead with $A_{\mathrm{V,los}}=0.5$ and $n_{\mathrm{H}}= 1.6\cdot 10^4\,\mathrm{cm}^{-3}$ or $n_{\mathrm{H}}=4.0\cdot 10^4\,\mathrm{cm}^{-3}$ are consistent with the data, whereas we estimated $A_{\mathrm{V,los}}=2.5$. The $A_{\mathrm{V,los}}=2.5$ line with $n_{\mathrm{H}}=4.0\cdot 10^4\,\mathrm{cm}^{-3}$ could fit the data within the extent of beam dilution. The $A_{\mathrm{V,los}}=5.0$ model lines cannot explain the data, even when taking into account beam-dilution effects. Another reason for the discrepancy might be dust depletion in this region decreasing the FIR emission. PDR1 is matched quite well by the curve for $A_{\mathrm{V,los}}=5.0$ and $n_{\mathrm{H}}= 3.0\cdot 10^3\,\mathrm{cm}^{-3}$, but the data points lie close to the $A_{\mathrm{V,los}}=2.5$ with $n_{\mathrm{H}}= 3.0\cdot 10^3\,\mathrm{cm}^{-3}$ line, as well. The latter also fits PDR2 quite well. The rest of the data points lie between these two curves and the curves for $A_{\mathrm{V,los}}=0.5$. Similar conclusions can be drawn for Fig. \ref{Fig.corr}b. The PDR data points are matched by the same model curves as before, with the remaining points lying in between these and the $A_{\mathrm{V,los}}=0.5$ curves.

The comparison of the data with the model predictions for the relation between [C\,{\sc ii}] and CO$\,(1\mhyphen 0)$ is less clear-cut. In Fig. \ref{Fig.corr}e, PDR1 and PDR2 lie close to the  $A_{\mathrm{V,los}}=5.0$ line. However, also the higher density lines agree with the data points in PDR1. The Horsehead points lie close to the $A_{\mathrm{V,los}}=2.5$ lines instead of the $A_{\mathrm{V,los}}=5.0\mbox{ or }0.5$ lines. This might reflect the fact that $A_{\mathrm{V,los}}=2.5$ in deeper layers of the Horsehead PDR, whereas it might be lower in the [C\,{\sc ii}] emitting surface layers.

In Fig. \ref{Fig.CplusCO}, model lines group according to their respective densities and there is comparatively little variation with $A_{\mathrm{V,los}}$. PDR1 lies on the model lines with $n_{\mathrm{H}}= 1.6\cdot 10^4\,\mathrm{cm}^{-3}$. This behavior might imply that the density deeper in the cloud in this region is increased. PDR2 lies on the $n_{\mathrm{H}}= 3.0\cdot 10^3\,\mathrm{cm}^{-3}$ lines. All the $A_{\mathrm{V,los}}$ lines for $n_{\mathrm{H}}= 1.6\cdot 10^4\,\mathrm{cm}^{-3}$ run through the Horsehead data points, though some data points lie closer to the model lines for higher or lower density.

The gas temperatures in the top layers of the PDRs are predicted by the models as $T\simeq 130\,\mathrm{K}$ for PDR1 and PDR2, and $T\simeq 115\,\mathrm{K}$ for the Horsehead PDR. The latter again is a deviation from the value we infer, $T\simeq 60\,\mathrm{K}$, and the temperature estimated by \cite{Habart2011}, which is $T\simeq 264\,\mathrm{K}$; it lies close to the value assumed by \cite{Goico09} for modeling [O\,{\sc i}] emission from the Horsehead PDR, $T\simeq 100\,\mathrm{K}$, and close to the value deduced from equating the photoelectric heating efficiency with the [C\,{\sc ii}]+[O\,{\sc i}] cooling efficiency, $T\simeq 100\,\mathrm{K}$ (cf. Sec. \ref{sec.heating}). For PDR1 and PDR2, the gas temperatures computed in Sec. \ref{sec.column} assuming excitation by C$^+$--H collisions, $T\simeq 90\,\mathrm{K,}$ are in good enough agreement with the value predicted by the model, especially considering the uncertainties. The derived temperatures would be substantially higher if we considered collisions with H$_2$ to dominate the excitation of C$^+$ ($T\simeq 160\mhyphen 200\,\mathrm{K}$); the contribution of C$^+$--H$_2$ collisions may be reflected in the underestimation of the gas temperature. Other reasons for slight discrepancies may be the uncertainties in gas density, the error margins of the spectral parameters that we used in the analysis, the uncertainty in the C$^+$ column, and, of course, uncertainties within the models.

In conclusion, we are quite capable of reproducing the observed data with model data within the given uncertainties. Our $A_{\mathrm{V,los}}$ and the gas densities can only be estimates, as is the geometry of the molecular cloud that probably has a corrugated surface and density gradients running through it.

\subsection{Comparison with OMC1 in the Orion A molecular cloud}
\label{sec:OMC1}

\begin{figure*}[!htb]
\begin{minipage}{0.49\textwidth}
\includegraphics[width=\textwidth, height=0.67\textwidth]{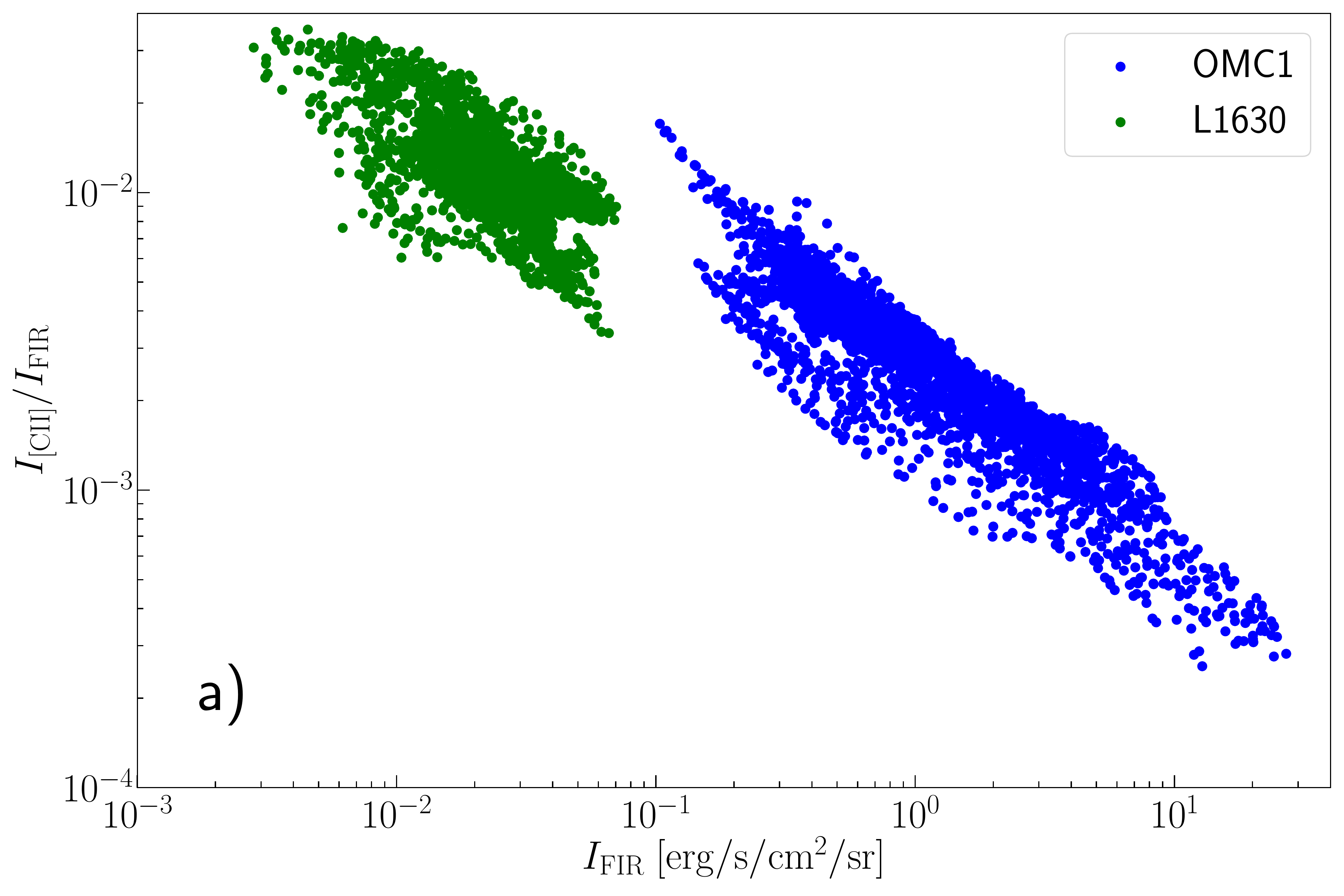}
\end{minipage}
\begin{minipage}{0.49\textwidth}
\includegraphics[width=\textwidth, height=0.67\textwidth]{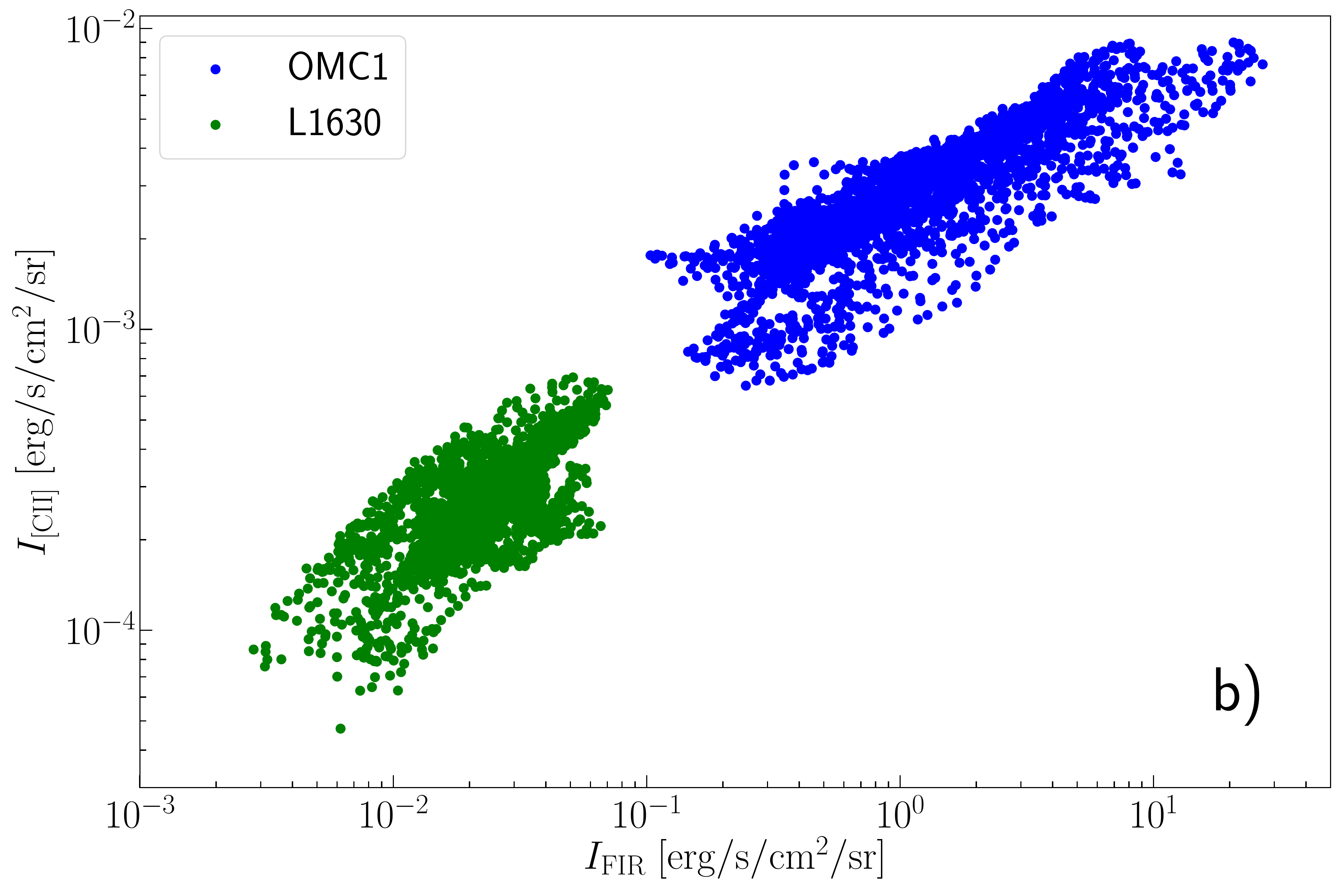}
\end{minipage}

\begin{minipage}{0.49\textwidth}
\includegraphics[width=\textwidth, height=0.67\textwidth]{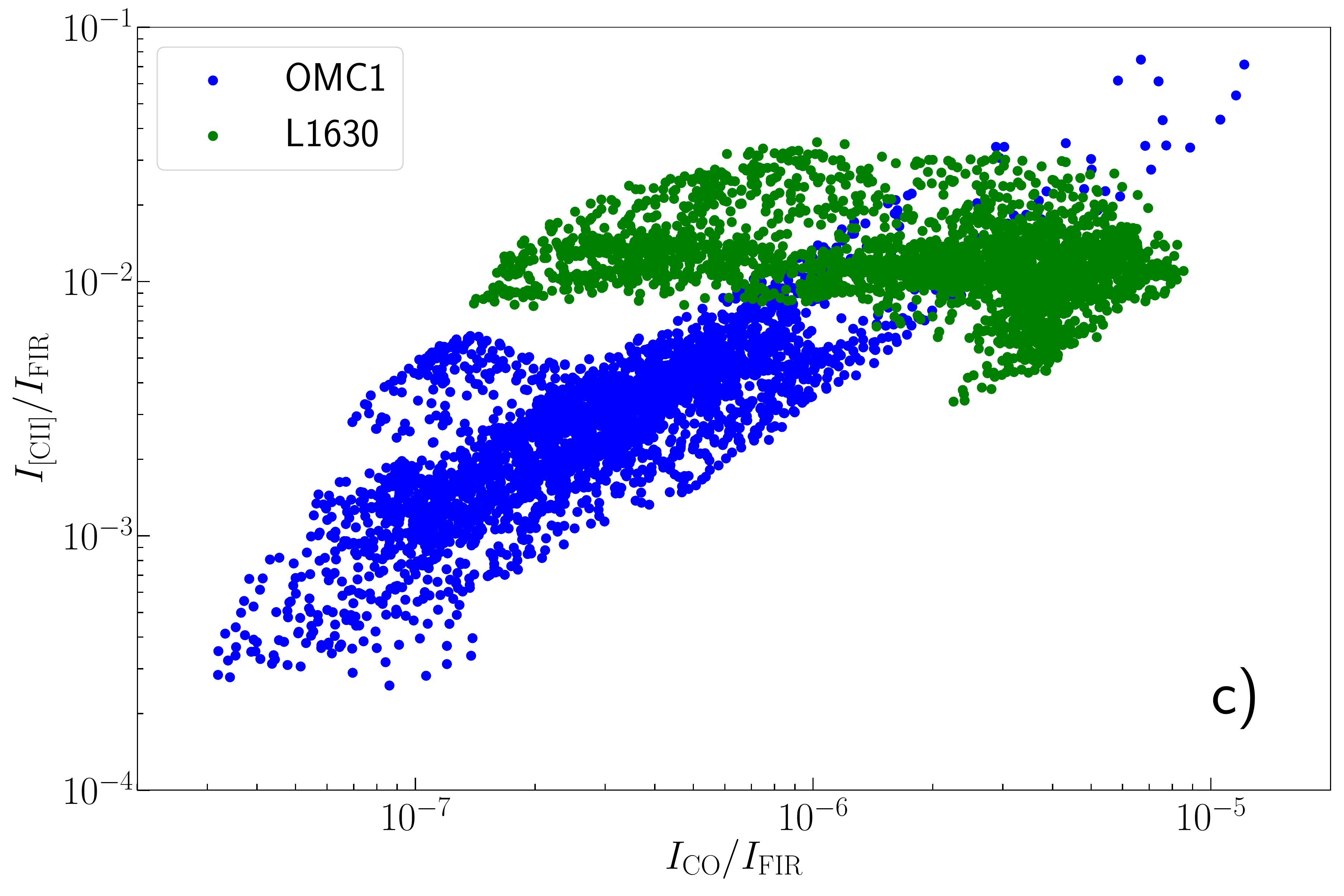}
\end{minipage}
\begin{minipage}{0.49\textwidth}
\includegraphics[width=\textwidth, height=0.67\textwidth]{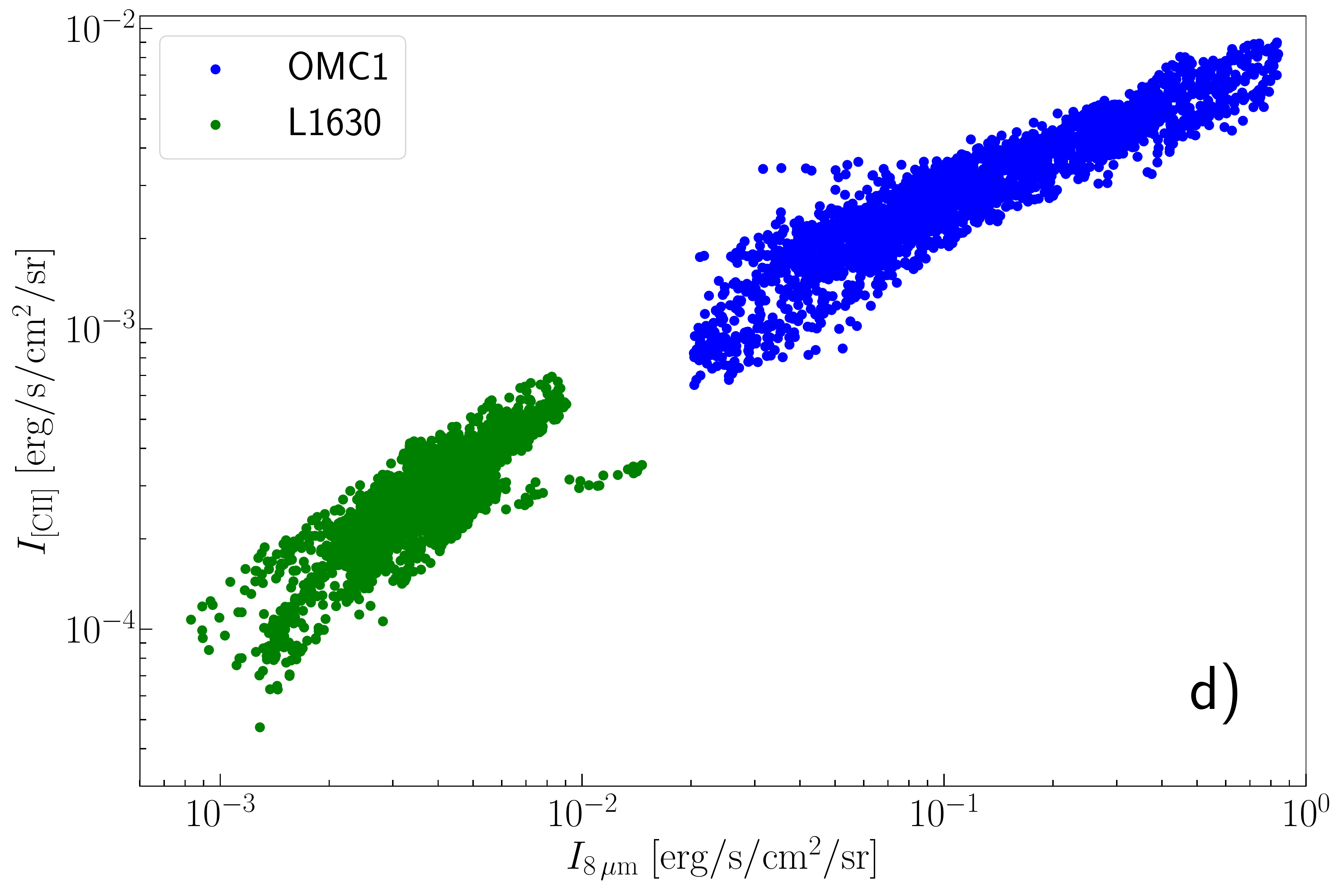}
\end{minipage}
\caption{Correlation plots for L1630 (Orion B) and OMC1 (Orion A); the OMC1 data are convolved to $25\arcsec$ resolution, L1630 data to $36\arcsec$ resolution. CO data are for the CO$\,(1\mhyphen 0)$ line in L1630 and for the CO$\,(2\mhyphen 1)$ line, divided by 8 (see Sec. \ref{sec:OMC1}), in OMC1.}
\label{Fig.corrAB}
\end{figure*}

One of the aims of the present study is to establish correlations of astrophysical tracers under moderate conditions (intermediate density and moderate UV-radiation field) and to compare those to correlations found under harsher conditions (higher density and strong UV-radiation field). An example of the latter conditions is OMC1 in the Orion A complex. Whereas L1630 has edge-on geometry, OMC1 can be approximated as a face-on PDR with respect to its UV-illuminating sources, the Trapezium cluster. In this section, we compare the correlations found in L1630 to those in OMC1. Data on OMC1 are from \cite{Goico}.

The OMC1 data seem to continue the trend found in our L1630 data very well in Fig. \ref{Fig.corrAB}a, b, and d. Both in OMC1 and in L1630, the $8\,\mu\mathrm{m}$ emission correlates well with $I_{\mathrm{[C\,\textsc{ii}]}}$. From a linear fit, we find $I_{\mathrm{[C\,\textsc{ii}]}} \simeq 4.9\cdot 10^{-2} I_{8\,\mu\mathrm{m}} + 9.0\cdot 10^{-5}\,\mathrm{erg\,s^{-1}\,cm^{-2}\,sr^{-1}}$ ($\rho=0.79$), which is similar to the result of \cite{Goico}: $I_{\mathrm{[C\,\textsc{ii}]}} \simeq 2.6\cdot 10^{-2} I_{8\,\mu\mathrm{m}} + 1.6\cdot 10^{-3}\,\mathrm{erg\,s^{-1}\,cm^{-2}\,sr^{-1}}$ ($\rho=0.91$). However, the slope of \cite{Goico} should be divided by 2.9, since they use a bandwidth of the IRAC $8\,\mu\mathrm{m}$ band of $1\,\mu\mbox{m}$, whereas it is really $2.9\,\mu\mbox{m}$. Thus, there is some discrepancy, which expresses itself as a flattening of the correlation curve at high $8\,\mu\mathrm{m}$ and [C\,{\sc ii}] intensity, comparable to the $I_{\mathrm{FIR}}$-$I_{\mathrm{[C\,\textsc{ii}]}}$ dependency (Fig. \ref{Fig.corrAB}a).

In Fig. \ref{Fig.corrAB}c, the $I_{\mathrm{[C\,\textsc{ii}]}}/I_{\mathrm{FIR}}$ versus $I_{\mathrm{CO}}/I_{\mathrm{FIR}}$ relation in L1630 and OMC1 is shown. CO-line intensities in L1630 are from the CO$\,(1\mhyphen 0)$ transition, whereas in OMC1 line intensities are from the CO$\,(2\mhyphen 1)$ transition; we divide the CO$\,(2\mhyphen 1)$ intensity in OMC1 by 8, the frequency ratio of the two lines cubed, which for optically thick thermalized CO emission as in OMC1 (cf. \cite{Goico}) gives a good estimate of the CO$\,(1\mhyphen 0)$ intensity. The data points form two patches with different slopes. Yet, the L1630 data in their entirety seem to show a continuation of the trend set by the OMC1 data. We note that L1630 is characterized by a much higher $I_{\mathrm{[C\,\textsc{ii}]}}/I_{\mathrm{FIR}}$ which reflects the decrease of photoelectric heating efficiency with increasing $G_0$ and the importance of [O\,{\sc i}] cooling for high $G_0$ and high density. The difference in behavior of the $I_{\mathrm{[C\,\textsc{ii}]}}/I_{\mathrm{FIR}}$ versus $I_{\mathrm{CO}}/I_{\mathrm{FIR}}$ relation reflects the difference in geometry. As demonstrated in the study of \cite{Goico}, OMC1 is well-modeled as a face-on PDR, while L1630 has edge-on geometry (cf. Fig. \ref{Fig.CplusCO}).

\cite{Goico} find a decrease of $I_{\mathrm{[C\,\textsc{ii}]}}/I_{\mathrm{FIR}}$ with dust temperature, which is notably different from our observation shown in Fig. \ref{Fig.corr}d; we find no significant slope. Also, we obtain a less good correlation of $I_{\mathrm{FIR}}$ with $\tau_{160}$ than do \cite{Goico} for OMC1 (cf. Sec. \ref{sec.corr}, Fig. \ref{Fig.FIR}a).
The figure corresponding to our Fig. \ref{Fig.Cplustau}, $I_{\mathrm{[C\,\textsc{ii}]}}/I_{\mathrm{FIR}}$ versus $\tau_{160}$, in \cite{Goico}, reveals that OMC1 matches a slab of constant foreground [C\,{\sc ii}] emission, that is, a face-on PDR geometry, much better than the studied region of L1630.

\section{Conclusion}
We have analyzed the velocity-resolved [C\,{\sc ii}] map towards the Orion B molecular cloud L1630 observed by upGREAT onboard SOFIA. We compared the observations with FIR photometry, IRAM $30\,\mathrm{m}$ CO$\,(1\mhyphen 0)$, IRAC $8\,\mu\mathrm{m}$, SCUBA $850\,\mu\mathrm{m,}$ and H$\alpha$ observations.

About 5\% of the total [C\,{\sc ii}] luminosity, $1\,L_{\sun}$, of the surveyed area stems from the H\,{\sc ii} region IC 434; the molecular cloud (not including the north-eastern corner with possible contamination from NGC 2023) accounts for 95\% , that is, $13\,L_{\sun}$. The bulk of the [C\,{\sc ii}] emission originates from PDR surfaces. The total FIR luminosity of the mapped area (without NGC 2023) is $1210\,L_{\sun}$, of which $1175\,L_{\sun}$ stem from the molecular cloud and $35\,L_{\sun}$ from the H\,{\sc ii} region. This yields an average [C\,{\sc ii}] cooling efficiency in the molecular cloud within the mapped area of 1\%. From the dust optical depth, we derive a total gas mass of $M_{\mathrm{gas}}\simeq 280\,M_{\sun}$. Most of the gas mass, $M_{\mathrm{gas}}\simeq 250\,M_{\sun}$, is contained in the CO-emitting molecular cloud. The [C\,{\sc ii}]-bright gas contributes $M_{\mathrm{gas}}\simeq 20\,M_{\sun}$, which is only about 8\% of the total gas mass in the mapped area. This is in close agreement with the PDR mass fraction traced by [C\,{\sc ii}] found by \cite{Goico} in OMC1, which also is 8\% (within a factor of approximately two). The mass of the H\,{\sc ii} region accounts for an additional $M_{\mathrm{gas}}\simeq 10\,M_{\sun}$.

The [C\,{\sc ii}] cooling efficiency is found to decrease with increasing $I_{\mathrm{FIR}}$, in continuation of the results from OMC1 \citep{Goico}. Its peak value is about $3\cdot 10^{-2}$, ranging down to $3\cdot 10^{-3}$. Highest values are obtained at the edge of the molecular cloud towards the H\,{\sc ii} region. The overall [C\,{\sc ii}] cooling efficiency of the mapped area, calculated from the total luminosities $L_{\mathrm{[C\,\textsc{ii}]}}/L_{\mathrm{FIR}}$, is $\sim 10^{-2}$; this compares to an average single-pixel [C\,{\sc ii}] cooling efficiency in PDR regions of $1.1\pm 0.3\cdot 10^{-2}$. We note that due to the edge-on geometry of the molecular cloud, $I_{\mathrm{FIR}}$ does not trace the incident UV radiation in general, for there may be deeper and colder cloud layers located along the line of sight. The [C\,{\sc ii}] intensity increases with FIR intensity. [C\,{\sc ii}] intensity and PAH $8\,\mu\mathrm{m}$ intensity are closely related, reflecting the predominance of gas heating through the photoelectric effect on (clusters of) PAHs and very small grains.

We derive gas densities of the molecular cloud in the range $n_{\mathrm{H}}\simeq 10^3\mhyphen 10^4\,\mathrm{cm}^{-3}$, with the Horsehead PDR having a slightly higher density, $n_{\mathrm{H}}\simeq 4\cdot 10^4\,\mathrm{cm}^{-3}$. Dust temperatures lie in the range $T_{\mathrm{d}}\simeq 18\mhyphen 32\,\mathrm{K}$. From the column densities, we derive an extent of the cloud along the line of sight of $l\simeq 0.5\mhyphen 1\,\mathrm{pc}$ at the edge of the cloud, and $l\simeq 1.5\,\mathrm{pc}$ at the eastern border of the studied area. The Horsehead Nebula scores low with $l\simeq 0.05\,\mathrm{pc}$. As discussed, these values are afflicted by significant uncertainties.

We estimated the [C\,{\sc ii}] optical depth and the excitation temperature towards three representative points in the mapped area. From our analysis we gather that the column density of the Horsehead PDR cannot be straightforwardly calculated from the dust optical depth. By deep observations of the brightest [$^{13}$C\,{\sc ii}] line, we can calculate the [C\,{\sc ii}] optical depth directly, implying [C\,{\sc ii}] emission to be optically thick: $\tau_{\mathrm{[C\,\textsc{ii}]}}\simeq 2$. The corresponding (beam-averaged) excitation temperature is $T_{\mathrm{ex}}\simeq 60\,\mathrm{K}$, which is basically equal to the gas temperature. From other studies we observe higher temperatures, which would imply a lower $\mathrm{C}^+$ column density. Also our models for lower $\mathrm{C}^+$ column density match the Horsehead data in the correlation diagrams, as opposed to those for the $\mathrm{C}^+$ column density inferred from [$^{13}$C\,{\sc ii}]. Towards PDR1 and PDR2, we obtain [C\,{\sc ii}] optical depths of $\tau_{\mathrm{[C\,\textsc{ii}]}}\simeq 1.5$ and excitation temperatures of $T_{\mathrm{ex}}\simeq 65\,\mathrm{K}$, which gives a gas temperature of $T\simeq 90\,\mathrm{K}$.

The observed [C\,{\sc ii}] intensity, in combination with the [O\,{\sc i}] intensity where appropriate, provides a direct measure of the heating efficiency of the gas. We have compared the observed heating efficiency with models for the photoelectric effect on PAHs, clusters of PAHs, and very small grains by \cite{BakesTielens1994}. Theory and observations show a very similar dependence on the ionization parameter $\gamma$, albeit that theory seems to be offset to a slightly higher efficiency. This may, for example, reflect a too high abundance of these species in the models, an issue we will revisit in a future study.

We have endeavored to establish the edge-on nature of the Orion B L1630 molecular cloud. The data suggest that there are multiple PDR fronts across the molecular cloud, implying that the cloud surface is warped and not a single edge-on bulk. The edge-on warped geometry makes it difficult to correlate different quantities with each other, since they may relate to offset layers of the molecular cloud and depend on the local inclination of the cloud surface as well as on the length of the emitting column along the line of sight. Our model predictions for edge-on PDRs are capable of replicating the observed correlations between $I_{\mathrm{[C\,\textsc{ii}]}}$ and $I_{\mathrm{FIR}}$; we can also interpret the model correlations between $I_{\mathrm{[C\,\textsc{ii}]}}$ and $I_{\mathrm{CO\,(1\mhyphen 0)}}$ .

Velocity-resolved line observations are an excellent tool to study gas dynamics and to identify distinct gas components by their different kinematic behavior. The line profile allows inference of the line optical depth, especially when line observations of isotopes (or isotopologues in the case of molecular lines) are available. In the present study, the [C\,{\sc ii}] line profile yields information on the origin of the emission, narrow in dense PDRs, broad in the H\,{\sc ii} region. In combination with other tracers, we can form a picture of the physical conditions prevailing in a molecular cloud. Several more tracers could be included to render the picture more comprehensive such as [O\,{\sc i}] and H\,{\sc i}.

\vspace{1cm}

\begin{acknowledgements}
JRG and EB thank the ERC for funding support under grant ERC-2013-Syg-610256-NANOCOSMOS, and the Spanish MINECO under grant AYA2012-32032. JP, FLP, ER, and EB acknowledge support from the French program "Physique et Chimie du Milieu Interstellaire" (PCMI) funded by the Centre National de la Recherche Scientifique (CNRS) and Centre National d'\'{E}tudes Spatiales (CNES). Studies of the ISM at Leiden Observatory are supported through the Spinoza Prize of the Dutch Science Foundation (NWO).
\end{acknowledgements}

\bibliographystyle{aa} 
\bibliography{article1_arxiv} 

\appendix
\section{Calculating [C\,{\sc ii}] optical depth and excitation temperature}
\label{app.1}

For optically thick [C\,{\sc ii}] emission, we can employ the following two equations to get an estimate of the optical depth and the excitation temperature\footnote{Neglecting background IR emission, $T_{bg}\simeq 2.5\,\mathrm{K}$ for $T_{\mathrm{d}}\simeq 25\,\mathrm{K.}$}:
\begin{align}
T_{\mathrm{P}} &\approx J(T_{\mathrm{ex}})=\frac{91.2\,\mathrm{K}}{e^{91.2\,\mathrm{K}/T_{\mathrm{ex}}}-1} \\
\tau_{\mathrm{[C\,\textsc{ii}]}} &= \frac{A\lambda^3}{4\pi b}N_{\mathrm{C}^+}\frac{e^{91.2\,\mathrm{K}/T_{\mathrm{ex}}}-1}{e^{91.2\,\mathrm{K}/T_{\mathrm{ex}}}+2} \\
&\approx\frac{A\lambda^3}{4\pi b}N_{\mathrm{C}^+}\frac{91.2\,\mathrm{K}/T_{\mathrm{P}}}{91.2\,\mathrm{K}/T_{\mathrm{P}}+3},
\end{align}
where $A=2.3\cdot 10^{-6}\mbox{ s}^{-1}$ and $b$ is the line width with $\tau_{\mathrm{[C\,\textsc{ii}]}}b\approx \int\tau_{\mathrm{[C\,\textsc{ii}]}} \mathrm{d}v$ (cf. eq. (5) in \cite{Gerin}). For an optically thin emission, we take into account the dependence of the peak temperature on the optical depth. Here, we assume that $\tilde{f}(\tau)=1-e^{-\tau}$ accounts for the increase in the peak temperature with optical depth.\footnote{From \cite{Tielens} we get $f(\tau)=\int\limits_0^{\tau} \beta(\tau')\,\mathrm{d}\tau'$ with eq. (2.44) for $\beta$. However, this expression takes into account opacity broadening of the line; the standard radiative transfer solution uses $\tilde{f}(\tau)=1-e^{-\tau}$, so we assume this to be the expression describing the increase in peak emission.} Hence, the above becomes
\begin{align}
T_{\mathrm{P}} &\approx J(T_{\mathrm{ex}})(1-e^{-\tau_{\mathrm{[C\,\textsc{ii}]}}}) \label{eq.Tp2} \\
\tau_{\mathrm{[C\,\textsc{ii}]}} &\approx \frac{A\lambda^3}{4\pi b}N_{\mathrm{C}^+}\frac{91.2\,\mathrm{ K}(1-e^{-\tau_{\mathrm{[C\,\textsc{ii}]}}})/T_{\mathrm{P}}}{91.2\,\mathrm{ K}(1-e^{-\tau_{\mathrm{[C\,\textsc{ii}]}}})/T_{\mathrm{P}}+3}.\label{eq.tau2}
\end{align}
Equation (\ref{eq.tau2}) has the form $\tau=g(\tau)$, which can be solved either graphically or by the fix-point method. In order to do so, we need to make some assumptions on $b$ and $N_{\mathrm{C}^+}$. In the following, we assume that opacity broadening is insignificant, that is, the lines are Gaussian such that $b\approx \Delta v_{\mathrm{FWHM}}$. The  $\mathrm{C}^+$ column density we take from the dust optical depth.

\section{Face-on calculation}
\label{sec.face-on}

\begin{figure*}[hb]
\includegraphics[width=1.0\textwidth, height=0.5\textwidth]{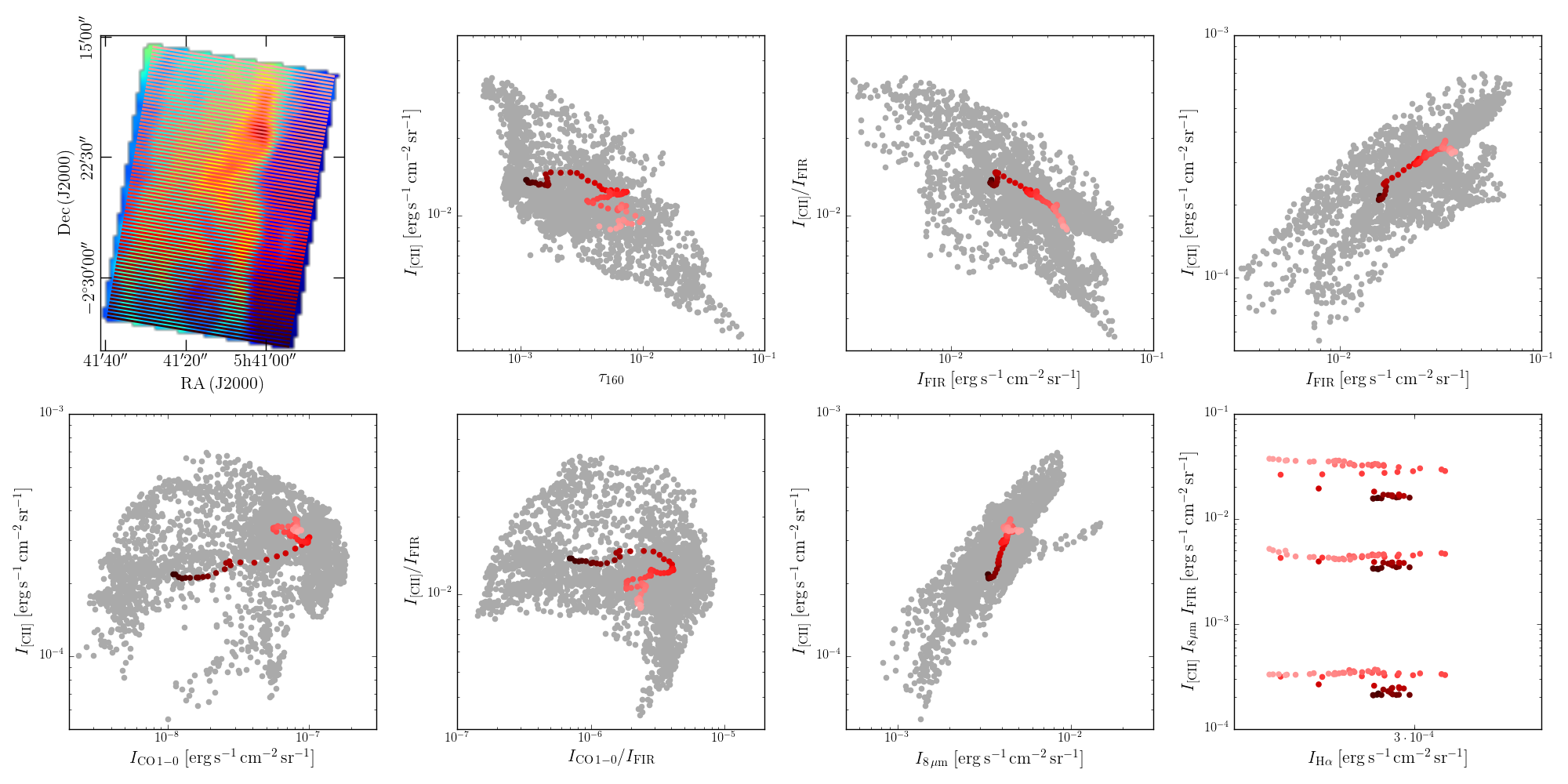}
\caption{Correlation plots. Red dots are the face-on integrated values; gray dots are the original data shown throughout the main text. The first panel shows the lines along which the data are integrated.}
\label{Fig.face-on}
\end{figure*}

In order to compare our observations with face-on PDRs, we integrate our, presumably edge-on, observations along the depth into the molecular cloud from the surface with respect to the incident FUV radiation. We normalize the measured intensity to the pixel size ($15\arcsec$), assuming a line-of-sight length of $l\simeq 1\,\mathrm{pc}$ and an average gas density of $n_{\mathrm{H}}\simeq 3\cdot 10^3\,\mathrm{cm}^{-3}$. We further assume that radiation is emitted isotropically and homogeneously along the line of sight. The correlation plots are shown in Fig. \ref{Fig.face-on}.

\end{document}